\documentclass[twocolumn]{aastex631}

\usepackage{graphicx}	% Including figure files
\usepackage{amsmath}	% Advanced maths commands
{\LARGE 
\usepackage{hyperref}
}
\usepackage{wrapfig}
\usepackage{subfigure}

\usepackage{savesym}
\savesymbol{tablenum}
\usepackage{siunitx}
\restoresymbol{SIX}{tablenum}
\newcommand{\ext}{{\color{blue}\textbf{EXT}}}
\newcommand{\band}{{\color{blue}\textbf{BAND}}}

\newcommand{\eazy}{{\tt{EAZY}}}
\newcommand{\bagpipes}{{\tt{Bagpipes}}}

\newcommand{\sextractor}{{\tt{SExtractor}}}

\newcommand{\solm}{$\mathrm{M}_\odot$}

\received{October 3, 2022}
\revised{October XX, 2022}
\accepted{October YY, 2022}

\submitjournal{ApJL}

\shorttitle{JWST EPOCHS High-z Sample}
\shortauthors{Conselice et al.}

\def\solm{M$_{\odot}\,$}

\def\solm{M$_{\odot}\,$}

\def\casgm20{CAS-G-M$_{20}\,$}
\def\m20{M$_{20}\,$}

\begin{document}  

\title{EPOCHS I. The Discovery and Star Forming Properties of Galaxies in the Epoch of Reionization at $6.5 < z < 18$ with PEARLS and Public JWST data}

\correspondingauthor{Christopher Conselice}
\email{conselice@manchester.ac.uk}

\author[0000-0003-1949-7638]{Christopher J. Conselice}
\affiliation{Jodrell Bank Centre for Astrophysics, University of Manchester, Oxford Road, Manchester M13 9PL, UK}

\author[0000-0003-4875-6272]{Nathan Adams}
\affil{Jodrell Bank Centre for Astrophysics, University of Manchester, Oxford Road, Manchester M13 9PL, UK}

\author[0000-0002-4130-636X]{Thomas Harvey}
\affil{Jodrell Bank Centre for Astrophysics, University of Manchester, Oxford Road, Manchester M13 9PL, UK}

\author[0000-0003-0519-9445]{Duncan Austin}
\affil{Jodrell Bank Centre for Astrophysics, University of Manchester, Oxford Road, Manchester M13 9PL, UK}

\author[0000-0002-8919-079X]{Leonardo Ferreira}
\affil{Department of Physics \& Astronomy, University of Victoria, Finnerty Road, Victoria, British Columbia, V8P 1A1, Canada}

\author[0000-0003-2000-3420]{Katherine Ormerod}
\affil{Jodrell Bank Centre for Astrophysics, University of Manchester, Oxford Road, Manchester M13 9PL, UK}
\affil{Astrophysics Research Institute, Liverpool John Moores University, 146 Brownlow Hill, Liverpool, L3 5RF}

\author[0009-0009-8105-4564]{Qiao Duan}
\affiliation{Jodrell Bank Centre for Astrophysics, University of Manchester, Oxford Road, Manchester M13 9PL, UK}

\author[0000-0002-9081-2111]{James Trussler}
\affil{Jodrell Bank Centre for Astrophysics, University of Manchester, Oxford Road, Manchester M13 9PL, UK}

\author[0000-0002-3119-9003]{Qiong Li}
\affil{Jodrell Bank Centre for Astrophysics, University of Manchester, Oxford Road, Manchester M13 9PL, UK}

\author[0009-0003-7423-8660]{Ignas Juodžbalis}
\affiliation{Jodrell Bank Centre for Astrophysics, University of Manchester, Oxford Road, Manchester M13 9PL, UK}
\affiliation{Kavli Institute for Cosmology, University of Cambridge, Madingley Road, Cambridge, CB3 OHA, UK}
\affiliation{Cavendish Laboratory - Astrophysics Group, University of Cambridge, 19 JJ Thomson Avenue, Cambridge, CB3 OHE, UK}

\author[0009-0008-8642-5275]{Lewi Westcott}
\affiliation{Jodrell Bank Centre for Astrophysics, University of Manchester, Oxford Road, Manchester M13 9PL, UK}

\author[0009-0005-0817-6419]{Honor Harris}
\affiliation{Jodrell Bank Centre for Astrophysics, University of Manchester, Oxford Road, Manchester M13 9PL, UK}

\author[0000-0002-7020-3079]{Louise T. C. Seeyave}
\affiliation{Astronomy Centre, University of Sussex, Falmer, Brighton BN1 9QH, UK}

\author[0000-0001-6395-4504]{Asa F. L. Bluck}
\affiliation{Stocker AstroScience Center, Dept. of Physics, Florida International University, 11200 S.W. 8th Street, Miami, FL 33199, USA}

\author[0000-0001-8156-6281]{Rogier A. Windhorst}
\affiliation{School of Earth and Space Exploration, Arizona State University, Tempe, AZ 85287-1404}

\author[0000-0003-0883-2226]{Rachana Bhatawdekar}
\affiliation{European Space Agency (ESA), European Space Astronomy Centre (ESAC), Camino Bajo del Castillo s/n, 28692 Villanueva de la Cañada, Madrid, Spain}

\author[0000-0001-7410-7669]{Dan Coe} %%% dancoe@gmail.com 
\affiliation{AURA for the European Space Agency (ESA), Space Telescope Science
Institute, 3700 San Martin Drive, Baltimore, MD 21218, USA}

\author[0000-0003-3329-1337]{Seth H. Cohen} %%% seth.cohen@asu.edu
\affiliation{School of Earth and Space Exploration, Arizona State University,
Tempe, AZ 85287-1404}

\author[0000-0003-0202-0534]{Cheng Cheng}
\affiliation{Chinese Academy of Sciences South America Center for Astronomy, National Astronomical Observatories, CAS, Beijing 100101, People's Republic of China}
\affiliation{2 CAS Key Laboratory of Optical Astronomy, National Astronomical Observatories, Chinese Academy of Sciences, Beijing 100101, People's Republic of China}

\author[0000-0001-9491-7327]{Simon P. Driver} %%% Simon.Driver@icrar.org
\affiliation{International Centre for Radio Astronomy Research (ICRAR) and the
International Space Centre (ISC), The University of Western Australia, M468,
35 Stirling Highway, Crawley, WA 6009, Australia}

\author[0000-0003-1625-8009]{Brenda Frye} %%% brendafrye@gmail.com 
\affiliation{University of Arizona, Department of Astronomy/Steward
Observatory, 933 N Cherry Ave, Tucson, AZ85721}

\author[0000-0001-6278-032X]{Lukas J. Furtak}
\affiliation{Department of Physics, Ben-Gurion University of the Negev, P.O. Box 653, Be'er-Sheva 84105, Israel}

\author[0000-0001-9440-8872]{Norman A. Grogin} %%% nagrogin@stsci.edu
\affiliation{Space Telescope Science Institute, 3700 San Martin Drive, Baltimore, MD 21218, USA}

\author[0000-0001-6145-5090]{Nimish P. Hathi}
\affiliation{Space Telescope Science Institute, 3700 San Martin Drive, Baltimore, MD 21218, USA}

\author[0000-0002-4884-6756]{Benne W. Holwerda}
\affiliation{Department of Physics and Astronomy, University of Louisville, 102 Natural Sciences Building, Louisville, KY 40292, USA}

\author[0000-0003-1268-5230]{Rolf A. Jansen} %%% rolfjansen.work@gmail.com 
\affiliation{School of Earth and Space Exploration, Arizona State University,
Tempe, AZ 85287-1404}

\author[0000-0002-6610-2048]{Anton M. Koekemoer} %%% koekemoer@stsci.edu 
\affiliation{Space Telescope Science Institute, 3700 San Martin Drive, Baltimore, MD 21218, USA}

\author[0000-0001-6434-7845]{Madeline A. Marshall} %%% madeline_marshall@outlook.com 
\affiliation{National Research Council of Canada, Herzberg Astronomy \&
Astrophysics Research Centre, 5071 West Saanich Road, Victoria, BC V9E 2E7,
Canada; \& ARC Centre of Excellence for All Sky Astrophysics in 3
Dimensions (ASTRO 3D), Australia}

\author[0000-0001-6342-9662]{Mario Nonino}
\affiliation{INAF-Osservatorio Astronomico di Trieste, Via Bazzoni 2, I-34124 Trieste, Italy}

\author[0000-0003-0429-3579]{Aaron Robotham} %%% Simon.Driver@icrar.org
\affiliation{International Centre for Radio Astronomy Research (ICRAR) and the
International Space Centre (ISC), The University of Western Australia, M468,
35 Stirling Highway, Crawley, WA 6009, Australia}

\author[0000-0002-7265-7920]{Jake Summers}
\affiliation{School of Earth and Space Exploration, Arizona State University,
Tempe, AZ 85287-1404}

\author[0000-0003-3903-6935]{Stephen M. Wilkins}
\affiliation{Astronomy Centre, Department of Physics and Astronomy, University of Sussex, Brighton, BN1 9QH, UK}

\author[0000-0001-9262-9997]{Christopher N. A. Willmer} %%% cnawillmer@gmail.com 
\affiliation{Steward Observatory, University of Arizona, 933 N Cherry Ave, Tucson, AZ, 85721-0009}

\author[0000-0001-7592-7714]{Haojing Yan} %%% yanhaojing@gmail.com 
\affiliation{Department of Physics and Astronomy, University of Missouri,
Columbia, MO 65211}

\author[0000-0002-0350-4488]{Adi Zitrin}
\affiliation{Department of Physics, Ben-Gurion University of the Negev, P.O. Box 653, Be'er-Sheva 84105, Israel}

%\author{Scott Tompkins} %%% satompki@asu.edu
%\affiliation{School of Earth and Space Exploration, Arizona State University, Tempe, AZ 85287-1404}

%\author{Jordan C. J. D’Silva}
%\affiliation{International Centre for Radio Astronomy Research (ICRAR) and the International Space Centre (ISC), The University of Western Australia, M468, 35 Stirling Highway, Crawley, WA 6009, Australia}




\begin{abstract}

We present in this paper the discovery, properties,  and a catalog of 1165 high redshift $6.5 < z < 18$ galaxies found in deep JWST NIRCam imaging from the GTO PEARLS survey combined with data from JWST public fields. We describe our bespoke, homogeneous reduction process and our analysis of these areas including the NEP, CEERS, GLASS, NGDEEP, JADES, and ERO SMACS-0723 fields covering a total of over 214 arcmin$^{2}$ imaged down to depths of $\sim 30$ mag. We give a description of our rigorous methods for identifying these galaxies, which involve the use of Lyman-break strength, detection significance criteria, visual inspection, and photometric redshifts probability distributions predominately at high redshift. Our sample is a robust and highly pure collection of distant galaxies from which we also remove brown dwarf stars, and calculate completeness and contamination from simulations. We include a summary of the basic properties of these $z > 6.5$ galaxies, including their redshift distributions, UV absolute magnitudes, standard stellar masses, and star formation rates. Our study of these young galaxies reveals a wide range of stellar population properties as seen in their observed and rest-frame colors which we compare to stellar population models. This mix of systems indicate a range of star formation histories, dust content, AGN and/or nebular emission. We find that a strong trend exists between stellar mass and $(U-V)$ color, as well as the existence of the `main-sequence' of star formation for galaxies as early as $z \sim 12$. This indicates that despite the complexities of galaxy formation,   stellar mass, or an underlying variable correlating with stellar mass, is driving galaxy formation, in agreement with simulation predictions.  We also discuss unusual and very high redshift candidates at $z > 12$ in our sample. Finally, we compare our galaxy counts in redshift to models of galaxy formation, finding a significant observed excess of galaxies at the highest redshifts compared to models at $z > 12$, revealing a tension between predictions and our observations.

\end{abstract}

%\keywords{high redshift galaxies, JWST, galaxy formation}

%% and \citet commands to identify citations. The citations are
\vspace{1em}
\section{Introduction} \label{sec:intro}

The study of high redshift galaxies has since the 1990s been one of the most active areas of astrophysical research, providing critical insights into the early universe and the formation and evolution of galaxies \citep[e.g.,][]{Adams2023, Finkelstein2022-CeersI, Castellano2022, Atek2023}. The earliest galaxies  likely formed within 100 million years of the big bang, and these objects represent the building blocks of the universe as they are the seeds of the structures we observe today. These distant galaxies are furthermore of particular interest because they are thought to have been shaped by different physical processes than those that govern the evolution of galaxies in the present-day universe. This can then lead to a new understanding for how the first large structures in the universe were assembled. However, finding these systems and separating them from contaminants has remained a major problem that has long plagued this field \citep[e.g.,][]{Adams2023,arrabal2023}.

Multiple research groups in the past 30 years have utilized the Hubble Space Telescope (HST) to identify galaxies at redshifts higher than $z \sim 6$ by employing the well-established technique of identifying absorption caused by neutral hydrogen through the Lyman-break method. Before JWST, astronomers identified tens of thousands of  galaxies beyond a redshift of $z = 4$ (corresponding to 10 per cent of the age of the Universe, $\sim1.5$~Gyr), and individual galaxy candidates were known to exist as early as $z \sim 10$ \citep[e.g.][]{Bouwens2011,McLeod2016,Bouwens2016,Oesch2018,Salmon2018,Morishita2018,Stefanon2019,Bowler2020, Harikane2022}. At redshift of $z \sim  8.5$ marks a significant threshold beyond which sources start to ``drop out" in the Hubble Space Telescope (HST) Y-band (F105W) and J-band (F125W) filters, making it a notable frontier for this type of work pre-JWST \citep[][]{Bouwens2011, Ellis2013, Oesch2014, Harikane2022}. 

A primary goal of \emph{JWST} is to push the redshift frontier and search for galaxies that host the first generation of stars when the Universe was less than 5 per cent of its current age. Since the launch of the \textit{James Webb Space Telescope} (JWST) on Christmas day of 2021 there have been many studies with claims for measuring and finding the most distant galaxies in the universe  at redshifts higher than the limit achievable with HST \citep[e.g.,][]{Donnan2022, Adams2023, Finkelstein2022-CeersI, Castellano2022, Atek2023, Yan2023, Harikane2023, Austin2023, Leung2023, Finkelstein2023, Willott2023, McLeod2023}. Whilst many of these galaxy candidates have yet to be verified with spectroscopy at their measured photometric redshifts, it is clear that we have entered a new epoch of extragalactic astronomy which may lead us to discover the first stars, black holes and galaxies \citep[e.g.,][]{trussler2022b, Nabizadeh2024}. Based on these galaxies we hope to be able to answer questions regarding the formation and evolution of the first objects, their dark matter halos, as well as potentially cosmological properties. 

Since the release of the first deep JWST images there are now many deep and independent fields in which to find the most distant galaxy candidates \citep[e.g.,][]{Adams2023, Carnall2022,Donnan2022}.
These early observations from JWST data suggest the possible presence of a significantly higher number of galaxies than initially anticipated, particularly during the epoch of reionization, or potentially even earlier at cosmic dawn. These very early results demonstrate that we are finding candidate galaxies at redshifts upwards of $z>12$ \citep[e.g.,][]{Adams2022,Castellano2022, Naidu2022, Atek2023, Yan2023, Donnan2022, Adams2023}. Some of these galaxies have possible confirmed spectroscopic redshifts \citep[][]{Curtislake2022, Wang2023, Castellano2024, Carniani2024} using NIRSpec observations, while others are convincingly shown to be contamination from lower redshifts \citep[e.g.,][]{arrabal2023}.

The \emph{JWST} clearly allows us to probe galaxies at a greater depth in the near and mid-infrared than previously missions  (e.g., \emph{Hubble Space Telescope}, the \emph{Spitzer Space Telescope}, and the VISTA telescope). The increased resolution, depth and general higher image fidelity of JWST allows features such as the Lyman-break and the rest-frame ultraviolet spectral energy distributions of galaxies with redshifts greater than $z > 9$ to be observed. This provides insights into not only the redshifts of these systems but also their stellar masses and star formation rates. However, to accurately study these galaxies we must ensure that we are finding and identifying correctly these systems with minimal contamination. 

Whilst the GTO teams \citep[e.g.,][]{Castellano2022, Eisenstein2023,Bunker2023,Hainline2023} and others have studied these fields for the most part in this way already, with the exception of the NEP field we discuss here as part of our GTO programme PEARLS, we have carried out this meta-analysis constructing the EPOCHS sample for a few reasons. One of these is to carry out a large area analysis with the same reductions, galaxy detection, and analysis processes. As different reductions can, and nearly alway do, lead to different high redshift galaxy samples constructed, a consistent method will allow for a better understanding of biases and the determination of random and systematic errors. We also combine all these data to limit the effects of cosmic variance, given that many of these fields are small, and thus any results derived from one or a few of them are to some degree biased by the narrow area  of the sky observed \citep[see e.g.][]{Moster2011,Jespersen2024}. 

In this paper we present the results of our search for $z > 6.5$ galaxies within 11 of the deepest JWST fields observed to date, including our GTO time as part of the PEARLS project \citep[][]{Windhorst2022}. This also includes the public GLASS, NGDEEP, JADES, CEERS, JADES and the SMACS-0723 fields. In this paper we explore the properties of 1165  galaxies which were discovered at these early epochs. We describe this sample which is used throughout the other papers in this series, including the construction of the UV luminosity and stellar mass functions \citep[][]{Adams2023b, harvey2024epochs}. As part of this goal, in this paper we describe the basic features of this large collection of distant galaxies in terms of their galaxy luminoisities, colors, star formation rates, as well as their redshift distributions and number densities. 

Other papers in the EPOCHS series include a measurement of the early UV luminosity function \citep[EPOCHS II,][]{Adams2023b}, the $\beta$ slopes and star formation rates for these systems \citep[EPOCHS III,][]{Austin2024}, and how stellar masses are distributed depending on certain stellar population models being used \citep[EPOCHS IV,][]{harvey2024epochs},  as well as morphology \citep[][]{Conselice2024}, and size evolution \citep[][]{Ormerod2024}. We have also studied the MIRI properties of some of our sample \citep[][]{Li2024}, as well as investigating how our samples varies as a function of environment, and how many of our systems are in overdensities \citep[][]{2024arXiv240517359L}. This particular paper is the introduction to this series and describes our methodology, our completeness calculations, and the basic properties of the $z > 6.5$ galaxies we have discovered. 

The ultimate understanding of the role of galaxies in the early Universe, including at the epoch of reionization, will require building up large samples at these redshifts. Studies such as these are the first step in this process with \emph{JWST}, which will ultimately address fundamental questions for how reionization occurred and when and how the first galaxies assembled. The advantage of our study is that we combine several of the deepest available fields to study the very first galaxies with NIRCam and the Hubble Space Telescope (HST). This allows us to determine how galaxy selection depends on field and filters as well as create a large sample for meta analyses. 
 
The structure of the remainder of this paper is as follows. In \autoref{sec:data}, we describe the PEARLS and public deep field observations and our observational program, focusing on the NIRCam observations which we have reprocessed, as well as the data products derived from this new data set. In \autoref{sec:method} we describe the selection procedure undertaken to define a robust sample of galaxies with redshifts greater than $z > 6.5$. We present an analysis of the completeness using our procedures and describe the properties of the galaxies we have found in \autoref{sec:results}. We present a discussion of this sample's properties in \autoref{sec:discussion}, whilst a summary of our findings is included in  \autoref{sec:conclusions}. Throughout this work, we assume a standard cosmology with $H_0=70$\,km\,s$^{-1}$\,Mpc$^{-1}$, $\Omega_{\rm M}=0.3$ and $\Omega_{\Lambda} = 0.7$ to allow for ease of comparison with other observational studies. All magnitudes listed follow the AB magnitude system \citep{Oke1974,Oke1983}.

\section{Data Reduction and Products} \label{sec:data}

\subsection{Surveys and Fields}

The data we use for this analysis originates from the Early Release Observations of CEERS, JADES, GLASS and SMACS~0723, alongside the PEARLS GTO Survey fields: El Gordo, Clio, MACS-0416 and the North Ecliptic Pole (NEP) \citep[][]{Windhorst2022} as well as the data from the NGDEEP survey. The data we use are mostly from observations taken with the \textit{Near Infrared Camera} \citep[NIRCam;][]{rieke05, rieke08, rieke15} of these various fields and pointings. A list of our fields and their properties are shown in detail in \autoref{tab:areas}. We generate datasets for each of these fields which are homogeneous and reduced and catalog in the same way for each field. We do this by consistently processing our data ourselves at all steps using a bespoke and refined method that maximizes our detections of faint galaxies and the accuracy with which we can measure their photometry. 

We call this collation of data the {\em EPOCHS} sample, and this paper is the introductory version of this series with succeeding papers describing various aspects of these galaxies and what they imply for galaxy evolution and formation (see appendix). The sample we describe here is the version 1 (v1) of the EPOCHS sample, whilst future studies will use different selections and increase our data using more JWST imaging and spectroscopy. 

We list in \autoref{tab:areas} the different fields in which this study and the other studies from EPOCHS are taken. We calculate depths for these fields by placing non-overlapping apertures in empty regions of each of the images  using \sextractor{} segmentation maps and our image masks. We then used 200 apertures to calculate the Normalised Mean Absolute Deviation (NMAD) of these measures to derive local depths for each individual source. 

\begin{table*}[]
\centering
\caption{List of our observed fields and the depths and areas of each, adapted from a version of this table from \citet{harvey2024epochs}. The values listed include the unmasked areas and depths of the observations for this paper which are also used in other EPOCHS papers. The depths listed are at 5$\sigma$ in AB magnitudes, measured in 0$\farcs$16 radius apertures. Where depths are tiered across mosaics (e.g. HST (ACS/WFC) observations in the Hubble Ultra Deep Field (HUDF) Parallel 2) we have listed the depths and areas separately. The four spokes of the NEP-TDF and ten CEERS pointings have uniform depths (within 0.1 mags) with the exception of CEERS pointing-9 (P9) which we list separately. Areas are given in arcmin$^2$ and measured from the mask to account for the masked areas of the image and unused cluster modules. Fields with a `*' indicate that we have excluded the NIRCam module containing a lensing cluster from our analysis.}
\label{tab:areas}

\setlength{\tabcolsep}{3pt}
\begin{tabular}{|l|l|ll|lllllllll|}
\hline
              & Area      & \multicolumn{2}{l|}{HST/ACS\_WFC} & \multicolumn{9}{c|}{JWST/NIRCam}                                       \\
Field         & (arcmin$^2$)  & F606W           & F814W           & F090W & F115W & F150W & F200W & F277W  & F335M & F356W & F410M & F444W \\ \hline
NEP           & 57.32 & 28.74           & -               & 28.50 & 28.50 & 28.50 & 28.65 & 29.15  & -     & 29.30 & 28.55 & 28.95 \\
El Gordo*      & 3.90  & -               & -               & 28.23 & 28.25 & 28.18 & 28.43 & 28.96  & -     & 29.02 & 28.45 & 28.83 \\
MACS-0416*     & 12.3  & -               & -               & 28.67 & 28.62 & 28.49 & 28.64 & 29.16  & -     & 29.33 & 28.74 & 29.07 \\
Clio*          & 4.00  & -               & -               & 28.12 & -     & 28.07 & 28.21 & 28.675 & -     & 28.91 & -     & 28.71 \\
CEERS         & 66.40 & 28.6            & 28.30           & -     & 28.70 & 28.60 & 28.89 & 29.20  & -     & 29.30 & 28.50 & 28.85 \\
CEERSP9       & 6.08  & 28.31           & 28.32           & -     & 29.02 & 28.55 & 28.78 & 29.20  & -     & 29.22 & 28.50 & 29.12 \\
SMACS-0723*    & 4.31  & -               & -               & 28.75 & -     & 28.81 & 28.95 & 29.45  & -     & 29.55 & -     & 29.28 \\
GLASS         & 9.76  & -               & -               & 29.14 & 29.11 & 28.86 & 29.03 & 29.55  & -     & 29.61 & -     & 29.84 \\
NGDEEP HST-S  & 1.28  & 29.20           & 28.80           & -     & 29.78 & 29.52 & 29.48 & 30.28  & -     & 30.22 & -     & 30.22 \\
NGDEEP HST-D  & 4.03  & 30.30           & 30.95           & -     & 29.78 & 29.52 & 29.48 & 30.28  & -     & 30.22 & -     & 30.22 \\
JADES Deep GS & 22.98 & 29.07           & -               & 29.58 & 29.78 & 29.68 & 29.72 & 30.21  & 29.58 & 30.17 & 29.64 & 29.99 \\ \hline

\end{tabular}
\end{table*}

The public data we use includes the CEERS, NGDEEP, JADES, GLASS and SMACS surveys and fields which have been discussed in previous papers \citep[e.g., ][]{Treu2022, Finkelstein2022-CeersI}. However, this is not the case for the PEARLS datasets which makes up a large fraction of our sample of high redshift galaxies.   As can be seen in Table~\ref{tab:areas}, the PEARLS area constitute about 38\% of the total area in which we take our survey data on distant galaxies form. 

Observations of the three PEARLS lensing fields of SMACS~0723, MACS-0416 and El Gordo,  are such that one of the two NIRCam modules in each pointing is positioned such that it is centered on the lensing cluster. The other module is  located approximately 3 arcminutes to the side, giving effectively a 'blank-field' view of the distant universe. Although we reduce both modules in these fields, we decided not to include sources found in the cluster module in this study. The high redshift galaxies directly behind these clusters will be presented in a future study, using methods similar to our own. By not including the cluster region in our analysis we simplify things such that we do not need to consider strong gravitational lensing and contamination from intra-cluster light (ICL), which are significant effects \citep[e.g.,][]{Griffiths2018, Bhatawdekar2021}. For these clusters only the NIRCam module which is not centred on the cluster is used in our analysis and thus high-z galaxies lensed behind magnifying clusters of galaxies are not included in this analysis.

\subsubsection{PEARLS Fields and Data}

The prime fields from which the EPOCHS sample is taken from include the \textit{Prime Extragalactic Areas for Reionization Science}  GTO Survey (PEARLS, PI: R. Windhorst \& H.Hammel, PID: 1176 \& 2738). PEARLS is unique amongst the early GTO programs in that it is concentrated on imaging new deep fields, including the North Ecliptic Pole (NEP) region, as well as examining distant galaxy clusters. Several of our PEARLS fields are observed with a cadence that allows for variability to be detected \citep{Yan2023}. This includes the discovery of one of the highest redshift supernova discovered to date \citep[][]{Frye2024}. A full and complete overview of the PEARLS survey can be found in \cite{Windhorst2022}.
   
\begin{figure}
\centering
\includegraphics[width=1.0\columnwidth]{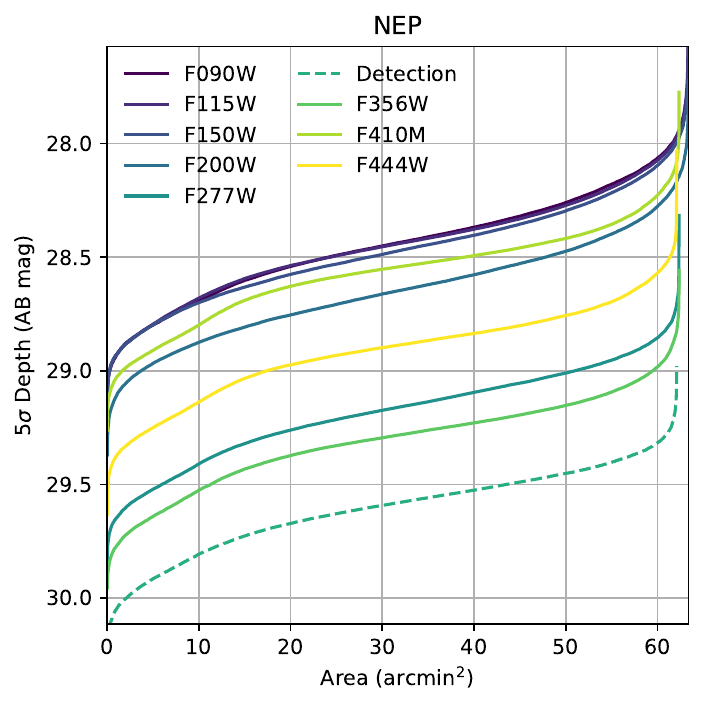}
\caption{Plot showing the cumulative distribution of unmasked area as a function of depth within the NEP field for the different filters in which this data was obtained. Also shown is the cumulative depth and area for our detection method which uses an inverse variance weighted stack of the three reddest wideband filters - F277W, F356W and F444W. This figure gives us an idea of the different depths reached and in which filter over the area of the NEP field. The other EPOCHS fields we use in this paper have similar patterns of depth and area. }
\label{fig:nep_area}
\end{figure}

As of writing, PEARLS has completed 4 series of observations. Four of these include targets in and around gravitationally lensing galaxy clusters and one within a blank field. The three clusters we include in this paper are MACS 0416, Clio, and El Gordo. The blank field is located at the North Ecliptic Pole (NEP). While most of this data has priority time, the first two NIRCam pointings of the NEP were made publicly available. The majority of PEARLS observations  we consider within this paper consist of 8 NIRCam photometric bands: F090W, F115W, F150W, F200W, F277W, F356W, F410M and F444W. However observations of the Clio cluster, a compact lensing cluster at z$=0.42$, do not include F115W or F410M \citep{Griffiths2018}. A typical distribution of depth vs. area in the Pearls fields is shown in Figure~\ref{fig:nep_area} for the NEP field. 

As described already briefly, for the lensing fields of Clio, SMACS~0723, MACS-0416 and El Gordo the observations include pointings with one NIRCam module centered on the lensing cluster, with the second module offset around 3 arcminutes in a `blank' region. While we reduce both modules in these fields, we do not include sources found in the module containing the lensing cluster in this study. MACS-0416 was observed across multiple visits for time domain science at 3 separate position angles, resulting in three parallel 'blank' regions which we incorporate. A further analysis of the high redshift galaxies that are lensed within this cluster components will be presented in future papers.

Where possible we incorporate complementary observations from HST, particularly optical observations with the ACS WFC instrument which cover a wavelength range below the NIRCam F090W filter and allow observation of the Lyman-break at $z < 7$. A list of the fields where ACS data has be utilised by us is included in \autoref{tab:areas}. Specifically, the NEP Time-domain field (TDF) was observed with HST ACS/WFC as part of programs GO-15278 (PI: R.~Jansen) and GO-16252/16793 (PIs: R.~Jansen \& N.~Grogin) between  October 1. 2017 and  October 31. 2022. A mosaic of the F606W observation, astrometrically aligned to Gaia/DR3 and resampled on 0\farcs03 pixels, were made available pre-publication by R.~Jansen \& R.~O'Brien \citep[][]{o2024treasurehunt}. %Other ACS imaging we use includes that taken in the CEERS field, NGDEEP as well as JADES. 

\subsubsection{SMACS-0723 Field and Data}

The first JWST data and imaging publicly released in July 2022 was the SMACS-0723 field, which contains a galaxy cluster, as well as a parallel blank module. Our early analysis of this field was presented in \citet{Adams2023} with further details in \citet{Adams2023b}. The observations of the SMACS-0723 galaxy cluster were part of the JWST Early Release Observations (ERO) programme \citep[PID: 2736, PI: K. Pontoppidan,][]{Pontoppidan2022}. 
 This cluster was observed in 6-band NIRCam photometry in the F090W, F150W, F200W, F277W, F356W, and F444W filters. However, SMACS-0723 is missing the critical F115W filter, which makes it difficult to identify galaxies at $8 < z < 10 $ with certainty. 
 
 Within the EPOCHS sample we include high$-z$ candidates from SMACS-0723 in our final catalogues, but only with care and attention to which redshifts are being used. For example, we do not use this field when measuring the UV luminosity function \citep{Adams2023b}. In more detail, the absence of the F115W photometric band leads to a very significant scatter in photometric redshift measures for galaxies within the redshift range of $7<z<10$. 
 However, higher redshift objects can be identified confidently \citep[e.g.,][]{Adams2023}. However, we do know that spectroscopic measurements for this sample prove that some of these galaxies are correctly identified as high$-z$ sources. \citep[][]{Trussler2022}. Thus while we can identify high$-z$ galaxies in this field, understanding which redshift bin they are in is more of a challenge for $z < 10$. However, we can still examine these systems and their properties as individual detections at more confident redshifts. 

\subsubsection{The GLASS Field and Data}

The GLASS observation programme focuses primarily on the Abel 2744 galaxy cluster with a selection of JWST instrumentation \citep[ID: 1324, PI: T. Treu,][]{Treu2022}. In parallel to these observations, the GLASS programme has generated one of the deepest NIRCam imaging sets pubically available. GLASS contains two overlapping parallel NIRCam observations  in seven filters: F090W, F115W, F150W, F200W, F277W, F356W and F444W. This field has already provided several strongly detected high-redshift candidates up to $z=12.5$ \citep[e.g.,][]{Castellano2022, Naidu2022}. %This study only makes use of the first epoch of GLASS NIRCam observations. This is because the second epoch was run in parallel to NIRSpec observations with a very small dither pattern, leading to issues with noisy long wavelength imaging. The processing of the GLASS field will be revisited in future work which will seek to fold in the NIRCam imaging provided by UNCOVER \citep{Bezanson2022}.

\subsubsection{The CEERS Field and Data}
    
This study also makes use of both observing runs (July 2022 \& December 2022) observed as part of the CEERS survey \citep[ID: 1345, PI: S. Finkelstein, see also][]{Bagley2023}. This consists of 9 NIRCam pointings with 7 photometric bands (F115W, F150W, F200W, F277W, F356W, F410M and F444W). This field subsequently provides the single largest area used in our study at 66.4 square arcminutes. However, the lack of observations in the F090W band limit the capabilities of the JWST observations within this field for identifying galaxies at $6.5<z<8.5$, so we incorporate HST ACS WFC F606W and F814W observations taken as part of the CANDELS program \citep{Grogin2011, Koekemoer2011} and re-released as part of the CEERS teams HDR1 data release. %Therefore we are limited in what we can learn from this field at this redshift range. 

\subsubsection{The NGDEEP Field and Data}

This study also makes use of the NGDEEP \citep[ID: 2079, PIs: S.\@ Finkelstein, Papovich and Pirzkal, ][]{Bagley2023} reduction and high$-z$ sample that was obtained in the work of \citet{Austin2023}. The data from this field follows the same reduction and selection procedures used for the other JWST fields we study. NGDEEP consists of NIRCam imaging that was run in parallel to JWST's Near Infrared Imager and Slitless Spectrograph (NIRISS) spectroscopy of the Hubble Ultra Deep Field (HUDF). The NIRCam imaging covers part of the HUDF-Par2 parallel field and consists of 6 broadband filters: F115W, F150W, F200W, F277W, F356W and F444W, all with average depths of $m_{\rm AB}>29.5$. It is subsequently the deepest dataset used in this study. For more details see \citet{Austin2023} and \citet{Bagley2023}. In addition to the JWST data we use for the NGDEEP field, we also include HST imaging in the F606W and F814W bands from v2.5 of the Hubble Legacy Fields project  \citep{Illingworth2016,Whitaker2019}. In fact, we find that when we include this imaging in the fits, we find that several galaxies no longer are identified as being high$-z$ when using just the JWST bands \citep[][]{Austin2023}. This is critical for our analysis and demonstrates the importance of using HST data when possible for exploring distant galaxies with JWST.

\subsubsection{The JADES GTO Data} 

In June 2023 the JWST Advanced Deep Extragalactic Survey (JADES) \citep[JADES, PID:1180, PI: D. Eisenstein][]{Eisenstein2023,Bunker2023,Hainline2023} team kindly released part of their data products, including the raw imaging which we use in this paper to find in an independent way, the highest redshift sources. This data was released in June 2023, including full mosaics using the pmap1084 calibrations \citep{Rieke2023}. This released data consists of 6 overlapping NIRCam pointings from the JADES DEEP observations within the  filters: F090W, F115W, F150W, F200W, F277W, F335M, F356W, F410M and F444W. This field is located around the Hubble Ultra-Deep Field (HUDF) in the GOODS-S area. As can be seen in \autoref{tab:areas} the depth of the JADES data within the JWST bands ranges from 29.58 to 30.21 in its deepest band in F277W. This field also has deep F606W data from ACS. 

For the EPOCHS analyses we rereduce this NIRCam data using our own version of the pipeline for consistency with our other fields. Our bespoke reductions have depths that are around 0.1 magnitudes shallower than the official JADES reductions, with one small region of the field affected by residual wisping in the F150W and F200W bands. We also use the same HST/ACS F606W mosaic of the wider GOODS-S region as we use for the analysis of the NGDEEP field.

\subsection{Reduction Process}

We reprocess all of our uncalibrated lower-level JWST data products and data using the methods outlined in \citet{Leo2022} and \citet{Adams2023b}. This includes reprocessing all of the NIRCam imaging from their lowest-level, raw form obtained from the MAST database using computers at the University of Manchester. We follow the same procedures as used in these published works, but include a series of minor improvements which we developed over the first year of handling JWST data involving innumerable experiments and trials of different reduction processes. 

Below we give a description of the pipeline and processes used to arrive at our final reduced imaging data. Our full pipeline can be summarised through the following steps. First,  we use version 1.8.2 of the official JWST pipeline and CRDS v1084 for the calibration files, which contains the most up-to-date NIRCam calibrations at the time of writing. These files also include the third round of post-flight calibrations, which are essential for achieving a reliable photometric calibration and flat fielding, an issue which has plagued the early analysis of JWST distant galaxy discoveries \citep[][]{Adams2023, Rigby2022}. After running `Stage 1' of the JWST pipeline, we subtract off templates of 'wisps'. These are large scale artefacts in the imaging caused by rogue light and these features affect the A3, A4, B3 and B4 NIRCam modules for the F150W and F200W imaging. The templates we use are the second generation templates released by STScI.\footnote{\url{https://jwst-docs.stsci.edu/jwst-near-infrared-camera/nircam-instrument-features-and-caveats/nircam-claws-and-wisps}} This method has been effective for most observing programs which we utilise, however the ultra-deep and small dithering nature of the NGDEEP observations result in some residual wisps which affect the final depths achieved in these two blue filters \citep[e.g.,][]{Austin2023}. We are presently developing a new series of improved wisp templates to solve these minor issues (Adams et al. in prep), which will be implemented in v2 of the EPOCHS dataset.

Next, our data goes through `Stage 2' of the pipeline and we apply a 1/f noise correction derived and  provided by Chris Willott.\footnote{\url{https://github.com/chriswillott/jwst}}  We then extract the sky subtraction step from `Stage 3' of the pipeline and run this independently first on each NIRCam frame. This allows for a rapid assessment of the background subtraction performance from which we fine-tune our process. We do this by conducting an initial flat background subtraction which is then followed by a 2-dimensional background subtraction utilizing the tool {\tt photutils} \citep{larry_bradley_2022}. 

The `Stage 3' process is then run on these background corrected frames and a final mosaic is produced such that we align the WCS of GAIA DR3 \citep[e.g.,][]{GAIADR3} using \texttt{tweakreg}, part of the DrizzlePac python package. We then ensure that each of our final images are aligned with each other by aligning them to the F444W image.\footnote{\url{https://github.com/spacetelescope/drizzlepac}} From this we pixel-match the images to match the F444W image using the method from \texttt{reproject}\footnote{\url{https://reproject.readthedocs.io/en/stable/}} with the final scale of the drizzled images at 0.03 arcseconds/pixel. 

Our reductions of all the frames and images in each field differ from the official PEARLS and public team reductions described in papers such as \citet{Windhorst2022}. However, we carry out our reductions in a systematic way across all our fields, to avoid problems with inhomogenous data quality, methodology, and systematics that can be present when comparing data in different fields. These issues can be seen from ground-bases surveys as well as in space-based data \citep[e.g.,][]{Conselice2022}. A further description of our reduction and pipeline process is provided in \citet{Adams2023b} and \citep[][]{harvey2024epochs}.

%We compare the photometry extracted from the different reduction pipelines in \S 2.4.

\subsection{A Robust Sample of Ultra-High Redshift sources} \label{sec:method}

This section describes how we identity our high redshift galaxy candidates which we later use to determine the properties of early galaxies, as well as for the sample which is used in other papers in this series to investigate the mass function \citep[][]{harvey2024epochs}, the UV luminosity function \citep[][]{Adams2023b}, as well as properties such as galaxy structure, AGN, star formation and dust content which are present in various degrees within our alaxy sample \citep[e.g.,][]{Ignas2023, Fu2024}. 

In this section, we first describe how we measure the photometric redshifts for our sample and then we describe how we use this information combined with the detection significance of our sample to create a high quality sample of $z > 6.5$ candidates for our detailed studies.

\subsubsection{Photometry and Detection}

To construct our catalogs of objects after our reduction procedure we utlize the software package {\tt SExtractor} \citep{Bertin1996}. Our analysis with source extraction runs in dual-image mode, using an inverse variance weighted stack of the F277W, F356W and F444W bands and by performing forced aperture photometry for multi-band measurements. 

We calculate photometry for each galaxy within circular apertures of 0.32 arcsecond diameter, including an aperture correction that is derived from simulated \texttt{WebbPSF} point spread functions for each band used \citep{Perrin2012,Perrin2014}. We chose this diameter to enclose the central and brightest $70-80$ percent of a point source's flux, but which is still small enough to avoid contamination. By doing so, we balance the use of high-signal pixels when computing fluxes and avoid dependence on a PSF model correction that is as high or higher than the measurement made. We found from experimentation that this is the best method for measuring the fluxes of our objects which obtains most of the light from galaxies without significant contamination from other sources. 

To determine the depth of our final images, we use circular apertures in regions of the image where no pre-existing sources are known to exist and have been identified within 1 arcsecond of the aperture's central coordinate. This allows us to derive an average depth for each field, as well as to calculate local depths across each field. In order to generate more realistic photometric errors, we calculate the final errors for each source using the normalized median absolute deviation (NMAD) of the nearest 200 empty apertures \citep[NMAD:][]{hoaglin2000understanding}. This process is necessary as {\tt SExtractor} is known to underestimate photometric errors, and these are critical for deriving the photometirc redshifts and other galaxy properties accurately. The average depths of each photometric band for each field are presented in \citep[][]{Adams2023b}, in the context of using this data to measure the UV luminosity function.

For each of our fields, we also carefully examine and mask out within the image `defect areas' such as diffraction spikes, remaining snowballs, and high-intensity intra-cluster medium. We do this within the NIRCam modules that include foreground clusters, as well as a buffer area around the edges of the images. We find that the edges of the images may be shallower due to the dithering patterns used in the JWST observations. This study only utilizes the total unmasked area, which is listed alongside the average depths of each field in \citet{Adams2023b}, \citet{harvey2024epochs} and \autoref{tab:areas}. This process also ensures that we do not include galaxies whose features in particular bands are influenced by noise properties. 

\subsubsection{Photometric Redshifts}

\begin{figure}
\centering
\includegraphics[width=1.0\columnwidth]{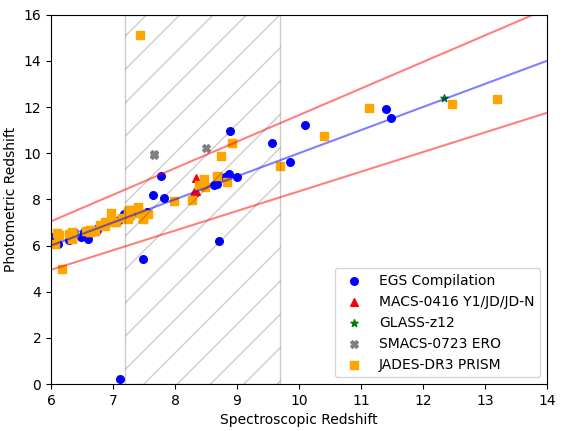}
\caption{The comparison between the photometric and spectroscopic redshifts for our sample, where available. We compare against recent results from JADES DR3 \citep{DEugenio2024} and a compilation of spectra from the EGS region featuring data from CEERS \citep{Arrabal2023b}, the follow-up DDT programme \citep[PID 2750][]{arrabal2023} and PID 2565 \citep{Glazebrook2024}. Also added are spectroscopic redshifts for GLASS-z12 \citep{Castellano2024}, recent results from the MACS-0416 field \citep{Ma2024}, and the SMACS-0723 ERO programme \citep{Pontoppidan2022}. The grey hatched region indicates the region where the lack of F115W in SMACS-0723 leads to larger expected uncertainties. Samples have been cut to show those which are 5$\sigma$ detected, unmasked and have strong spectroscopic flags. This shows the quality of our results and methods in finding secure and robust redshifts.}
\label{fig:specz}
\end{figure}

With the imaging and cataloging complete, we calculate the NIRCam derived spectral energy distributions for all sources identified in order to derive photometric redshifts. We also include a detailed discussion of photometric redshfits for this sample in other papers such as \citet{Adams2023b, harvey2024epochs, duan2024}. We give another description here which is focused on our particular sample of galaxies used within this paper. 

 %Our approach for this within this first generation EPOCHS sample is to use two different codes and to require consistency in redshifts between these two codes before the galaxy is included in our ultimate sample. 
 
 %The catalogs of redshifts we use are generated using the photometric redshift code \eazy, while we utilize BC03 template sets with a Chabrier initial mass function for our analyses, as presented in Bruzual and Charlot (2003) and Chabrier (2000), respectively. These template sets encompass both exponential and constant star formation histories, with 10 characteristic timescales ranging from $0.01<\tau<13$~Gyr, and 57 different ages spanning 0 to 13 Gyr. The redshift range allowed in our study is $0<z<25$, and we account for dust attenuation following the prescription of Calzetti et al. (2000). To account for potential contamination from dusty, mid-redshift ($2<z<5$) galaxies, we allow for $E(B-V)$ values up to 3.5. Our code also incorporates treatment for emission lines and applies the intergalactic medium (IGM) attenuation derived from Madau (1995).

To measure photometric redshifts we utilize the \eazy \ photometric redshift code \citep{brammer2008eazy} with the default pipeline set to ``tweak\_fsps\_qsf\_v12\_v3'', which were generated using \citet{conroy2010fsps} models. We also incorporate Set 1 and Set 4 of the templates presented in \citet{Larson2022}, which provides bluer-rest frame colors and high equivalent width emission lines, due to their young ages (between 10$^6$ and 10$^7$ yr). These templates have been found to better reproduce the observed colors of some high$-z$ galaxies at redshifts greater than $z = 8$. The redshift range allowed in our study is $0<z<25$, and we take the maximum likelihood draw (z$_{\textrm{ml}}$) of the redshift PDF distribution as our redshift estimate, although we note that in the vast majority of cases this is very close to or the same as as the median of the PDF. We do not incorporate any optional prior on apparent magnitude.

%as well at those which are presented by the JADES Team \citep[][]{Rieke2022}. These templates build upon one of the default template sets and incorporate galaxies that exhibit bluer colors and stronger emission lines, which are expected to be more appropriate for modeling the spectral energy distributions (SEDs) of galaxies a
   
We have refrained from employing techniques for fine-tuning the zero points of the photometric bands in EAZY, as the NIRCam modules consist of multiple individual chips (8 in the blue and 2 in the red), each with their own independent calibrations and photometric zero point offsets. Applying zero point modifications on a chip-by-chip basis, instead of on the final mosaic, would be necessary due to the small field of view covered by each chip, which results in a limited number of objects with spectroscopic redshifts within each chip. This approach could easily introduce potential bias towards certain galaxy colors, depending on the types of spectroscopically confirmed galaxies within each module. These photometric band offsets are however used by others to optimize their photometric redshfits for high$-z$ JWST galaxies \citep[e.g,.][]{Hainline2024, Hainline2023}. 

Discussions with members of the community have indicated that residual zero point errors are anticipated to be around 5 percent. Therefore, we have implemented a minimum 10- percent error on the measured photometry to account for potential zero point issues within the NIRCam reduction pipeline in addition to other error sources such as minor imperfections in the template sets or PSF corrections.

We identify our sample using the fact that distant galaxies at $z > 6.5$ have very distinctive spectral features in the rest-frame ultraviolet. This includes the fact that the spectrum bluer than the Lyman-limit at 1216\AA{} is more-or-less absent due to the absorption of the light from neutral hydrogen gas. These features can produce galaxies seen as drop-outs, and this has been a traditional way to find the most distant galaxies for over 30 years 
\citep[e.g.,][]{Steidel1992}. Despite this, these apparent `drop-outs' can result from the Balmer break, dust absorption, and nebular lines in lower$-z$ systems. Thus far, all of the highest redshift galaxies observed with NIRSpec do not detect emission lines at $z > 10$, and thus likely our redshifts for the highest redshift galaxies will be based on the continuum shape and the Lyman-break presence vs. Balmer break \citep[e.g., Fig. 4 of][]{Curtislake2022, Fujimoto23, Carniani2024}.

Overall, we can quantify the photometric redshift quality in a few ways. One of these to to measure the difference between the photometric and spectroscopic redshifts. For our sample, we show this comparison in \autoref{fig:specz}.  Figure~\ref{fig:specz} demonstrates the agreement between our photometric redshifts and the spectroscopic redshifts from sources described below.  We can further quantify the quality of our photometric redshift quality by the value of the NMAD for the photometric redshift difference with the spectroscopic data.  This quantity is defined by:

  \begin{equation}
    \sigma_{\text{NMAD}} = 1.48 \times \, \mathrm{median} \left( \frac{|\Delta \mathit{z}|}{1 + \mathit{z}_{\text{spec}}} \right) ,
  \end{equation}
  
\noindent such that $\Delta z = z_{\rm phot} - z_{\rm spec}$ is the difference between photometric redshifts and spectroscopic redshift.  Note that the normalizing factor of 1.48 in the $\sigma_{\rm NMAD}$ equation is such that the NMAD expectation value is equivalent to the standard deviation of a normal distribution.

\noindent We carry out this comparison by using spectroscopic redshifts for galaxies within our EPOCHS v1 sample. At the time of writing this paper there are a few published spectroscopic redshifts within our fields. 
 This includes those from: the JADES DR3 \citep{DEugenio2024} release, a compilation of spectra and redshifts from from the Extended Groth Strip (EGS) region from CEERS \citep{Arrabal2023b}, including a follow-up DDT programme \citep[PID 2750][]{arrabal2023} and PID 2565 \citep{Glazebrook2024}. We also include spectroscopic redshifts for the GLASS-z12 object \citep{Castellano2024}, including recent results from the MACS-0416 field \citep{Ma2024}, and the SMACS-0723 ERO programme \citep{Pontoppidan2022}. 
 
 Using these spectroscopic redshifts and those which we calculate with the photometric redshifts, we find that the NMAD values are 0.021 for redshifts $z>6.5$ which are all within our redshift range. In terms of the outlier fraction, defined as those  with a dz $> 15$\%, we find that 9 out of 86 objects have redshifts with differences this high at $z>6.5$. When we omit the cluster SMACS-0723 (due to lack of the F115W making some redshifts quite uncertain at $7.5 < z < 9.5$) the fraction of outliers is then about $\sim 7$ \% (6/83). The result is a photometric redshift sample of high quality determined by this comparison with galaxies with confirmed spectroscopic redshifts.

\subsubsection{The Final EPOCHS Sample}

To select a robust sample of high redshift galaxies, we employ a series of selection criteria which we outline below, although see \citet{Adams2023b} and \citet{harvey2024epochs} for a further description of this process. The list of criteria and the process for finding these galaxies is given below. 

Within the photometric redshift code we have a measurements of the location of the likely Lyman-break, which is then used as the pivot point to test whether the data is detected at enough significance for us to include in our samples. The process for doing this is described below.

\begin{enumerate}
    \item To be included in our sample, the galaxy must be detected at $> 5\sigma$ significance in the two bands immediately redward of the estimated Lyman break position based on the photometric redshift with \eazy{}, and with non-detections, or less than $3\sigma$ detections, in all bands (minimum of a single band) bluewards of the Lyman-break. This is to ensure that our galaxies are well detected in the redder bands, and are not detected below the inferred Lyman-break and to remove all obvious Balmer break sources at lower redshifts. If the candidate is a F200W dropout, we increase this to a 7$\sigma$ and 5$\sigma$ requirement, as we have observed a number of spurious sources (similar to the z$\approx$ 16 candidate in \citep{Donnan2022}, which appeared only in the long wavelength observations, but was later found, via spectroscopy, to to be a $z=5$ interloper \citep[][]{arrabal2023}.
    
    \item We use the integrated probability density function of our photometric redshifts, PDF($z$), to determine the likelihood of the galaxy being at a particular redshift. We require that the integration of the PDF within the range of 10\% of the peak photometric redshift PDF value must include at least 60\% of the total PDF integral. We do this to remove galaxies which have strong bimodal solutions, particularly where one solution is at low redshift. Other descriptions of this method used for finding galaxies in pairs and in groups can be found in \citet{Mundy2017, Duncan2019ApJ, Sarron2021}.

    \item For each galaxy we perform an additional fit with \eazy\, runs with a maximum redshift of $z=6$ allowed in the fits. This allows us to obtain the best 'low-redshift' solution for each galaxy. We require that the difference in the $\chi^2$ between the high$-z$ and low$-z$ solutions be $\leq -4 $.
    %\item We also require that the same procedure applied to any secondary peaks should have an integral which is 50 per cent of the primary peak. The idea here is that this will remove confusion between Lyman and Balmer breaks.

    \item From the samples that pass these above tests we define a "robust" sample and a "good" sample depending on the quality of the fit. We define the "robust" sample as those galaxies with redshift fits with $\chi^2_{red} < 3$ and "good" as those with a fit giving $\chi^2_{red} < 6$. We do not distinguish between these two sets further in this paper and both are included in our EPOCHS v1 sample. 
    
    \item We compare the candidate's half-light radii to model PSFs as a way to remove likely hot pixels. Objects with sizes significantly smaller than a PSF (half light radius $<$ 1.5 pix.) in the long wavelength detectors are removed as likely artefacts. Note that we do not remove objects that are only close to the size of the PSF without other criteria not being met. 
    %\item Candidates are then further considered only if they have a consistent high-redshift solution when cross-checked using the same criteria as listed above in the secondary photo$-z$ pipeline utilizing EAZY. This ensures that the candidates meet the required criteria for high-redshift galaxies as determined by our analysis, and adds an additional level of confirmation through cross-validation using an alternative photometric redshift code.
    
    \item To ensure the quality and reliability of the sample, all objects are subjected to visual inspection by multiple authors to identify and remove any artifacts or contaminated sources. This meticulous vetting process involves careful examination of each object's characteristics, such as its morphology, brightness, and consistency with expected high-redshift galaxy features. Any objects that are deemed as artifacts or contamination are removed from the final sample to ensure the integrity of the results. This process ensures that only genuine high-redshift galaxies are included in the final sample. We have endeavoured to implement as many of our selection cuts as specific criteria in order to increase the reproducibility and fidelity of our sample. We remove $\leq$10\% of our total sample by eye, which is significantly lower than some comparable studies. 
    
\end{enumerate}

In \citet{Adams2023b} we include a discussion of how well our results compare with previous studies in the context of using these results for deriving properties of galaxies, including the UV luminosity function. However, that paper did not include an analysis of all galaxy detections, as it only carried out corrections for the luminosity function for galaxies bins in UV luminosity which were greater than 50\% complete. This means that many of the fainter and lower mass systems would not have been included in that analysis. As opposed to this, we include and analyze all significantly detected galaxies that pass our criteria within this paper.

\begin{table*}
    \centering
    \caption{Column names, units and descriptions for the EPOCHS v1 catalog, including column names, units, descriptions and column shape. A ``$^{\star}$" indicates that the column has been corrected for any flux associated with the galaxy which falls outside the extraction aperture. A full description of the catalogue is provided at: URL.}
    \begin{tabular}{ccc}
    \hline
        \large \textbf{Column Name} & \large \textbf{Unit} & \large \textbf{Description}  \\ \hline \hline
        \multicolumn{3}{c}{\textbf{IDs, Positions, Fluxes and local depths}} \\ \hline
        ID &  &  Unique catalogue ID, consisting of number and fieldname \\
        ALPHA\_J2000 & \si{degree} & Right ascension \\
        DELTA\_J2000 & \si{degree} & Declination \\
        FIELDNAME & & Field/pointing the galaxy is in \\
        FLUX\_APER\_\band{} & \si{nJy} & Aperture corrected flux in 0.16 arcsec radius apertures \\
        FLUXERR\_APER\_\band{} & \si{nJy} & Local-depth derived flux error from NMAD of 200 nearby empty apertures \\
        sigma\_\band{} &  & SNR of detection in 0.16 arcsec aperture \\
        local\_depth\_\band{} & AB Mag & 5$\sigma$ local depth from NMAD of flux in 200 nearby empty apertures  \\
        unmasked\_\band{} & Boolean &  Whether galaxy is masked in \band{} \\
        auto\_corr\_factor\_\band{} & & Correction factor in \band{} for flux outside 0.16 arcsec aperture \\
        \hline
        \multicolumn{3}{c}{\textbf{Photometric Redshifts and Selection}} \\
        
        \hline
        zbest &  & Photometric redshift using \eazy{}  \\
        zbest\_l1 &  & -1$\sigma$ photometric redshift uncertainty using \eazy{} \\
        zbest\_u1 &  & +1$\sigma$ photometric redshift uncertainty using \eazy{}  \\
        chi2\_best &  & $\chi^2$ of \eazy{} fit  \\
        PDF\_integral\_eazy & & $\int_{0.94 \times {\rm zbest}}^{1.06 \times {\rm zbest}} {\rm PDF}(z) dz $ - Integral of \eazy{} posterior redshift PDF  \\
        zbest\_lowz & & Photometric redshift using \eazy{}, with $z_{\rm max} = 6$   \\
        chi2\_best\_lowz  &  & $\chi^2$ of \eazy{} fit, with $z_{\rm max} = 6$    \\
        \hline
        \multicolumn{3}{c}{\textbf{UV Properties}} \\
        \hline
        M\_UV$^{\star}$ & AB Mag & Absolute UV mag  in 100\AA{} tophat at 1500\AA{} rest-frame flux at redshift zbest  \\
        M\_UV\_u1 & AB Mag &  \\
        M\_UV\_l1 & AB Mag &  \\
        BETA\_UV & & UV slope $f \propto \lambda^\beta$ (see \citet{Austin2024}).\\
        BETA\_UV\_l1 & & \\
        BETA\_UV\_u1 & & \\
        SFR\_UV$^{\star}$ & M$_\odot$ yr$^{-1}$ & \\
        SFR\_UV\_l1 & M$_\odot$ yr$^{-1}$ & \\
        SFR\_UV\_u1 & M$_\odot$ yr$^{-1}$ & \\
        \hline
        \multicolumn{3}{c}{\textbf{Sample identifiers}} \\
        \hline
        certain\_by\_eye & Boolean & Visual inspection of cutout and SED boolean \\
        EPOCHS\_II & Boolean & Used in EPOCHS II (UV LF) \\
        EPOCHS\_III & Boolean & Used in EPOCHS III (UV $\beta$ and dust) \\
        EPOCHS\_IV & Boolean & Used in EPOCHS IV (SMF) \\
        % Any more EPOCHS papers that we want to include boolean columns for?
        \hline
\end{tabular}
\end{table*}

\subsection{Galaxy Properties from Bayesian SED Fitting}
\label{sec:bagpipes}

We measure galaxy physical properties using the Bayesian SED-fitting code \bagpipes{} \citep{Carnall2018_Bagpipes, carnall2019galaxy}. \bagpipes{} allows flexibility in the choice of models, priors and star formation histories, which can have a large impact on derived galaxy properties \citep[e.g.][]{carnall2019galaxy, pacifici2023art}. A complete analysis of the range of physical parameters derived for our galaxy sample given different SED fitting tools, star formation histories and priors are published and discussed in great detail in EPOCHS~IV \citet{harvey2024epochs}. 

Whilst we have many different possible stellar masses to use, computed through different parametric and non-parametric methods,  in this paper we present results of our fiducial \bagpipes{} run. These runs are based on a log-normal star formation history with logarithmic priors on age, dust extinction, and metallicity. The log-normal star formation history was chosen to represent the predicted 'rising' star formation rate of high$-z$ galaxies \citep[e.g.][]{Madau2014}. Dust, metallicity, and age in particular are difficult to constrain based on photometry alone, and this choice of prior favours low ages, low dust extinction and low metallicity, which is predicted by simulations and confirmed by spectroscopy (CITE). We assume \cite{Calzetti2000} dust emission, \cite{Madau1995} ISM extinction and \cite{Bruzual2003} stellar population models. We use an informative redshift prior based on our \eazy \ results, with a Gaussian centered on the median of the \eazy \ redshift posterior, and standard deviation based on the average of the 16 and 84th percentiles of the PDF, and capped at $\pm3\sigma$.

\subsection{Brown Dwarfs}

%Help from Tom here

Low-mass stars within the Milky Way, particularly L and T-type brown dwarfs, can masquerade as high$-z$ galaxies due to an apparent Lyman-break like dropout in their broadband SEDs. In order to ensure our sample is not contaminated with brown dwarfs, we fit synthetic brown dwarf templates using a least-squares fitting routine. We use the Sonora Bobcat and Cholla templates \citep{marley2021sonora, karalidi2021sonora}. For each of the SEDs predicted by both sets of Sonora templates we calculate mock photometric measurements for HST/ACS WFC and JWST/NIRCam filters and fit the model photometry to our observed photometry, varying only the best-fitting normalization. 

We flag an object as a possible brown dwarf by comparing the $\chi^{2}$ values of brown dwarf vs. galaxies templates. This is such that an object is identified as a brown dwarf when the $\Delta \chi^2$ between the best-fitting brown dwarf template and the best-fitting \eazy \ galaxy template is less than 4, which is the same  $\Delta \chi^2$ criteria we apply between our low$-z$ and high$-z$ \eazy \ fits. We also additionally require that the galaxy appears compact, as we expect brown dwarfs to appear as point sources, so we require that the 50\% encircled flux radius (as measured by \sextractor) is smaller than the FWHM of the PSF in the F444W band. In total, across all of the fields, we flag 59 objects as possible brown dwarfs, which is $\sim 4.6\%$ of the full sample. Whilst the brown dwarf candidates presented in \cite{hainline2023brown} are not within our initial sample of high$-z$ galaxies, due to not meeting other of our criteria, we test our methodology on their candidates and recover and identify all of them as brown dwarfs.

\subsection{Galaxy Detection and Completeness}

One of our goals within this paper is to have a high degree of completeness and purity within our sample of galaxies. There are various ways to determine this, although understanding this exactly is difficult to impossible without deep spectroscopic studies that are complete to certain magnitude of flux depths. However, one way we can investigate this is to use simulations and mock catalogs of our fields and determine how many distant galaxies witin our redshift range of interest we would detect using our methods and criteria for finding distant galaxies.

We compute this using the \texttt{JAGUAR} simulation \citep{Williams2018}. \texttt{JAGUAR} is a  novel phenomenological model that is designed and meant to describe the evolution of galaxy number counts, morphologies, as well as  spectral energy distributions across a broad range of redshifts ($0.2 < z < 15$) and at stellar masses log (M/M$_{\odot}) > 6$. \texttt{JAGUAR} essentially creates  mock catalogs that reproduce the properties of various deep JWST surveys from which we create  mock catalogs for each of our deep fields. It has previously been shown before JWST that the output from  \texttt{JAGUAR} matches well with observed stellar mass and luminosity functions for both star-forming and quiescent galaxies, and can accurately replicate the redshift evolution of colors, sizes, star formation rates, and chemical properties of the galaxy population. It does this by including a self-consistent treatment of stellar and photoionized gas emission and dust attenuation, utilizing the BEAGLE tool. Thus the \texttt{JAGUAR} simulation produces a list of simulated galaxies, with characteristics such as their redshifts, stellar masses, star formation rates, and other physical properties including the fluxes of each galaxy in each band. 

To use \texttt{JAGUAR} effectively for a given field, we use the known average depths of each of our fields and apply a Gaussian scatter to each galaxy's photometry accordingly. We thus create a new catalog for the \texttt{JAGUAR} sources, such that the photometry is now adjusted to represent how these galaxies would have been observed within each different field. This new photometric catalog is then run through our  \texttt{EAZY} SED fitting and selection procedure. This is done to determine whether, and thus what fraction, of actual galaxies in the \texttt{JAGUAR} catalog would still remain detected after going through our selection. 

We furthermore apply \texttt{\bagpipes{}} fitting to the simulated data to determine how the stellar masses and star formation rates would have changed due to the limited depth of each field. We thus apply this process to our entire \texttt{JAGUAR} catalog which allows us to determine the fraction of true high redshift galaxies in our final sample (completeness) and the number of low-redshift interlopers (contamination). These values vary depending on the field, as both the filters and depths differ across fields. We can parameterize the completeness in terms of known variables (stellar mass, $\mathrm{M}_\mathrm{UV}$, apparent magnitude), by categorizing the completeness and contamination into bins. A more detailed explanation is outlined in previous EPOCHS papers \cite{harvey2024epochs, Austin2024} and we get a further description of this in the appendix. 

\section{Results} \label{sec:results}

In this section we discuss our sample of galaxies and their properties, including their apparent evolution and properties. We first discuss the overall trends of magnitude and luminosity for our sample and then we explore some of the more detailed properties of our galaxies, including their observed and rest-frame colors, as well as star formation histories and how the number of galaxies we find compares with theoretical models as a new test of the excess galaxy problem. 

\subsection{Redshift Distributions}
\label{sec:z_dist}
\begin{figure}
\centering
\includegraphics[width=1\columnwidth]{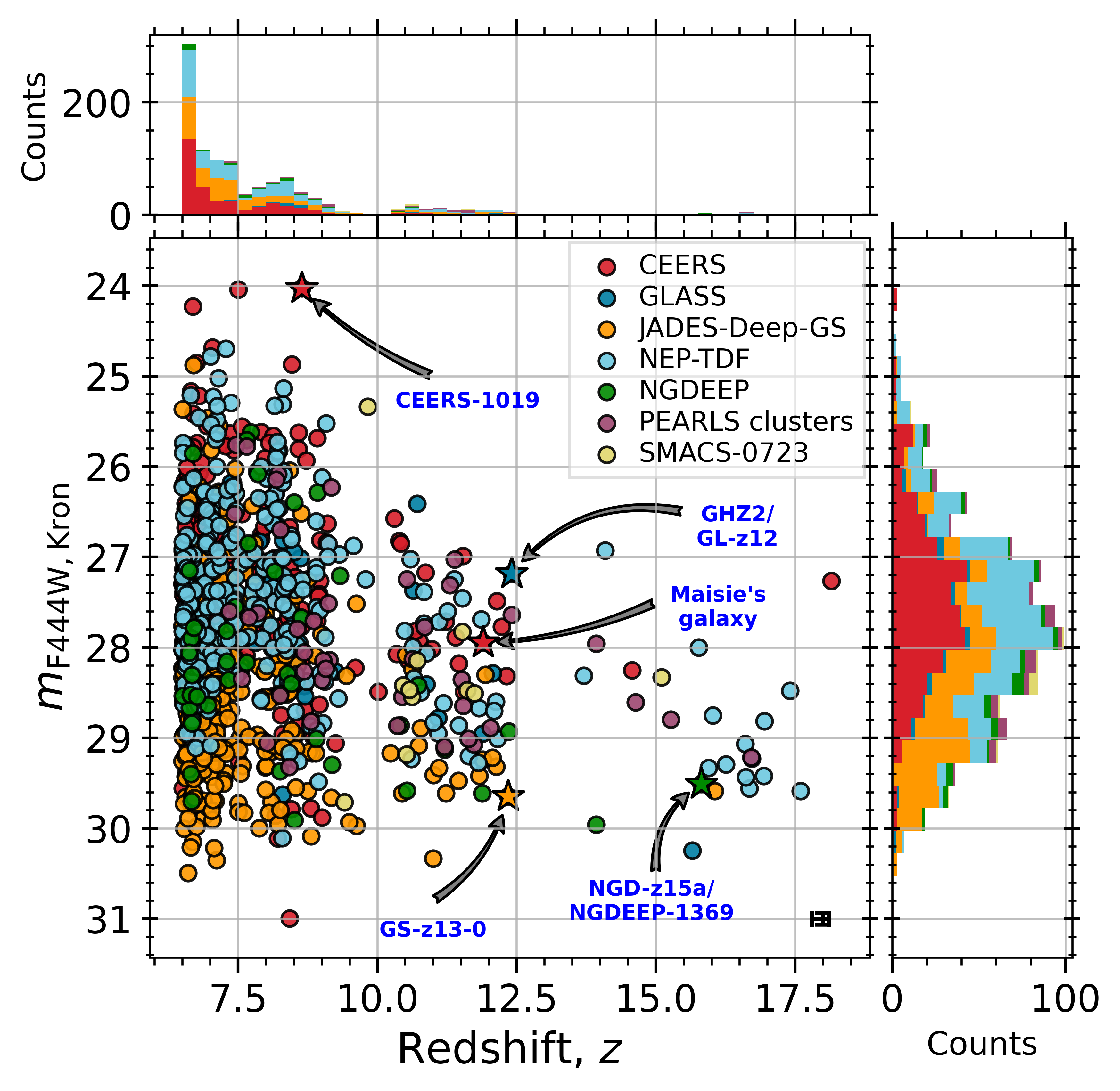}
\caption{Evolution of \sextractor{} F444W ``MAG\_AUTO'' apparent magnitudes \citep[measured in Kron apertures,][]{Kron1980} with {\tt{EAZY-py}} redshifts $z$ for our ``certain'' galaxy sample. The colors indicate the survey of origin, where the purple points in the PEARLS clusters are from the blank parallel fields of the El Gordo, MACS-0416 and Clio clusters. The median $1\sigma$ error for the sample is shown in the bottom right. Highlighted are notable spectroscopically confirmed galaxies from JADES \citep[GS-z13-0,][]{Curtislake2022}, CEERS \citep[Maisie's galaxy,][]{Finkelstein2022-Maisies}, and GLASS \citep[GHZ2/GL-z12,][]{Naidu2022, Castellano2022, Castellano2024}. The $z=8.679$ CEERS AGN from \citet{Larson2023a} (CEERS-1019), and $z\simeq15.6$ galaxy candidate from NGDEEP \citep[NGD-z15a/NGDEEP-1369,][]{Austin2023, Leung2023} are also shown.}
\label{fig:magz}
\end{figure}

\autoref{fig:magz} shows the redshift distribution for our sample, in terms of the observed F444W NIRCam magnitude of each of our sources, as labelled by the field in which they are discovered within. These are "mag-auto'' magnitudes as measured by SExtractor. There are a couple of major features that can be seen in this figure. The first is that our selection produces galaxies with a relatively high abundance up to $z \sim 12$, but fewer galaxies at higher redshifts. This is likely due to the fact that there are a limited number of galaxies which can be found with JWST imaging at these higher redshifts, as seen in previous work \citep[e.g,][]{Adams2023,Austin2023}. There are very few bright galaxies in our $z > 6.5$ sample compared to lower redshift galaxies or those found with HST, and we find that there are galaxies as faint as mag$\sim 30$ from our deepest pointing in NGDEEP \citep[][]{Austin2023}. It also remains to be seen if these $z > 12$ galaxies remain as viable ultra high redshift galaxies once NIRSpec data on this sample is obtained. 

We can also see that certain surveys favour different redshift ranges. For example, the NEP field is good at finding galaxies at the lower redshift range of our survey, whilst the CEERS field is superior at finding slightly higher redshift galaxies. This is due to the filter set within each of the observed fields, which differs slightly between the various fields. These different filters probe the SEDs of galaxies in different ways, most notably through the location of the Lyman-break. Some filter combinations make it difficult to find lower redshift galaxies, as in the case of CEERS due to the absence of the F090W band. This limitation gives us a higher certainty on finding more distant galaxies, and less certainty on others, depending on the exact field which is being observed. This also shows the necessity and benefits of combining data from various fields which contain data at not only different areas and depths but different filters, which strongly limits at which redshifts we can find reliable distant galaxy candidates. 

\begin{figure}
\centering
\includegraphics[width=1\columnwidth]{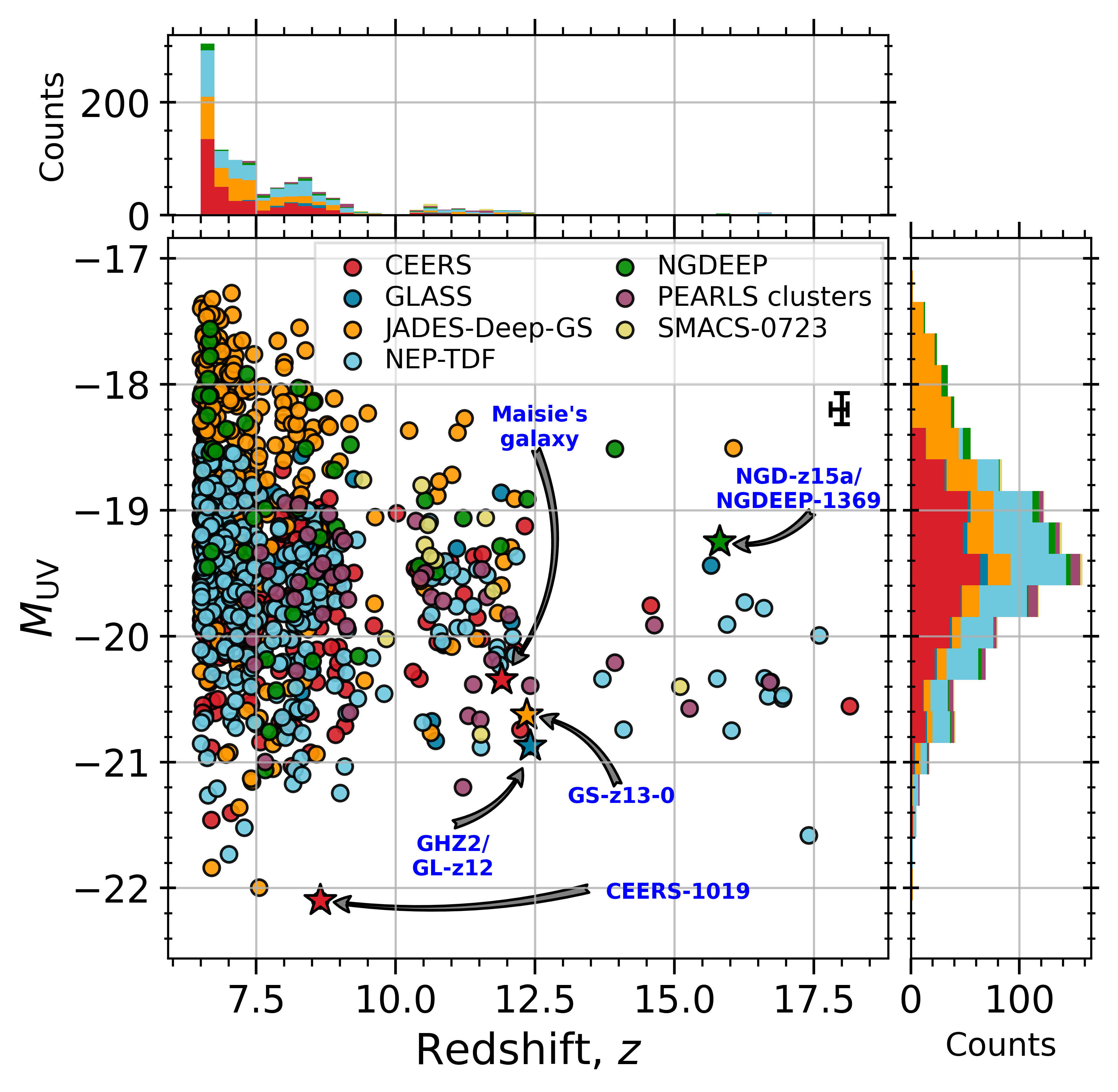}
\caption{Rest-frame absolute UV magnitudes $M_{\mathrm{UV}}$ as a function of {\tt{EAZY-py}} redshift, where the colors and highlighted galaxies are as above as also in \autoref{fig:magz}. The median $1\sigma$ error for the sample is shown in the upper right. We also show the distribution of redshifts as the upper histogram and the distribution of the absolute UV magnitudes in the histogram to the right. }
\label{fig:UVvsZ}
\end{figure}

Another feature is that the depths of the various datasets differ considerably, and it can be seen that the NGDEEP field finds both the faintest galaxies, both in terms of the F444W magnitude, as well as in absolute UV luminosity as characterized by the values of M$_{\rm UV}$ (\autoref{fig:UVvsZ}). Again we can see that the faintest galaxies in the rest-frame UV are found in the NGDEEP field, given its depth and despite its small field of view, relative to the other JWST pointings. We furthermore find several dozen galaxies at the highest redshifts $z > 12.5$ where there are still few to no spectroscopic confirmations. The properties of these galaxies are described later in this paper, but in general these are fairly bright systems, with the exception of a few galaxies found in the NGDEEP field at $z \sim 15$. These are the faintest galaxy candidates known at such low luminosities found at early times. 

%  The upper histogram shows the distribution of the photometric redshifts for our sample and the right histogram shows the magnitude distribution with the colors labelled by the fields these sources are found in. 

\subsection{Distribution and Evolution of UV Luminosity}

We have previously described in detail the UV luminosity function evolution in \citet{Adams2023b} and here we describe other features of our galaxy sample in terms of the distribution and evolution of UV luminosities for individual galaxies. In \citet{Adams2023b} we only include galaxies in the LF calculation if the completeness in any given bin is $>50\%$, and thus many galaxies in the EPOCHS sample are missing from that analysis. Here we discuss the full range of UV luminosities of our EPOCHS v1 sample. 

In \autoref{fig:UVvsZ} we show the distribution of our galaxies in terms of absolute UV magnitude as a function of redshift. As discussed in \cite{Austin2024}, our $M_{\mathrm{UV}}$ calculations are derived directly from the photometry, based on the \eazy\, photo-$z$, using the flux between 1450\AA{} and 1550\AA{} in the rest-frame.  Therefore these measurements are direct and not based on the SED fits which the other features of our galaxies are derived from, such as the stellar mass. 

This includes the trends and ranges of UV luminosities for our sample. As can be seen, we find that there is a large diversity of galaxy UV brightness consistent with an evolving LF through the first 500 Myr of cosmic time \citep[][]{Adams2023b}. We find very luminous galaxies up to $z \sim 18$, which reveals no obvious observational evolution in the upper limit of UV brightness, which otherwise might decline at higher redshifts. While we do find a gradual trend for galaxies to appear brighter at the highest redshifts, we still find objects with M$_{UV} \sim -18$ at $z \sim 11$, with some candidate systems this faint at even higher redshifts. These correspond to the bluest systems which we describe later in this paper.

 % fitting templates. Therefore these are not direct measurements of the UV luminosity, but inferred from which template is the best fit to the data. There are various systematics associated with this, although the fits we obtain are usually a good representation of the data at UV wavelengths \citep[e.g.,][]{Adams2023b, harvey2024epochs}.

\subsection{Color-Color Plots}

Another observational clue that we can use to determine the nature of the high redshift  EPOCHS galaxy population is to use color-color diagrams. This provides a purely observational picture of how the SEDs of these galaxies are distributed, which can relate to various features, such as their star formation histories, as well as the dust content, and to a lesser degree the metallicities of the underlying stellar populations. 

\begin{figure}
\centering
\includegraphics[width=1\columnwidth]{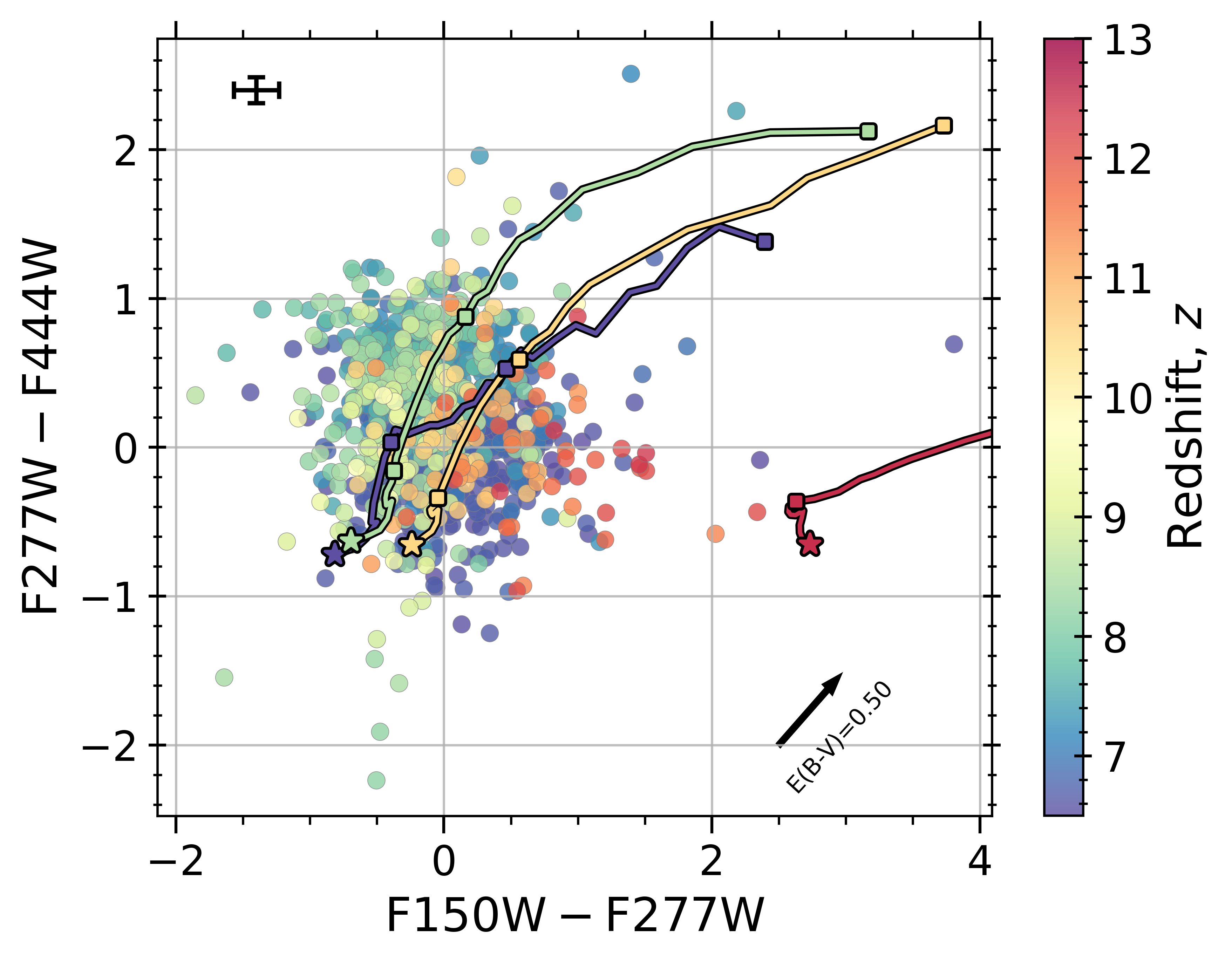}
\caption{The observed color-color diagram for the galaxies in the EPOCHS sample at $6.5 < z < 13$. We color the points by their {\tt{EAZY-py}} redshift and the median $1\sigma$ error for the sample is shown in the upper left. Overplotted are color-color tracks using v2.3 of the Binary Populations and Spectral Synthesis \citep[BPASS;][]{Eldridge2017-BPASS, Stanway2018-BPASS, Byrne2022-BPASS} SED models at redshifts $z=\{6.5, 8.5, 10.5, 12.5\}$ from rouhgly left to right, assuming a metallicity $Z=0.01$, an alpha enhancement of $\Delta \log_{10}(\alpha/\mathrm{Fe})=+0.6$, and with an IMF with slope $\Gamma=1.35$ and a high-mass cutoff of $M_{\star}=300~\mathrm{M}_{\odot}$. These models range in age from $1$~Myr to $1$~Gyr since an initial burst of star-formation, with the starred points showing results at $1$~Myr and subsequent squares showing colors at $\log_{10}(\mathrm{age/yr}) = \{7,8,9\}$. This demonstrates that our sample of galaxies is quite heterogeneous at these early times, and that there is a variety of star formation histories present beyond the simple burst we assume here. Since these models do not include attenuation by dust, we additionally show the impact of dust with the arrow in the lower right, assuming a \citet{Calzetti2000} attenuation law with $\mathrm{E(B-V)}=0.5$ at $z=8.5$.}
\label{fig:SEDs}
\end{figure}

To approach this issue we use a demonstrative sample of galaxies between $6.5 < z < 13$ and examine a color which for most galaxies in this range spans the Balmer break (F150W-F277W), and another color which reveals the properties of the galaxy's stars through the spectral shape redward of the Balmer break. This distribution is shown in \autoref{fig:SEDs} with models superimposed. This is somewhat crude representation of the SEDs of these galaxies, but still allows us to examine these properties without having to rely on detailed SED fits or other methods that are interpretative. Overall, for the bulk of this sample the Balmer break measures give us a rest-frame wavelength through the redshift range of $8.5 < z < 10.5$ of 0.13-0.15 $\mu$m to 0.24 - 0.29 $\mu$m. For the longer wavelength we are comparing 0.24 - 0.29 $\mu$m up to 0.39 - 0.47 $\mu$m. 

Immediately,
we can see that the Balmer break colors differ in magnitude spanning $\approx$ 2 mags for galaxies at $8.5 < z < 10.5$. If this is a relative measurement of age, it implies that our galaxies have a wide diversity of ages, which we later derive through the SED fitting discussed in \S 2.4. The UV color of our galaxies, more representative of recent star formation histories, also span about 2 magnitudes in color, again showing a wide diversity in ages and ongoing star formation rates for our galaxies.

Unlike for the magnitude distribution, we do not, in general, see any bias of the type of galaxies we are obtaining in our different fields, at least in terms of their colors. Thus, whilst the different JWST bands will allow only certain redshifts to be measured within a given filter set/field combination, this does not create any detailed biases in the underlying types of galaxies which are being retrieved from these fields. 

We also overplot on this diagram stellar population models of different ages. This demonstrates in a comparative way that our sample of galaxies has a diversity in star formation histories and that galaxies at high redshifts have a range of when their star formation began, and for how long and in what manner the formation histories have been ongoing. In the next section we consider these observations in more detail through SED modelling and examining the rest-frame colors of our galaxy sample. 

\begin{figure}
%\gridline{\fig{UVJ-epochs.png}{0.53\textwidth}{}
%\fig{uvj_models2.png}{0.47\textwidth}{}}
%\fig{EPOCHS_UVJ.pdf}{\columnwidth}{}
\fig{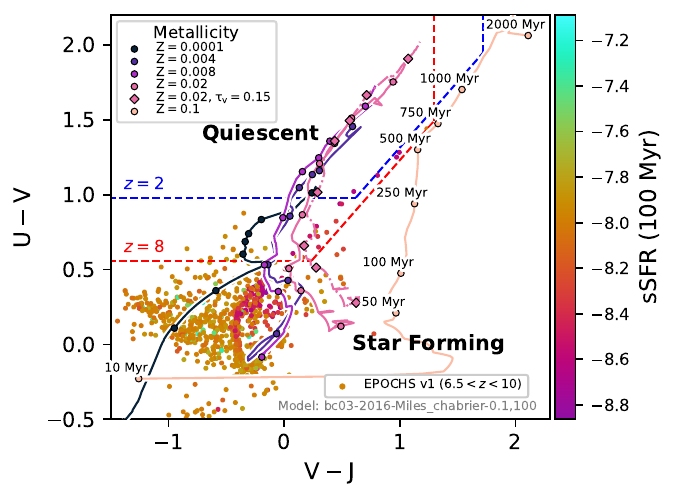}{1\columnwidth}{}
\caption{The distribution of our EPOCHS sample at $6.5 < z < 10$ in terms of the UVJ diagram, colored by specific star formation rate (sSFR). As can be seen we find a range of colors for our sample, yet we find few galaxies which are consistent with being passive, but do find some which are near the passive range.  We plot on this figure modelled evolution of stellar populations within the UVJ plane, showing models of galaxies of different metallcities (Z) within time-scales from from 10 Myr to 2 Gyr after a burst of star formation. This version shows \citet{Bruzual2003} stellar population models with 2016 Miles stellar population synthesis models, with a Chabrier IMF. Similar models using BPASS models \citep[][]{Eldridge2017-BPASS} with a Chabrier IMF shows comparable trends. For the most part, these models include the intrinsic stellar spectra, with no dust included. The time-markers are (from the bluest, bluest point redward) 10, 50, 100, 250, 500, 750, 1000, and 2000 Myr. Also shown are the z = 2 and z = 8 quiescent defined regions. }
\label{fig:uvj}
\end{figure}

Another traditional way to examine this problem is to plot the UVJ diagram for galaxies and to determine where our sample of $z > 6.5$ galaxies are located in this parameter space. We show this UVJ diagram for our sample in \autoref{fig:uvj}, with the marker color showing the \bagpipes{} derived specific star formation rate. The majority of our galaxies are in the star formation region of this parameter space, with very few of the systems approaching the area for passive galaxies. However, this is in general what is seen for galaxies at this epoch, with even galaxies with recently quenched star formation or "smouldering" galaxies found bluer than the passive region of the UVJ parameter space \citep[][]{Trussler2024}. It is also the case that galaxies do not enter the passive region of the UVJ region until $z \sim 1.5$. These systems  are mostly  compact and elliptical in morphology \citep[][]{Conselice2024}. 

\begin{figure*}
\gridline{\fig{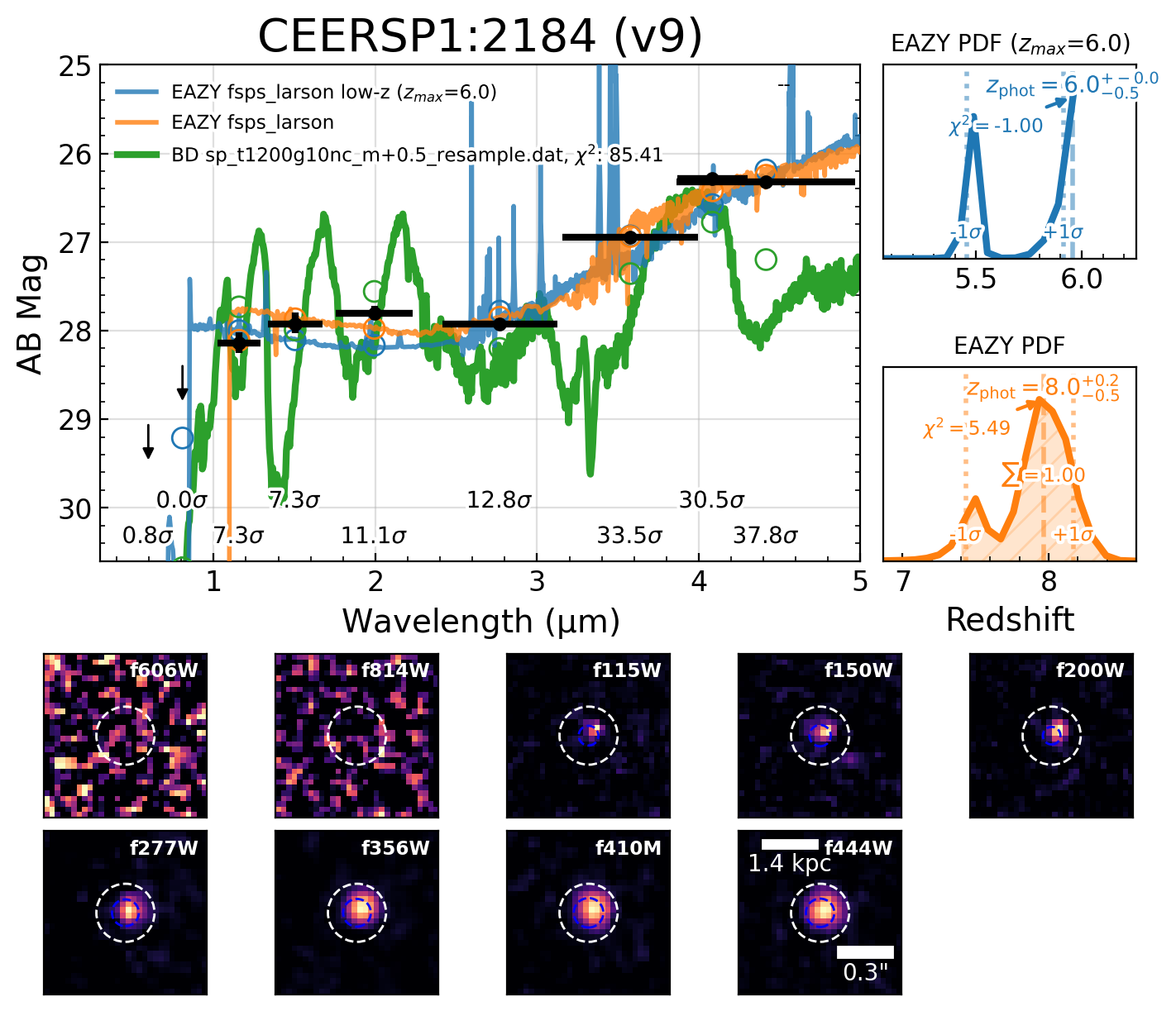}{0.47\textwidth}{}
          \fig{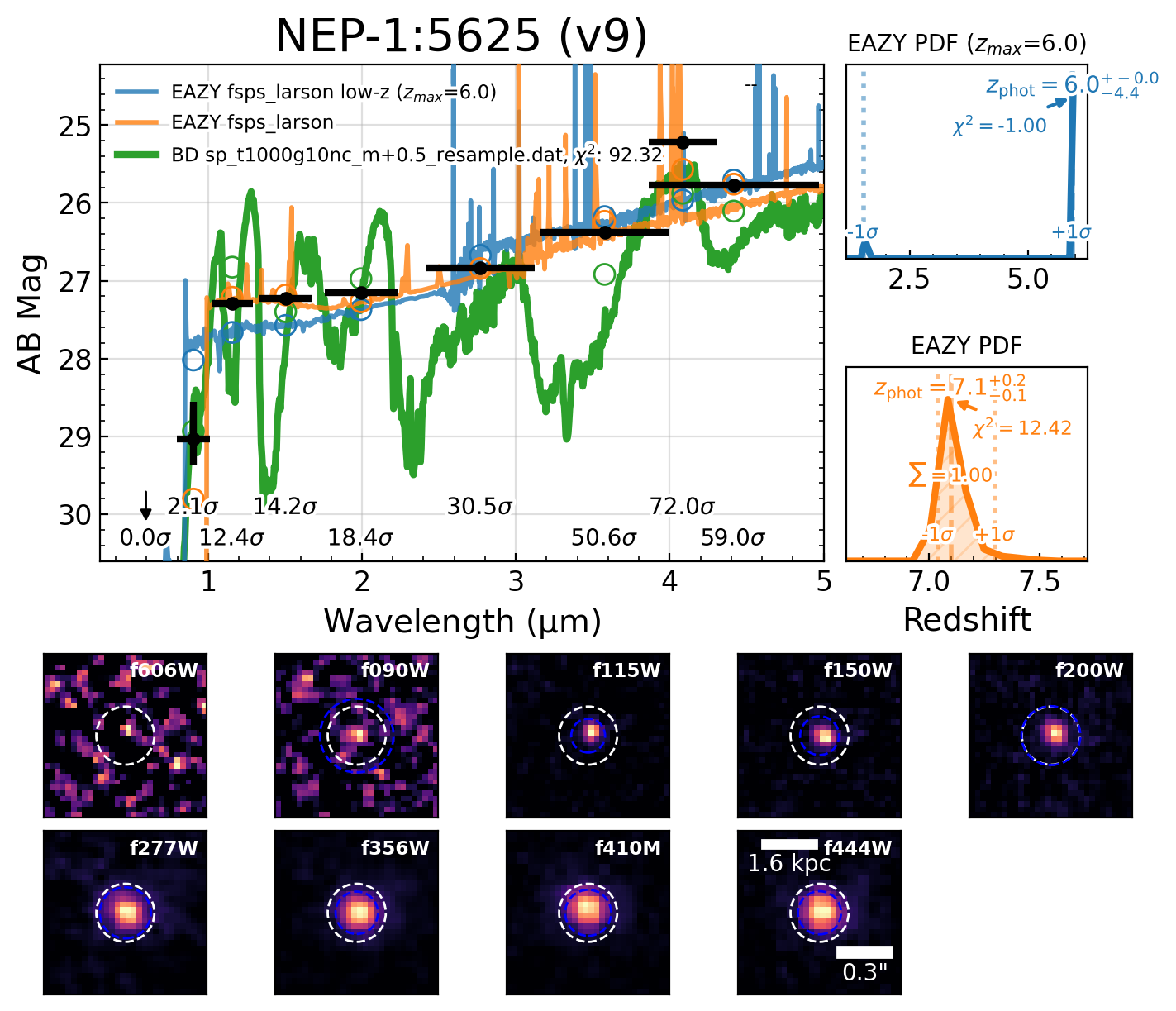}{0.47\textwidth}{}}
\gridline{\fig{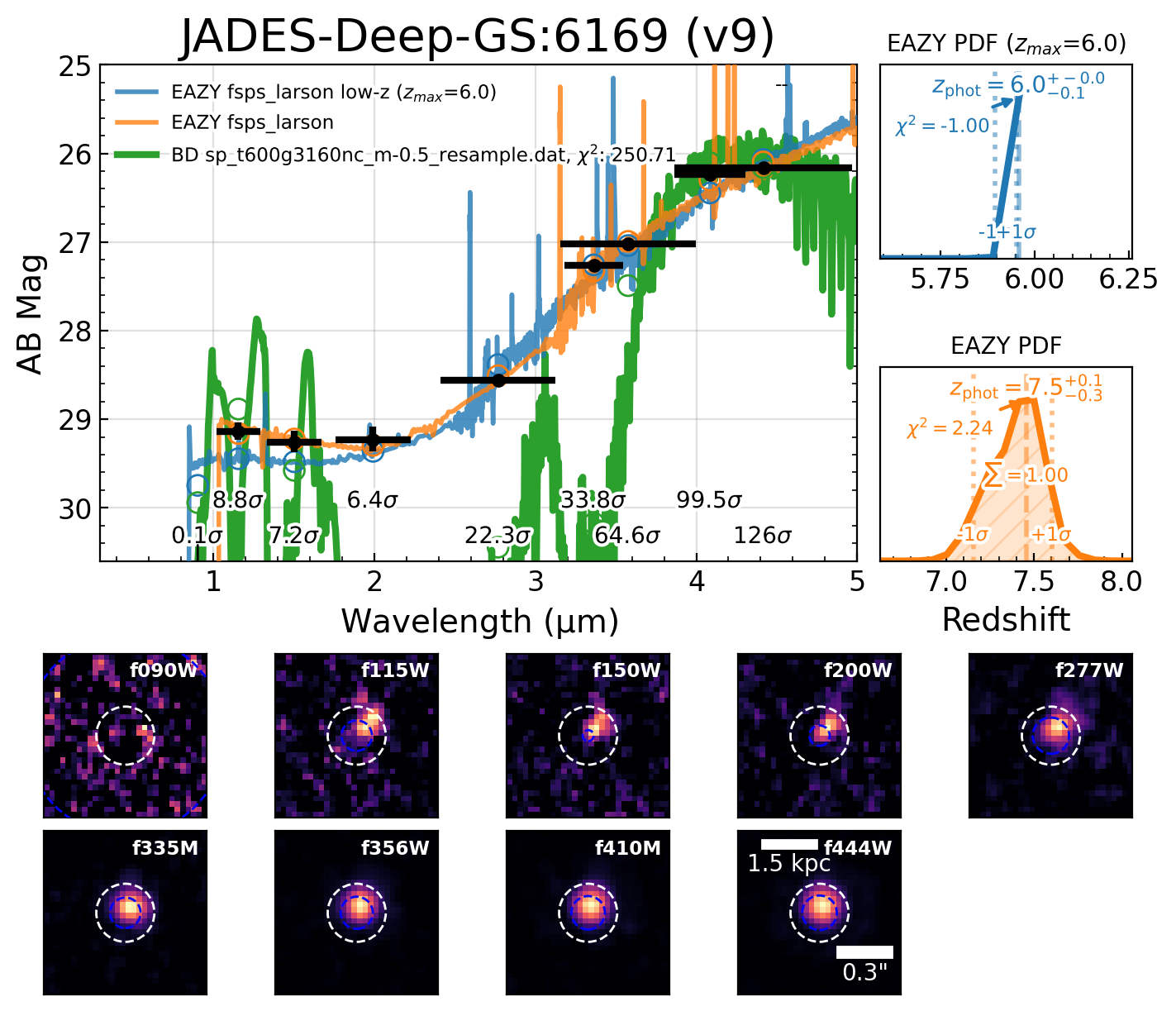}{0.47\textwidth}{}
          \fig{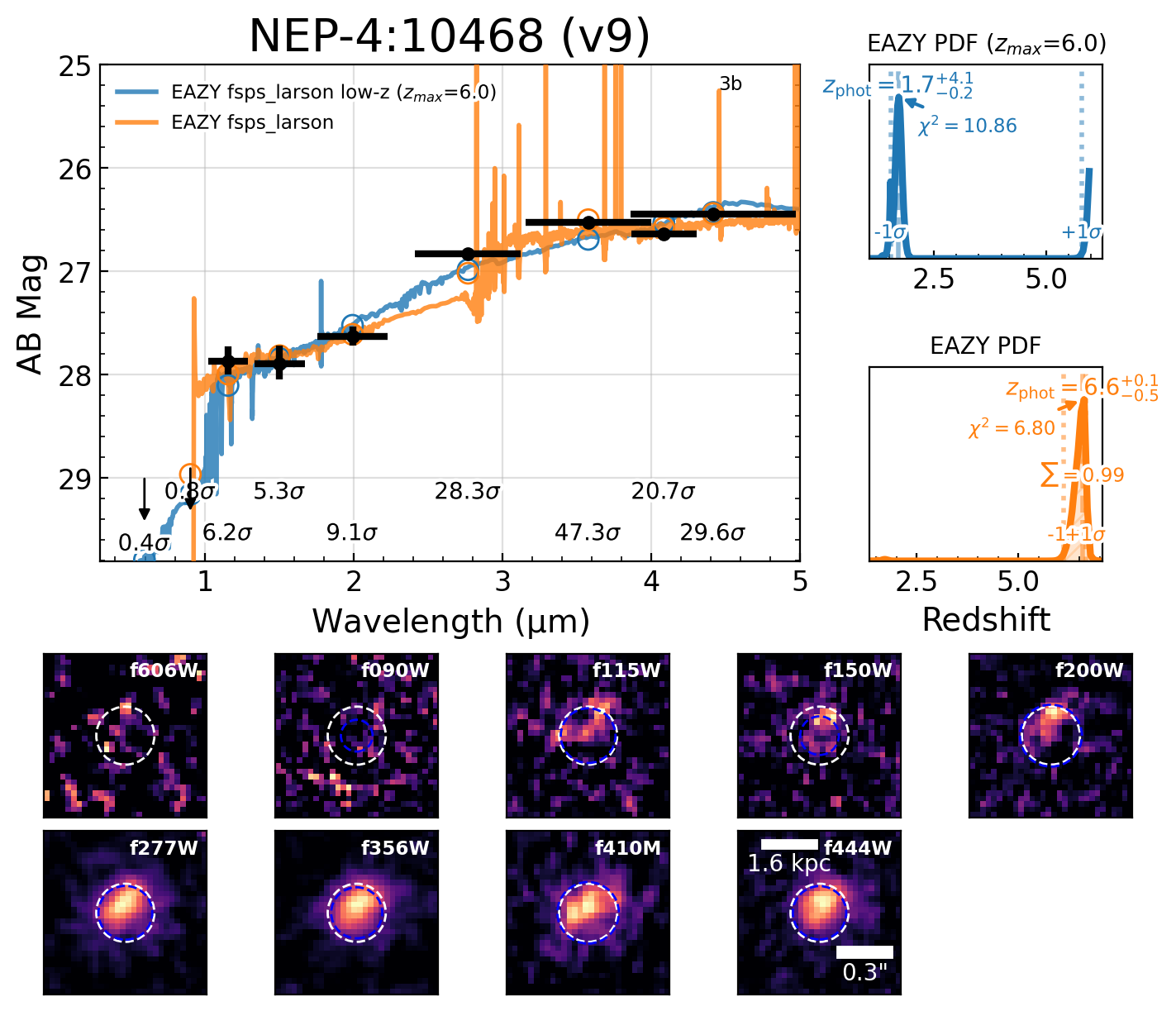}{0.47\textwidth}{}}
%\includegraphics[width=\linewidth]{sed_plots/2184.png}
%\end{subfigure}
%\begin{subfigure}{0.47\textwidth}
%\includegraphics[width=\linewidth]{sed_plots/5625.png}
%\end{subfigure}
%\begin{subfigure}{0.49\textwidth}
%\includegraphics[width=\textwidth]{sed_plots/6169.png}
%\end{subfigure}
%\begin{subfigure}{0.49\textwidth}
%\includegraphics[width=\textwidth]{sed_plots/10468.png}
%\end{subfigure}
\caption{Spectral energy distribution fits for redder galaxies in our sample. These various panels show the redshift PDFs and photometry for four representative high$-z$ galaxies with red colors for $z > 6.5$ galaxies. These systems are at $z \sim 7-9$ where we have the largest range of galaxy properties given our large sample. The measured photometry for the NIRCam observations are shown in black, with best-fitting EAZY SEDs shown in blue, and in orange for a low-redshift prior at $z < 6$ and a free fit to the redshift,  respectively. The green line shows the best fit to templates of brown dwarfs. Overlaid on the redshift PDFs are the selection statistics, including the photometric redshift estimates. The bottom section shows the cutouts of these galaxy candidates in the NIRCam photometric bands, on a log color scale. Note that these galaxies are from different fields which contain different observed bands. }
\label{fig:SEDs_red}
\end{figure*}

\begin{figure*}

\gridline{\fig{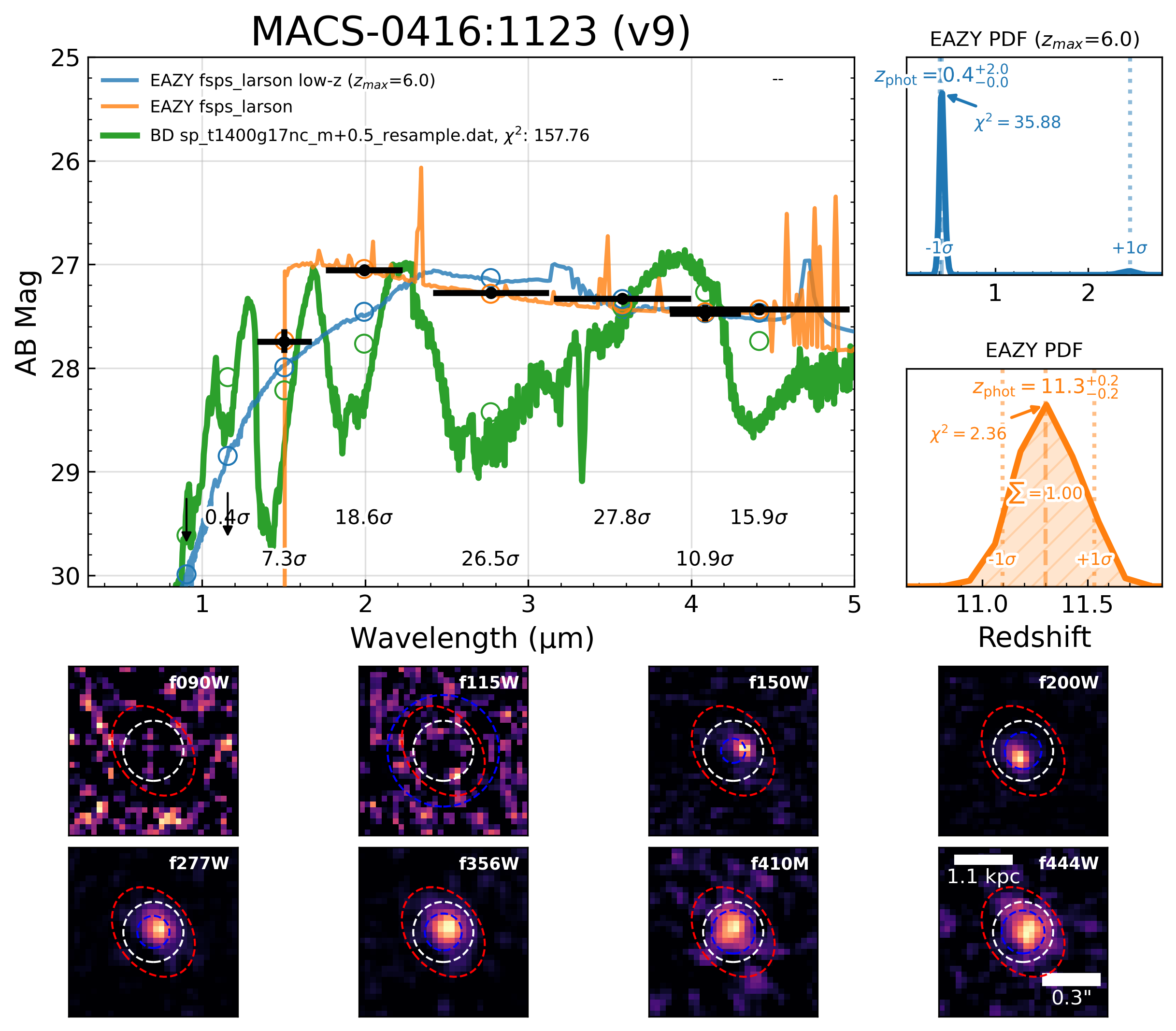}{0.47\textwidth}{}
          \fig{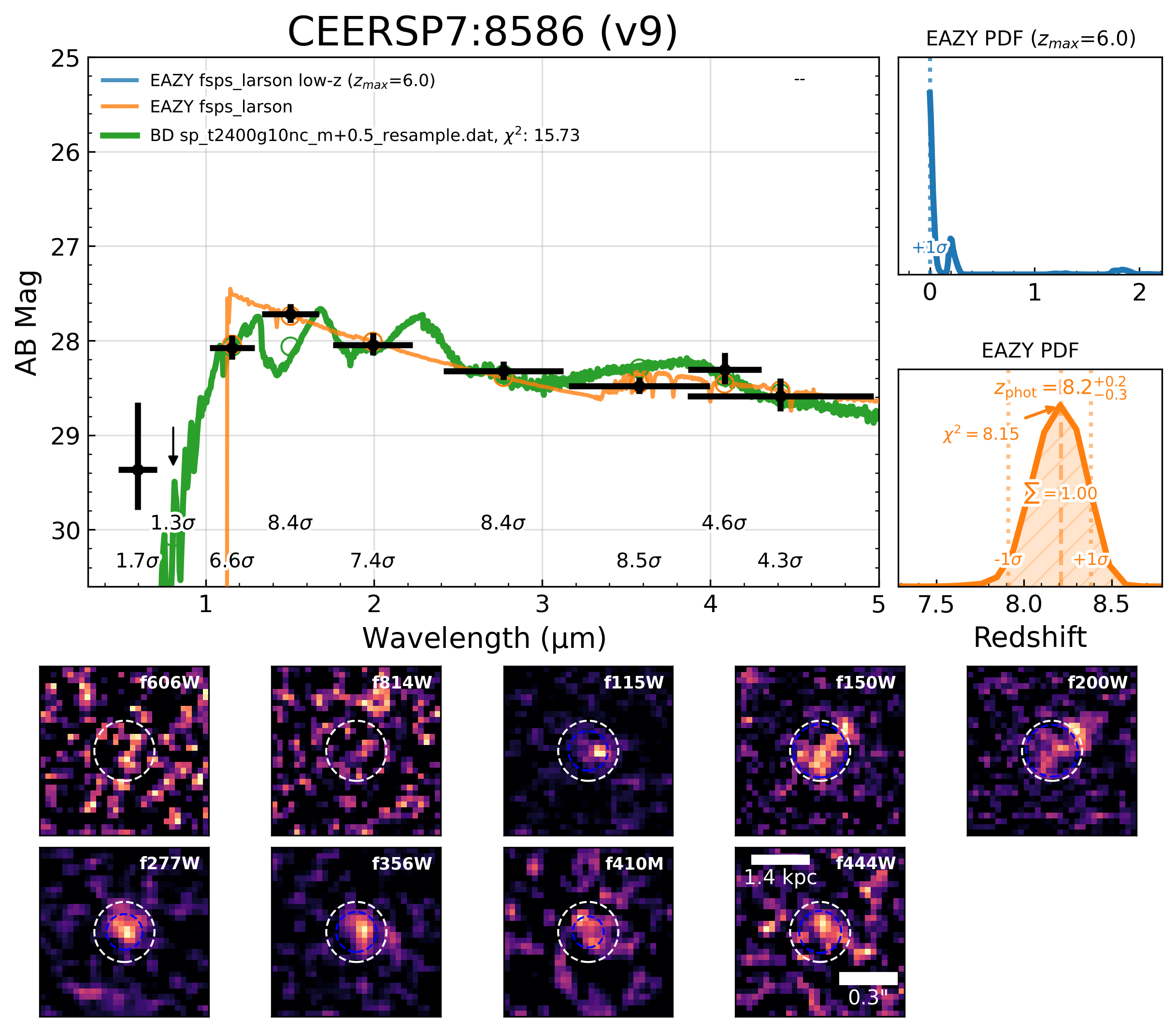}{0.47\textwidth}{}}

\gridline{\fig{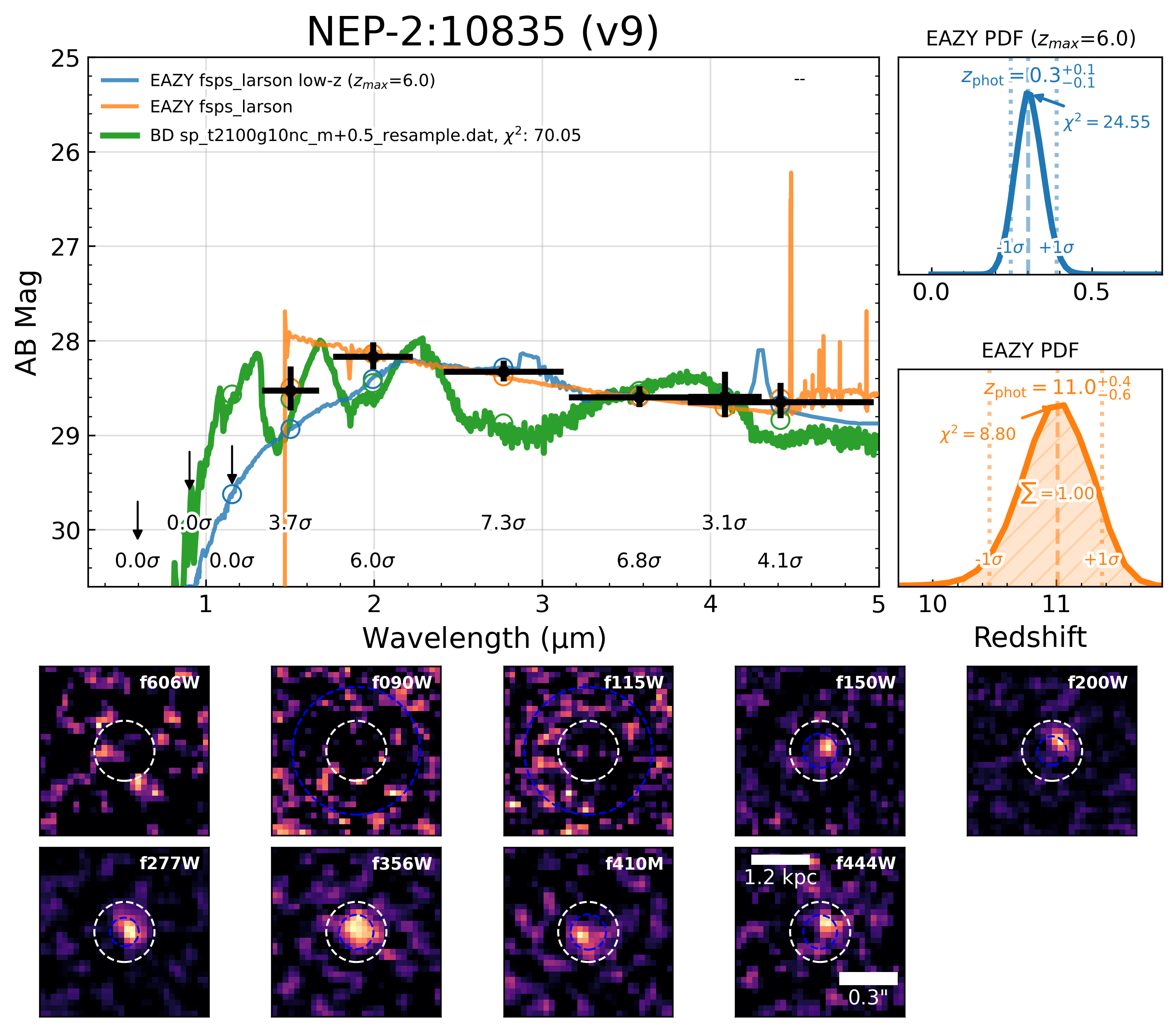}{0.47\textwidth}{}
          \fig{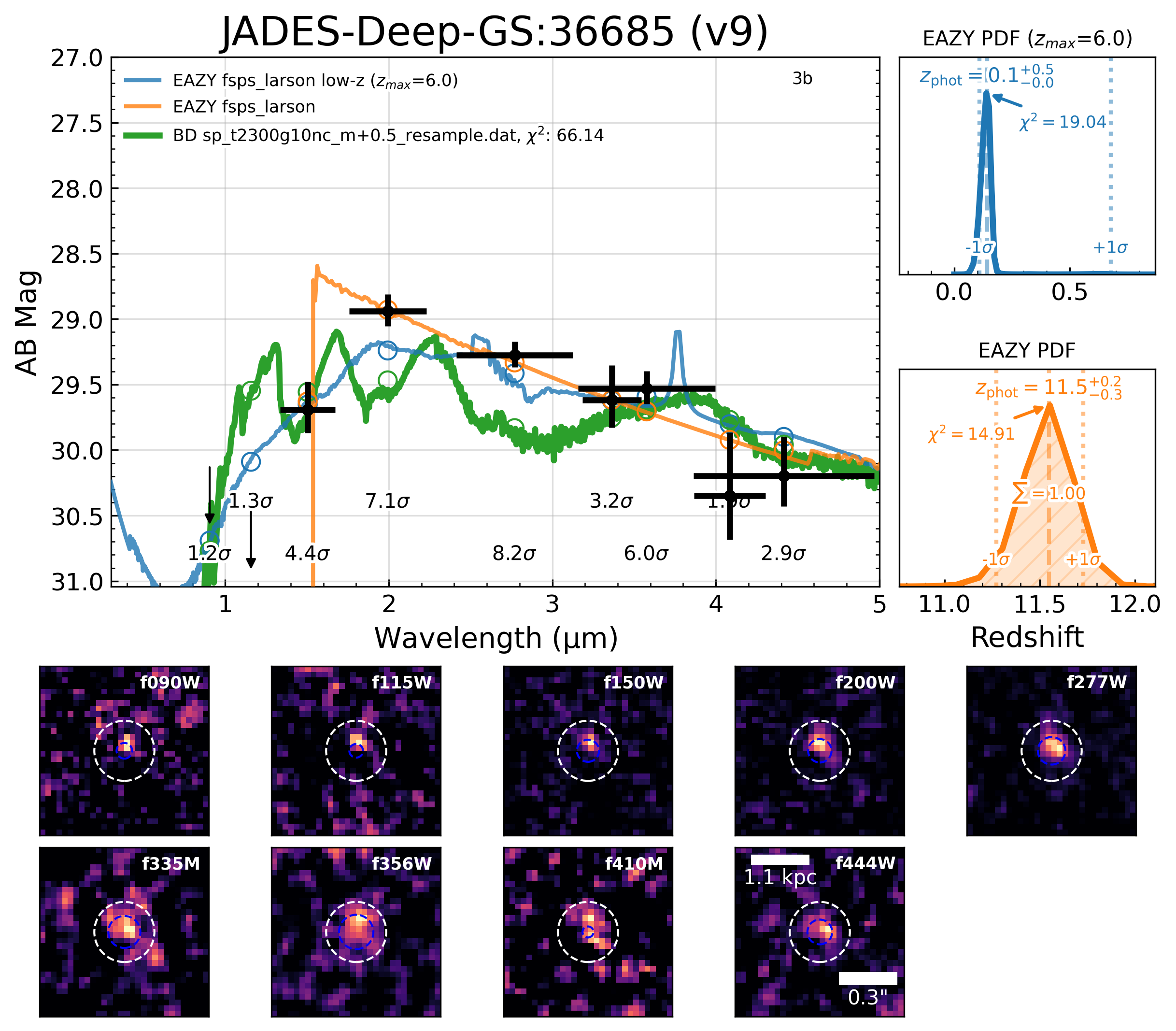}{0.47\textwidth}{}}
          
%\begin{subfigure}{0.47\textwidth}
%\includegraphics[width=\linewidth]{sed_plots/1123.png}
%\end{subfigure}
%\begin{subfigure}{0.47\textwidth}
%\includegraphics[width=\linewidth]{sed_plots/8586.png}
%\end{subfigure}
%\begin{subfigure}{0.49\textwidth}
%\includegraphics[width=\textwidth]{sed_plots/10835.png}
%\end{subfigure}
%\begin{subfigure}{0.49\textwidth}
%\includegraphics[width=\textwidth]{sed_plots/36685.png}
%\end{subfigure}
\caption{Similar to the previous plots in \autoref{fig:SEDs_red}, but for bluer examples in our sample. These panels show the redshift PDFs and photometry for four representative high$-z$ galaxies with blue colors from our $z > 6.5$ sample. The features of the galaxies displayed here are the same as in \autoref{fig:SEDs_red}. These blue systems are those found at $z \sim 8-11$ where we have the largest range of galaxy properties within our large sample. The measured photometry for the NIRCam observations are shown in black, with the  best-fitting EAZY high-redshift SEDs shown in orange, with the best $z<6.5$ solution shown  in blue. For comparison in green we also show the best-fitting brown dwarf template, taken from fitting all the Sonora Bobcat and Cholla synthetic brown dwarf templates \citep{marley2021sonora, karalidi2021sonora}. Overlaid on the redshift PDFs are the selection statistics, including the photometric redshift estimates. The bottom section shows the cutouts of these galaxy candidates in the NIRCam photometric bands, on a log color scale. Note that these galaxies are from different fields which contain different bands for observations. }
\label{fig:SEDs_blue}
\end{figure*}

\begin{figure}
\centering
\includegraphics[width=1\columnwidth]{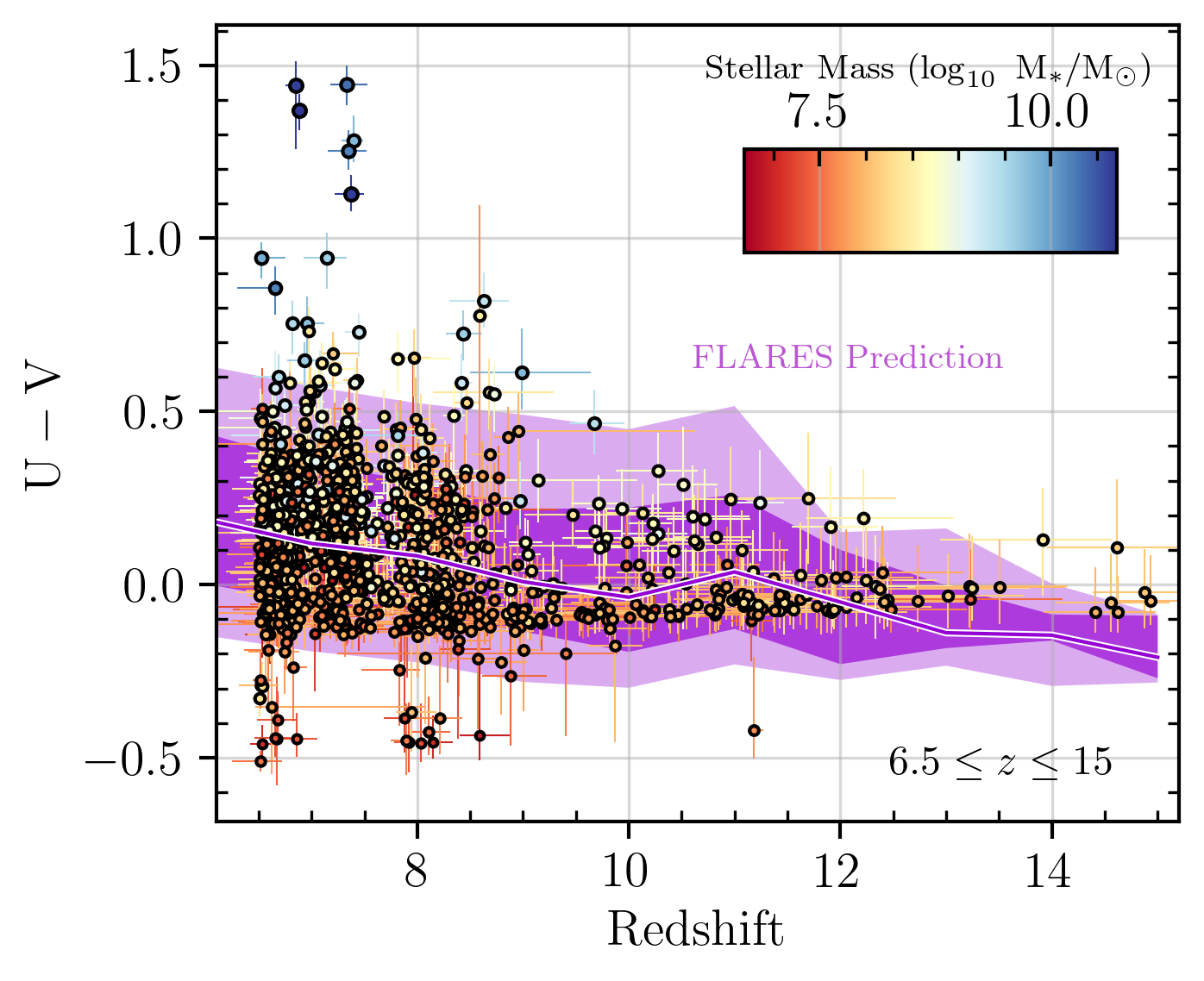}
\caption{Plot of the rest-frame (U$-$V) color for our sample vs. redshift, colored by the stellar mass as defined in the upper right shading. UV colors, redshift and stellar masses are taken from the \bagpipes{} SED fitting, as discussed in \autoref{sec:bagpipes} Uncertainties are taken from the 16 and 84th percentiles of the \bagpipes{} posterior distributions. We compare on this diagram simulation output from the FLARES simulation, showing the predicted color evolution for galaxies. The FLARES simulations are however only comparable down to that simulation's resolution limit which is M$_{*} = 10^{8}$ \solm. }
\label{fig:UVRedshift}
\end{figure}

\begin{figure*}

\gridline{\fig{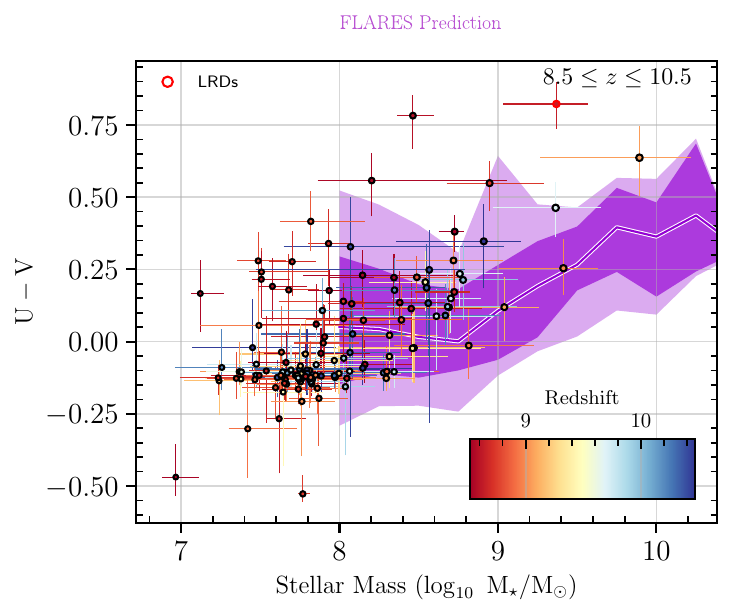}{0.47\textwidth}{}
          \fig{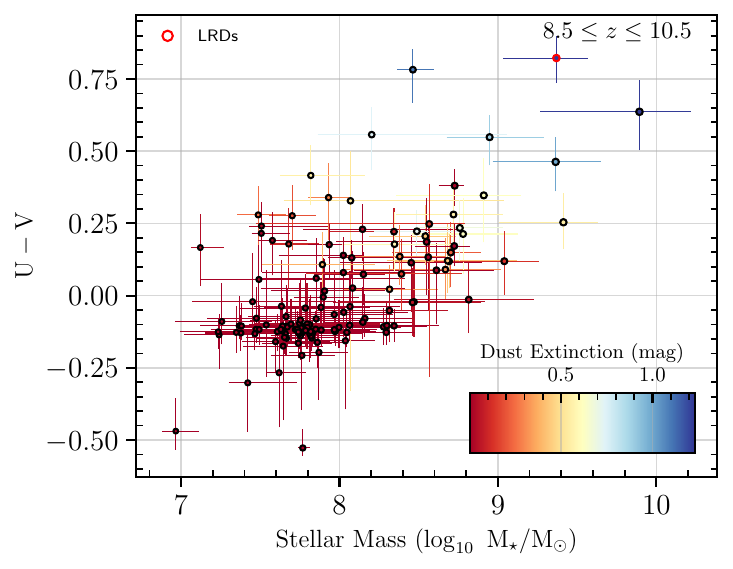}{0.47\textwidth}{}}

\gridline{\fig{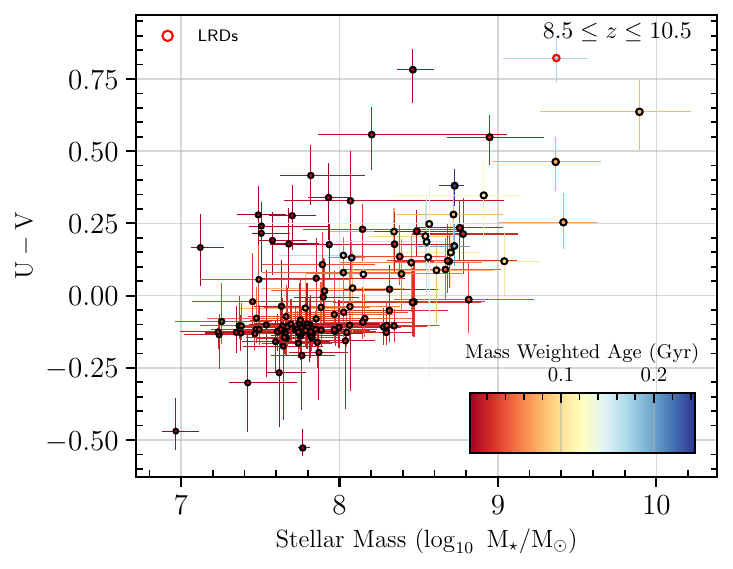}{0.47\textwidth}{}
          \fig{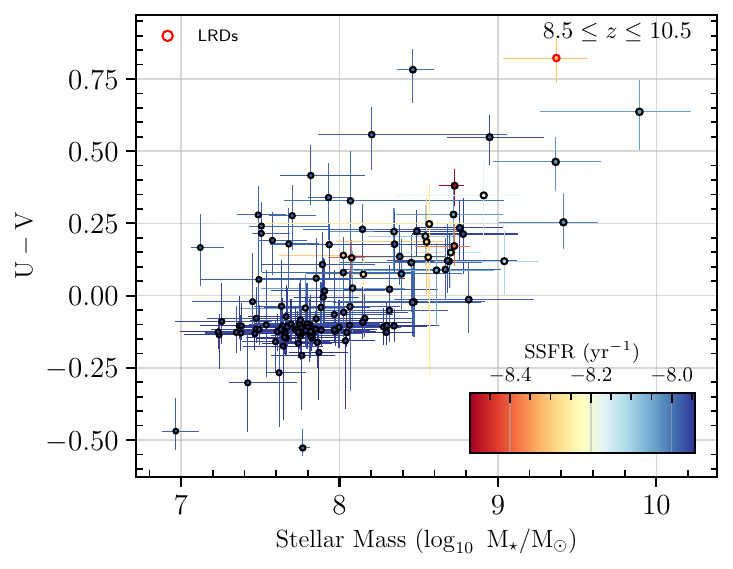}{0.47\textwidth}{}}
          
%\centering
%\includegraphics[width=2.2\columnwidth]{colorvsmass.png}
%\begin{subfigure}{0.49\textwidth}
%\includegraphics[width=\textwidth]{EPOCHS_8.5_z_10.5_stellar_mass_50_UV_colour_50_credshift_50_zgauss.png}
%\end{subfigure}
%\begin{subfigure}{0.49\textwidth}
%\includegraphics[width=\textwidth]{EPOCHS_8.5_z_10.5_stellar_mass_50_UV_colour_50_cdust:Av_50_zgauss.png}
%\end{subfigure}
%\begin{subfigure}{0.49\textwidth}
%\includegraphics[width=\textwidth]
%{EPOCHS_8.5_z_10.5_stellar_mass_50_UV_colour_50_cmass_weighted_age_50_zgauss.png}
%\end{subfigure}
%\begin{subfigure}{0.49\textwidth}
%\includegraphics[width=\textwidth]{EPOCHS_8.5_z_10.5_stellar_mass_50_UV_colour_50_cssfr_50_zgauss.png}
%\end{subfigure}
\caption{The relation between the stellar mass and (U-V) rest-frame colors of the EPOCHS v1 sample at a relatively narrow redshift range of $8.5 < z < 10.5$ as measured with \bagpipes{}. From the top left the markers are colored by redshift, dust extinction $A_{\rm V}$, mass-weighted age, and specific star formation rate (sSFR). As can be seen, there are strong trends here between mass and color, and within these redder/bluer colors there are further correlations with the star formation rate as well as the age of the stellar populations. The shaded purple region in the upper left panel shows the FLARES prediction for these two quantities. }
\label{fig:UVmass}
\end{figure*}

\begin{figure*}
\gridline{\fig{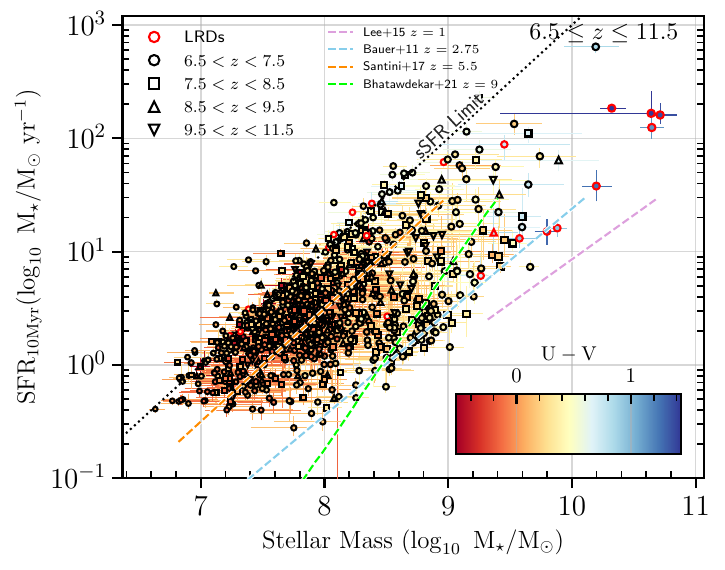}{0.47\textwidth}{}
\fig{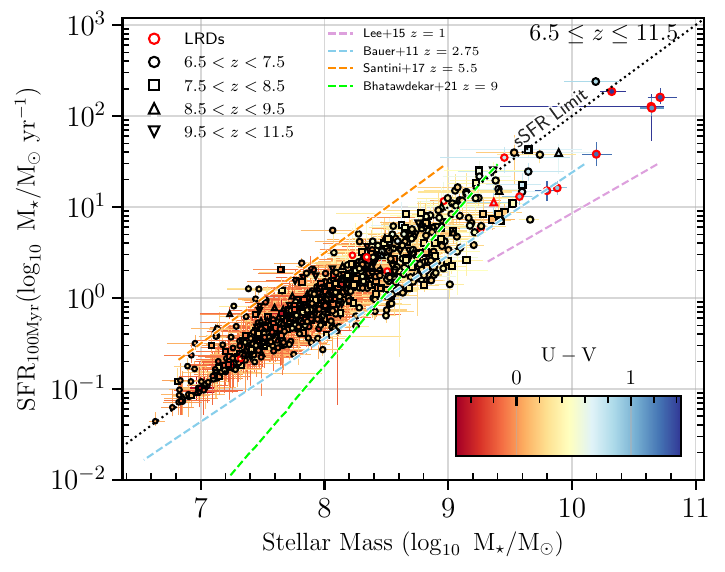}{0.47\textwidth}{}}
\caption{Star forming main sequence from our fiducial \bagpipes{} run, shown with comparison to \cite{Bauer2011, Lee2015, Santini2017, Bhatawdekar2021}. We show average star formation on both a 100 Myr (left) and 10 Myr (right) timescale. The stellar masses and star formation rates shown have been corrected from their aperture-derived values by scaling the quantities by the ratio of the aperture-derived flux to the flux within a elliptical Kron aperture enclosing the full galaxy. We use the band closest to the rest-UV to correct the SFR, and the F444W to correct the stellar masses for aperture effects. We color the points by their rest-frame U-V color, and the marker shape distinguishes the redshift bin of the galaxy. Galaxies which meet the little red dot (LRD) 'red2' criteria of \citet{kokorev2024census} are shown with a red border, to highlight that the nature of these sources is uncertain and the SF and stellar mass estimates shown are highly uncertain as these systems could very well turn out to be dominated by AGN. }
\vspace{0.5cm}
\label{fig:sf_mainsq}
\end{figure*}

\subsection{Rest-frame Color Evolution}

While the color-color plots are demonstrative of broad features, they are limited in that they span different rest-frame wavelengths depending on the redshift of the object and are degenerate to some degree between ages, star formation rate, and dust content. To understand these sources better we carry out detailed SED fitting using the \bagpipes{} code introduced in \autoref{sec:bagpipes}. Some examples of these fits are shown in \autoref{fig:SEDs_red} for systems which have red colors, while those shown in \autoref{fig:SEDs_blue} are examples for blue galaxies. Also shown below these fits are the images of the galaxies as imaged in the different bands in which they are imaged and where the photometry is measured. 

Based on these fits we can retrieve the rest-frame colors of our sample based on these best fits, as shown in the SED figures. We thus calculate the rest-frame colors of our objects using the best-fitting templates and measure the rest-frame (U-V) colors or these galaxies. This is an indicative color as it straddles the Balmer break, and thus gives us a good representation concerning how blue or red our galaxies are within this wavelength difference.  

We show the rest-frame (U-V) colors of our sample in \autoref{fig:UVRedshift} plotted as a function of redshift. This figure shows a range of colors at all redshifts, but with a gradual change in average color, such that  galaxies at higher redshifts are bluer in their rest-frame (U-V) colors. In addition to this, we also find that there are a collection of galaxies which appear quite blue. We also compare our results in this figure to the models from the FLARES simulation, demonstrating that there is a relatively good agreement between the data and model, however, our sample is on average slightly redder than the models at progressively lower redshifts.

To understand the origin of our sample in more detail, we examine the relationship between the rest-frame (U$-$V) color and stellar mass in \autoref{fig:UVmass}. We limit this part of our analysis to galaxies at redshifts $8.5 < z < 10.5$, as this is the redshift where we have the largest sample of galaxies, as well as the most reliable in photometric redshifts. When we examine this relationship at this narrow redshift range, we find that there is already present, at such early times, a strong correlation between this color, which straddles the Balmer break at these redshifts, and the stellar mass. This correlation is such that more massive galaxies  are redder in their rest-frame colors. Therefore, what we see is a population of relatively massive red galaxies already established at this early point in the universe's history. 

This also shows that while there is a diversity of SEDs forms and shapes for the high redshift sample, there is a trend of color with stellar mass. We discuss the origin of this trend in \autoref{sec:discussion}, including comparing these data to simulations to understand their nature. This result is however unlikely to be due to selection effects. The reason for this is that we would have easily seen any blue bright massive galaxies at these early types. It is clear that galaxies which are relatively red dominate the massive galaxies sample at  $z > 8.5$. 

We can also see in the upper left plot of \autoref{fig:UVmass} that there is a redshift evolution within these diagrams, such that the higher redshifts appears to form a bluer sequence of points, while the lower redshift galaxies appear to form a similar slope, but redder offset sequence plotted about the higher redshift sequence we see. What remains to be determined from this is how our measurements of color depend on the measurement of the star formation rate for our galaxies, which we examine the next section. 

\subsection{The Early Galaxy Main Sequence}

Using our data and results we can probe the formation of the main-sequence of galaxies. This main sequence is such that there exists at lower redshifts a well-defined correlation between galaxy stellar mass and star formation rate \citep[e.g.,][]{Noeske2007,Bauer2011, Santini2017, Bhatawdekar2021}. This correlation is such, that galaxies with a higher stellar mass have a higher star formation rate, which is independent of how the star formation rate is measured \citep[e.g.,][]{Noeske2007}.

We show the galaxy main sequence for our EPOCHS v1 sample in \autoref{fig:sf_mainsq} from our fiducial \bagpipes{} results, on both a 10 Myr and 100 Myr average star-formation timescale. We show here the star formation vs. stellar mass relation for our galaxies using two different time-scales at 10 Myr (on the right) and over the past 100 Myr (on left). These star formation time-scales show the average star formation rate over this time-period. One thing to immediately notice is that we see a trend such that, on average, systems with a high stellar mass have a higher star formation rate. The exact trend of this is difficult to quantify as the exact value will depend on the form of the assumed star formation history in which these star formation measures are taken from.  However, we find that there is very little evolution in the form of the agreement between these two quantities within redshift or within different star formation time-scales. 

Clearly the scatter for points is larger in the case of the longer time-scales, meaning that the star formation rates for galaxies narrows at a given stellar mass for more recent star formation events in the past 10 Myr. The scatter in the shorter time-scale (10 Myr) can range over a factor of $\sim 30$, whilst for the longer time-scale (100 Myr) the range is over a factor of $\sim 5$. 

We also find that the star formation rates are higher compared to most studies, including SFRs that are a factor of $\sim 5$ higher than the $z = 9$ main sequence from HST observations \citep[][]{Bhatawdekar2021}. In general we see a higher star formation rate at a given mass for nearly all comparisons, with the exception of \citet{Lee2019} who is roughly similar to our values, in particularly the 10 Myr star formation rate measures. 

We can also see in the sSFR panel (bottom right) of \autoref{fig:UVmass}, the red `massive' galaxies' generally have the lowest specific star formation rates, as well as the oldest ages. Whilst there is still ongoing star formation rate in these systems with values up to a few \solm yr$^{-1}$, the relative proportion of star formation within the existing stellar masses of these galaxies is much lower than for the bursty lower mass galaxies. The ages are also older, which is a sign that the bulk of the star formation for these red massive galaxies is much further back in time than the bluer lower mass systems that we have in our sample. This is a firm indication that these galaxies should certainly be forming at even higher redshifts than $z = 12$, a topic which we investigate in the next section concerning the star formation histories of our sample.

\subsection{Star formation history and stochasticity}

A proper analysis of the main sequence of the distant star forming galaxies in our sample requires knowledge of the star formation histories (SFHs) in our sample. Predictions from both numerical simulations \citep[e.g.][]{Sun2023, Dome2024} and recent observational evidence from JWST \citep[e.g.][]{Endsley2023, Dressler2024} now support the hypothesis that star-formation becomes burstier at high redshift. This is in the sense that galaxies do not have a gradual declining or increasing star formation rate with time, but that star formation happens in `bursts' whereby the star formation rate becomes rapidly higher for a relatively short period of time before declining again. This process is in terms of time a random one, or stochastic.

The fact that galaxies may appear to be burstier at high redshifts could be a selection bias.  To reason is that we expect galaxies towards the detection limit of our sample to be increasingly bursty in the last 10 Myr, as these events enhance their UV luminosities, making them easier to detect and thus are included in our sample. In addition to bursty SFHs, several mini-quenched (or smouldering) galaxies with a ``lull'' in star-formation activity with weak emission lines have been observed with JWST \citep{Strait2023, Looser2023, Trussler2024, Looser2024} up at redshifts $z\sim7$, perhaps caused by radiation-feedback driven outflows \citep{Ferrare2022}, bulge-formation \citep{Lu2021}, AGN feedback \citep{Nelson2021}, or environmental quenching processes \citep{Williams2021}.

\begin{figure*}
\centering
\includegraphics[width=1.98\columnwidth]{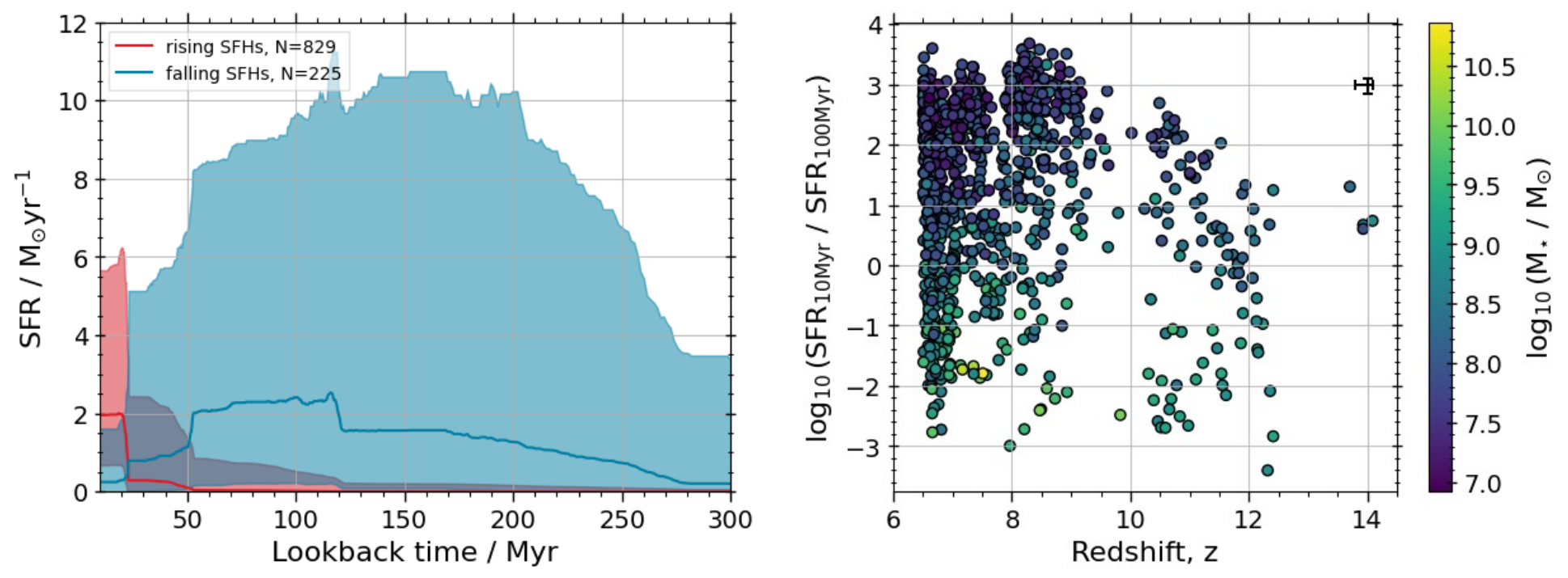}
\caption{Plots showing the different ways in which to represent our EPOCHS sample's properties. On the left we show the distribution of SFR for two defined samples - those that have a rising star formation (in pink) and those with a falling star formation rate (in blue). The shaded areas show the 86\% and 14\% distribution of these values. The right panel shows the distribution of the ratio of the star formation rates within two different time-scales (10 Myr) and (100 Myr) as a function of redshift. The scale shows the value of the stellar mass. A clear trend whereby the lower mass galaxies have a burstier star formation history than the higher mass galaxies in our observed sample. }
\label{fig:SFburstiness}
\end{figure*}

We investigate the star-formation histories of our EPOCHS v1 sample by re-running through the \bagpipes{} Bayseian SED-fitting code \citep{Carnall2018_Bagpipes, carnall2019galaxy} with our fiducial setup, but this time adopting the non-parametric ``continuity bursty'' SFH model of \citet{Leja2019} used in both \citet{tacchella2022stellar} and \citet{harvey2024epochs}. Since the wideband photometric data alone does not necessarily constrain the SFHs of individual galaxies without spectroscopic information, in \autoref{fig:SFburstiness} we stack the SFHs of galaxies with both recently ``rising'' and ``falling'' SFHs, split by star-formation burstiness parameter ($\phi$), which we define as:

\begin{equation}
\phi=(\mathrm{SFR}_{10\mathrm{Myr}}~/~\mathrm{SFR}_{100\mathrm{Myr}}), 
\end{equation}

\noindent with the time-scales those prior to their observed epoch. 
 There are two ways in which we use this burstiness parameter $\phi$. The first is that we examine the star formation history of our sample divided up into systems based on the values of $\phi$, such that $\phi > 1$ are considered bursty, and $\phi < 1$ are considered non-bursty. This simple division gives us an idea of a population which has had a recent burst vs. those whose star formation was higher in the relatively distant past compare with more recent star formation. What we in fact find when categorizing galaxies this way is that most of our sample are bursty galaxies. We find that 829 of the systems are bursty while 225 are non-bursty. It is unclear the interpretation of this, as we might be biased by finding systems which are bursty, but at the same time most galaxies at these higher redshifts are undergoing increasing amounts of star formation. What is likely is that we would not detect using our existing JWST data all of the progenitors of our bursty galaxies at higher redshifts, earlier than at the epoch than we observe them. 

 First, we show the star formation history of each of the bursty and non-bursty systems over the time period 10-300~Myr in the left panel of \autoref{fig:SFburstiness}. What this shows is that the bursty systems have most of their star formation occurring later in their history, whilst the non-bursty galaxies have a more drawn out history.
 
 We also show in \autoref{fig:SFburstiness} the burstiness parameter as a function of both redshift and stellar mass.  This distribution allows us to determine which systems are bursty, as we define this above, and which have had their star formation more dominate in the relatively distant past.  This figure allows us to see how this parameter changes for galaxies at different redshifts and stellar masses.  What is interesting is that in terms of redshift we find that the most bursty events are all towards the lower end of our redshift range.  This is another indication that samples of galaxies selected with JWST are biased towards more bursty systems.  Another feature to note is that the least massive galaxies exhibiting burstier SFHs. This is the case over all redshifts, such that the higher mass galaxies are those that have a more expanded history, and this hold at all redshifts.  Whatever is producing this trend, we find that it is present up to the highest redshifts where we can find galaxies.

\subsection{Ultra High-z Galaxy Properties at $z > 12$}

One of the interesting features of our sample is that even after we check and remove the vast majority of potentially high redshift galaxies, we are still left with a number of high quality candidate ultra-high redshift galaxies at $z > 12$. While there are now confirmations of galaxies at these high redshifts through spectroscopy \citep[][]{Curtislake2022, Carniani2024}, very few candidate galaxies at these redshifts have spectroscopic confirmation. The spectra of these systems so far are such that they generally lack emission lines or other features, beyond having a Lyman-break. Thus, even with spectroscopy it is different to learn much about these systems, and a photometric approach using the SED fitting we have discussed is an important method for understanding these systems and how they may have formed when the universe was less than $\sim 350$ Myr old. 

\begin{figure*}
\centering
\includegraphics[width=1.0\columnwidth]{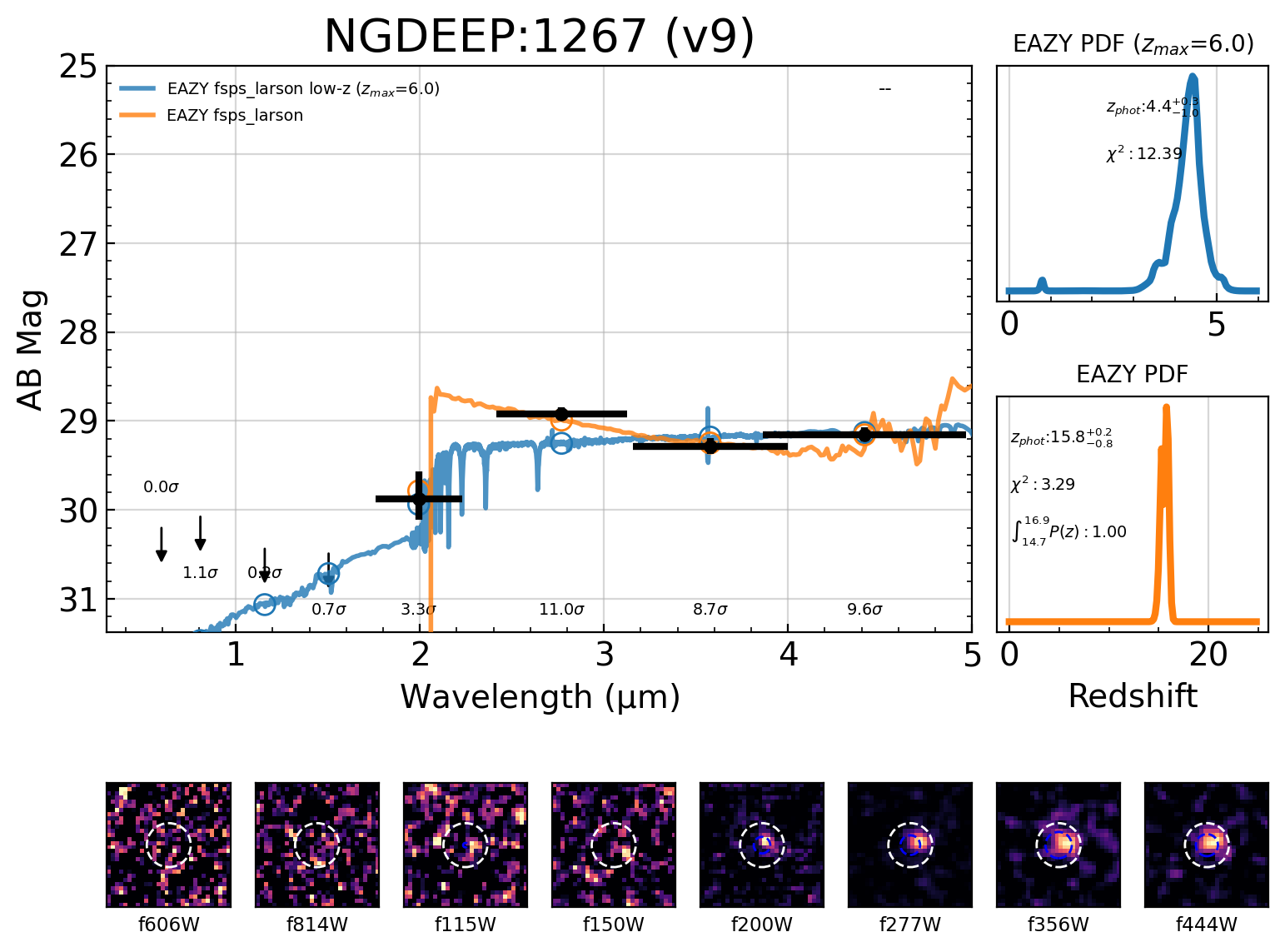}
\includegraphics[width=1.0\columnwidth]{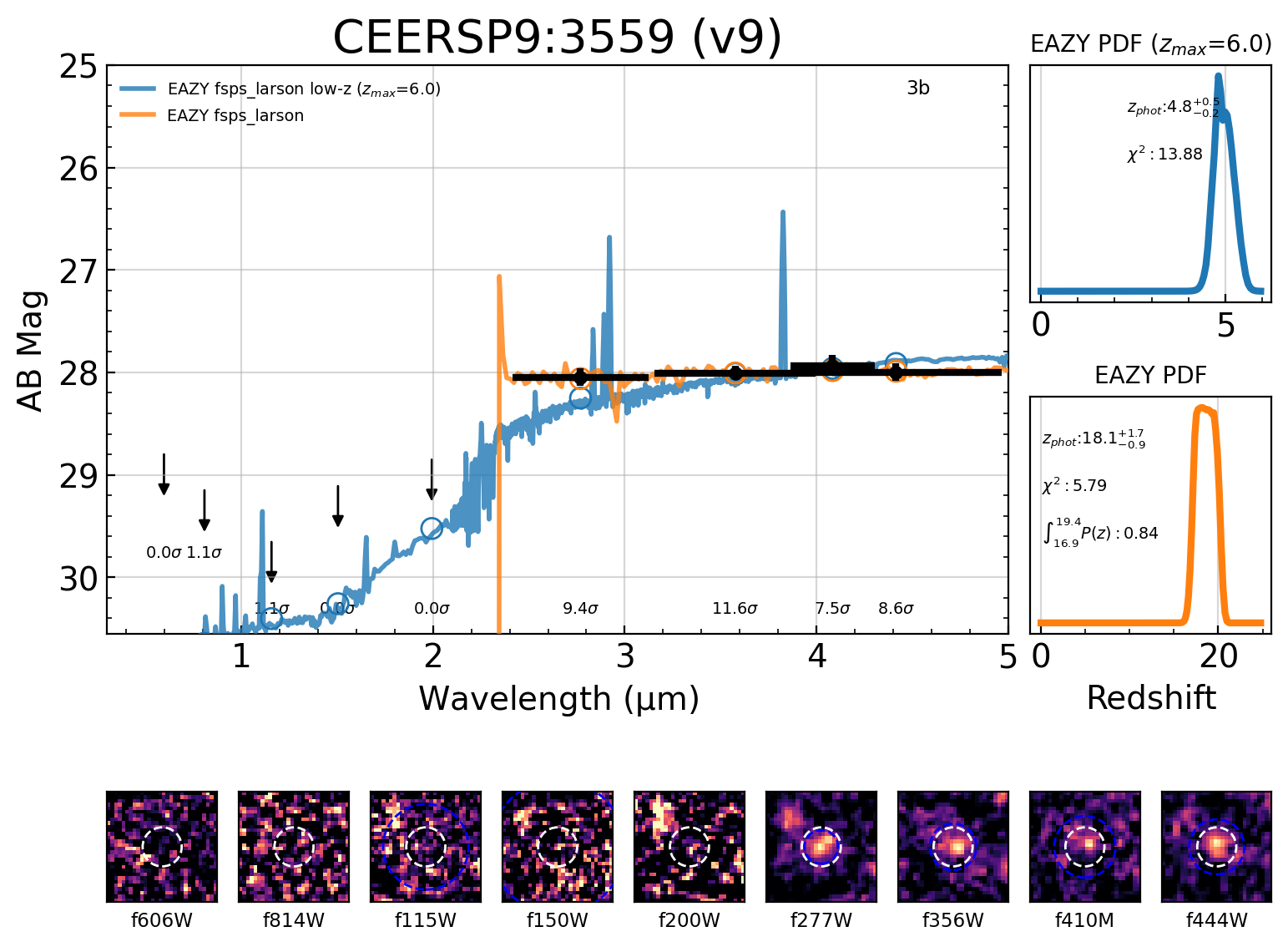}
\caption{Example SED fits for our high redshift galaxy candidates. Shown is the distribution and best fits, as in the previous figures for the lower redshifts of our sample. The left object is the NDGDEEP z = 15.6 galaxy discovered and first published in \citet{Austin2023}. }
\label{fig:seds_highz}
\end{figure*}

We present properties of two of these ultra high redshifts galaxies in \autoref{fig:seds_highz}. These show that the SEDs as measured in broad-band filters are quite flat after the supposed Lyman-break. These SEDs in terms of the filters observed is not dissimilar to the spectra for similar redshift galaxies, where the continuum for these objects is flat and there are no bright emission lines \citep[e.g.,][]{Curtislake2022, Carniani2024}.    These SEDs are typical for our $z > 12$ candidates, with very few showing any evidence for line emission.

At $12 < z < 15$, with the fiducial Bagpipes run, the average stellar mass for these systems is $\sim$ 10$^{8}$ M$_{\odot}$ and the average SFR using a 10 Myr timescale is $\sim 3.8$ M$_{\odot}$ year$^{-1}$. Using a 100 Myr time-scale the star formation rate is 0.9 M$_{\odot}$ year$^{-1}$.  These values are similar to what we have measured for slightly lower redshift galaxies. It thus remains likely that many of these objects are galaxies at the edge of our current observable universe, which future spectroscopy will confirm and allow us to study in more detail. 

\subsection{Galaxy Overabundance at High Redshifts}

There was a great detail of excitement after the first data from JWST were analyzed showing a possible excess of distant massive and bright galaxies in comparison to simulations based on the $\Lambda$CDM framework. If indeed there are more bright and massive galaxies than expected at the highest redshifts this could be the result of a few effects. This includes the possibility that our sample of galaxies have unusual stellar populations, a very low stellar mass to light ratio, implying that even though they have a relatively low mass, they are very luminous for their mass. One possible way to accomplish this is to have stellar populations which are dominated by high mass stars with a top-heavy initial mass function. Alternatively, if these galaxies have stellar populations similar to lower redshift galaxies, and if there is indeed a real excess then there may in fact be an issue with the simulations and models that predict a higher number of systems than what we observe.

We must also consider and cannot rule out, of course, that there is the possibility that the photometric redshifts are somehow wrong, but the vast number of galaxies with confirmed spectroscopic redshifts makes this last option more unlikely at this point \citep[e.g.,][]{Curtislake2022, arrabal2023, Carniani2024}.
We included some discussion of the comparison of observations with theory and simulations in terms of the UV LF (distribution of UV luminosities) within \cite{Adams2023b}, although this comparison is largely based on the estimates of the measurement of this particular quantity which has its own biases and incompleteness. In our study we are able to compare directly with observations of galaxy counts, as opposed to a derived LF with all the issues that go into constructing this accurately. We provide a more detailed study of this comparison between the data for high redshift galaxy counts, as a function of redshift, and what the theory predicts. We thus discuss here our version of the claimed excess seen in other observations \citep[e.g.,][]{Labbe2023}.

\begin{figure*}
\centering
\includegraphics[width=1.12\columnwidth]{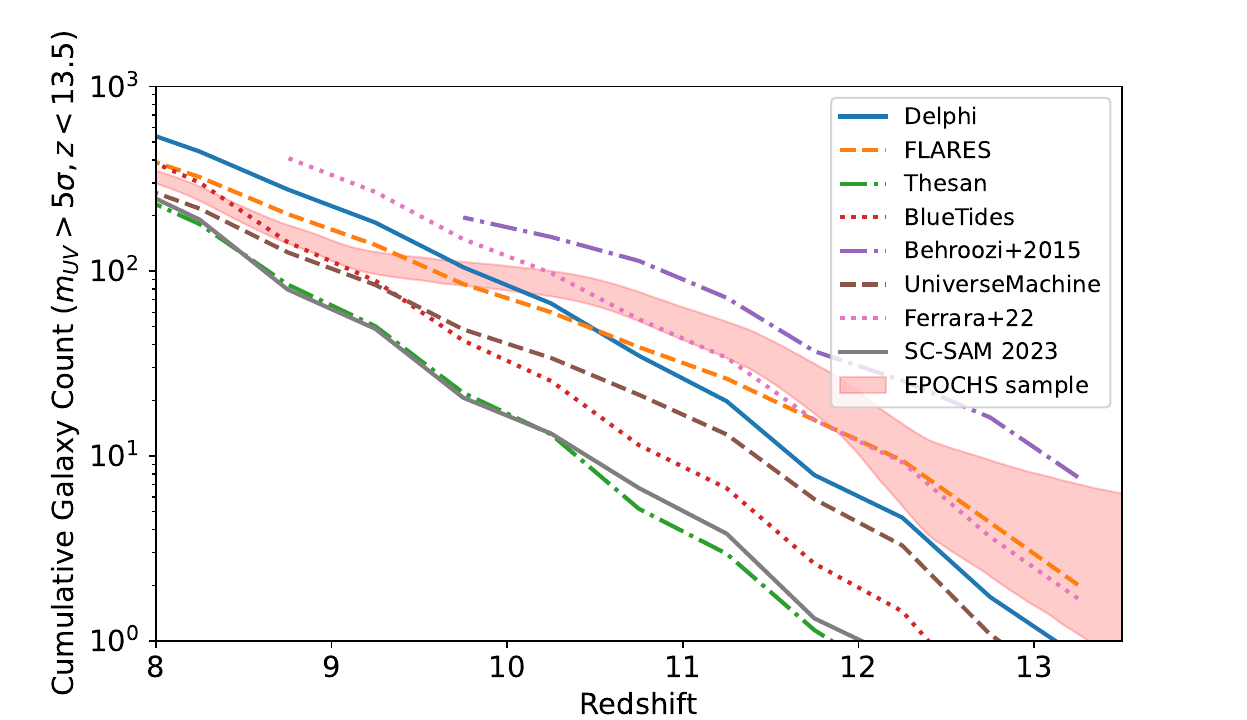}
\includegraphics[width=0.98\columnwidth]{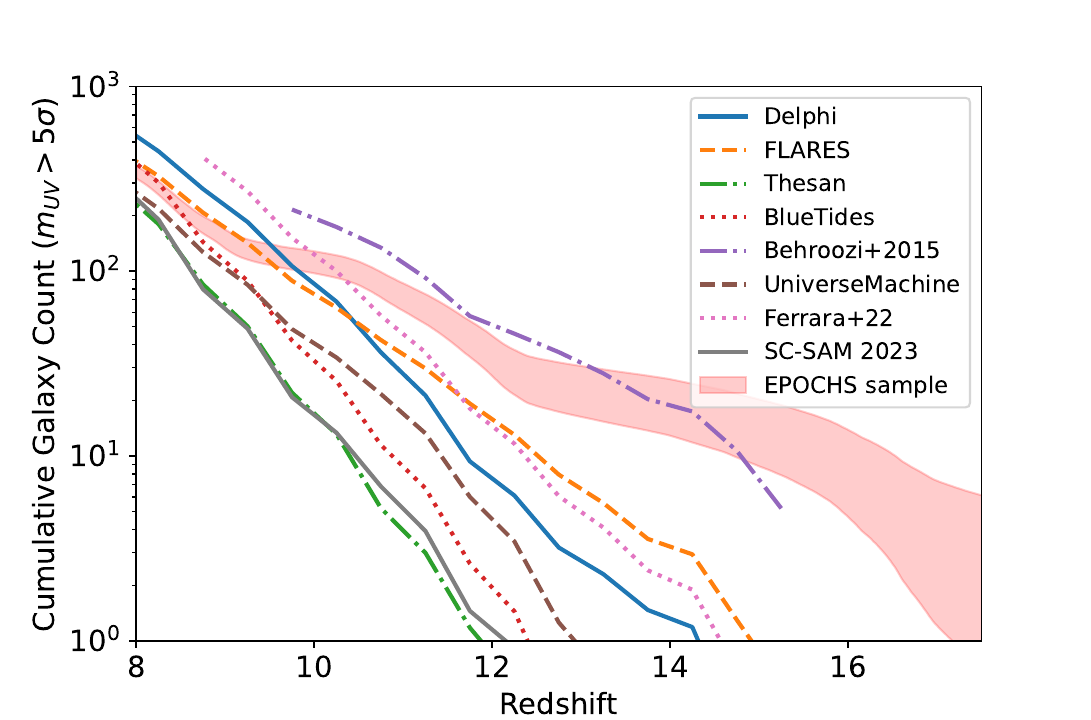}
\caption{The cumulative number counts of very-high-redshift galaxies at $z>8$. Here, we sum the number of galaxies detected in our fields at $5\sigma+$ in the rest-frame ultraviolet up to $z<13.5$ (left), to match the regime of highest spectroscopic confirmations and the redshift limit employed in our other works, as well as for our full sample (right). The width of the curve derived from observations (shown in the pink shaded region) was determined by bootstrap sampling of the redshift PDF's of the sources in our sample. The shape of our observed distribution is very similar to that found in \citet{Finkelstein2022-Maisies} for the CEERS field, however CEERS only makes up a third of our volume. This shape may subsequently be a systematic caused by the choice of templates or due to the discrete wavelength sampling of JWST filters.}
\label{fig:sims_comparison}
\end{figure*}

In order to provide additional context for the over abundance problem, we compare our observations and fits to predictions from a variety of recent simulations. To do this, we integrate the predicted ultraviolet luminosity function from various simulations: Bluetides \citep{Feng2016}, Delphi \citep{dayal2014,dayal2022}, DREAM \citep{Drakos2022}, Thesan \citep{Kannan2022}, the Santa Cruz semi-analytic Model \citep{Yung2023}, UniverseMachine \citep{Behroozi2020}  and FLARES \citet{Lovell2020,Vijayan2021,Wilkins2023}. Additional models also include those from \citet{Behroozi2015} and \citet{Ferrare2022}. The comparison of these simulations is illustrated in \autoref{fig:sims_comparison}. Integration of the luminosity functions encompasses the rest-frame magnitude range accessible by our observations with an apparent depth of $m=29.5$, which corresponds to the average depth of our F440W imaging. 

The comparison of these models with our galaxy counts is shown in \autoref{fig:sims_comparison}. The first thing to note that is that the galaxy counts with redshifts at $z < 13.5$ are higher than most models, but that a few of these, including the predictions from FLARES, Delphi, \citet{Ferrare2022} and the UniverseMachine do rather well in reproducing these galaxy numbers. The other models underpredict the number of galaxies compared to what we find, with only good agreements for most models at the lower redshifts at $z < 9$. 

 While these simulations tend to agree more with each other and with observations at these relatively lower redshifts, there is an increasing spread in their predictions towards higher redshifts. Specifically, the Delphi and FLARES simulations tend to predict greater numbers of UV faint galaxies, while Thesan and the Santa Cruz semi-analytic model predict fewer of these galaxies relative to the other models. Some simulations do better than others at higher redshifts.  At $z\geq10$, the simulations begin to diverge in their predictions, but the observational errors also increase in this regime, making it difficult to confidently favor one physical model over another.

As  can also be seen in this figure is a clear excess number of galaxies in our sample compared with models at $z > 12$. This is such that we are finding over an order of magnitude more galaxies in the areas of the sky we probe, than what we find in models of galaxy formation. For example at redshfit $z = 14$ we would need to have a factor of $\sim 10$ fewer galaxies than what we observe to match even the highest predeictions from FLARES. Within the area that we observe this would imply that we likely would have found statistically no galaxies at these redshifts, although spectroscopy is showing that there certainly are galaxies at these epochs \citep[][]{Carniani2024}. 

This observation has been seen before when comparing galaxy numbers to models \citep[][]{Adams2023b}, however, we now have a large sample and one with robust photometric redshifts and a consistent selection methodology. Whilst we cannot determine how significant this excess is in terms of alternative cosmological models or unusual star formation properties of distant galaxies, it does indicate that we might be seeing a tension with models of galaxy formation. This is consistent to some degree with other observations, such  as the rapid build-up of galaxy stellar mass early in the universe \citep[][]{Bhatawdekar2021,harvey2024epochs}. Future observations using JWST spectroscopy are needed to confirm our sample and determine how many of the galaxies that make up this excess are real. 

%Plot showing the relation between the number of galaxies, in a culumulative counts at redshfits $z > 8.5$ as a function of redshift, compared with different models of galaxy formation which predict the number of galaxies we should be seeing at these highest redshifts. The different lines show the differnet models, which the black solid line our observations. As can be seen, no models match the number of distant galaxies we see at $z > 12$, and even at lower redshifts we find that only the models which predict the largest number of galaxies appear to match the data well. }

\section{Discussion} \label{sec:discussion}

One of the major conclusions from our study is that with a high fidelity sample of distant galaxies, we are able to show that there exists a great diversity in star formation histories based on the photometry of these galaxies. Galaxies at $z > 6.5$ are not homogeneous at high redshift, and  this is even true when we consider the effects of stellar mass, which at lower redshifts is typically the driving observational feature for other galaxy properties \citep[e.g.,][]{grutzbauch2011}. Within this paper we also discuss a possible excess in the numbers of distant galaxies, particularly at the highest redshifts. We can interpret this excess in terms of an incorrect basis or interpretation of theory and/or observational biases. 

We discuss  some of these issues and what the origin of these galaxies possibly are, and what we have to look forward to once spectroscopy for these samples are obtained. One of the things to take away from this study is that not only do we find a large number of high redshift galaxies, but that the systems at redshifts $z \sim 10$, and higher, display  a great deal of diversity in their colors and stellar population properties.

We demonstrate this in several ways, including finding within a narrow redshift range, where most of our galaxies are found, that there is a great diversity of observed colors in (F277W-F444W) vs. (F150W-F277W). These colors are selected such that one is probing the location of the Balmer break within this redshift range, while the other probes the color of the galaxy redward of this break. From this we can see that the `color' term spans a range of values from $-1$ to 1.5, showing a large range in the colors of our galaxies at these redshifts, which can also be seen visually when examining the SEDs of these galaxies individually (\autoref{fig:SEDs_red} \& \autoref{fig:SEDs_blue}). What this means is that the galaxies we are observing  a few hundred Myr after the big bang contain a range of properties in their SEDs.  This implies that there is a great diversity in their star formation and merging histories \citep[e.g.,][]{Duan2024b}, and within their dust content.

To understand this issue in more detail we examine these properties of our galaxies at the same redshift range in \autoref{fig:SEDs}. We can see in this plot that indeed there is a great diversity in dust content (as fit from the SED fitting using \bagpipes{}) as well as in the ages and the sSFR values for sample at this narrow redshift range. All of this suggests that indeed there is a diversity in star formation history, which is also seen in other properties, as described earlier and throughout this paper. While a morphological study at high redshift for these galaxies is not yet published, we know from lower redshifts at $z < 7$ that there is a great diversity of star formation histories that is not accountable by overall morphology (as classified as disks, ellipticals, peculiars), but which does have a trend with stellar mass \citep[e.g.,][]{Conselice2024}. 

When we examine our sample in terms of stellar mass, we in fact do find some trends that suggest stellar mass is regulating the main aspects of the formation of these galaxies. We show the correlation between the stellar mass of the EPOCHS sample vs. the rest-frame (U-V) colors of these galaxies in \autoref{fig:UVmass}. As can be clearly seen, there is a trend, such that the more massive galaxies are redder in this color, which can be a probe of a number of physical properties and effects, including, age and dust absorption. However, this trend is not as obvious or significant at masses M$_{*} < 10^{8.2}$ \solm or so. As well, many of these more massive systems would be considered balmer break galaxies \citep[e.g.,][]{Trussler2024}.   This implies that these massive systems have already undergone a significant amount of star formation and that there is a natural break at the stellar mass of about $10^{8.2}$ \solm.  This is such that systems lower mass than this have a more chaotic formation history as the correlation of SFR with mass breaks down at this limit.   In terms of the agreement with the FLARES simulation, we can use this to infer what may produce the correlation between colour and stellar mass.  Within FLARES this correlation is present due to a higher dust content and higher dust density than lower mass galaxies, due to a longer star formation history, which is also backed up by our observations.  

However, what we do find from our {\em \bagpipes{}} fits is that the most massive galaxies which are red tend to have the largest dust extinctions with $A_{\rm V}$ values approaching $A_{\rm V} \sim 1$. This shows that one reason the most massive galaxies are redder is due to the presence of dust in these systems with significant amounts of extinction. These is opposite in some sense to idea that some massive galaxies will be blue due to blowout of dust and other feedback effects within these galaxies. If massive galaxies become blue at these redshifts, then this phase must be short lived  \citep[e.g.,][]{Ferrare2022, Fiore2023}. These massive galaxies also appear to have on average lower sSFR values than galaxies at lower masses.  This implies that the bulk of the star formation for these systems, as we observe them, occurs early in the history of the universe. This is also another indication for galaxy downsizing, such that the more massive galaxies undergo star formation earlier than lower mass systems. 

Overall, we also find that the model predictions, beyond the counts of the number of galaxies, agree reasonably well with the data. As can be seen in \autoref{fig:UVmass} there is a good agreement between one simulation, FLARES, and that of our data for the distribution of U-V color with stellar mass. The only difference is that at a given stellar mass we find that the average value of the color is larger than that predicted, but that these are still within the range of the model. These are redder galaxies than those predicted, and thus it is likely that effects which create redder systems are present in the observations but not as much in the models.

\section{Conclusions} \label{sec:conclusions}

Understanding galaxy formation at the earliest times is one of the main science drivers of JWST, dictating its design and final properties. As such, there has been a considerable amount of deep imaging and spectroscopic data collected to date on both GTO and public data sets since JWST's first data releases in July 2022. In this paper we collate our GTO PEARLS data with a large fraction of public imaging data to construct a sample of distant galaxies at $z > 6.5$ selected with extreme care and consistently across fields, which we call the EPOCHS v1 dataset. This paper describes this v1 EPOCHS dataset and we discuss the basic properties of these galaxies, focusing on their star formation rates and histories. 

This EPOCHS sample is one of the largest collected from JWST observations for $z > 6.5$ galaxies to date \citep[see also][]{Donnan2022,Harikane2023,McLeod2023,Donnan2024}, representing one of our best opportunities to study the physical properties of early galaxies. In this paper, which is the introduction to a series of papers discussing these targets and their properties \citep[e.g.,][]{Adams2023b, harvey2024epochs, Li2024, Austin2024}, we include a general description of their discovery and features. This includes the methods by which we ensure that this sample of 1165 galaxies is robustly identified with minimal contamination. 

Our results clearly show a great diversity in galaxy properties even amongst the most distant galaxies that JWST has discovered to date. This is shown in the vast ranges of colors, both observed as well as when comparing the rest-frame UV, for these systems. We quantify the star formation history of our objects, showing that there is a large range in the SFR at 100 Myr ago vs. 250 Myr ago. We find a general trend of downsizing, such that the most massive galaxies at $8.5 < z < 10.5$ have the lowest specific star formation rates, the oldest ages, and the highest masses. 

We also find that there a well defined star forming main sequence for galaxies up to $z \sim 11.5$, such that on average the star formation rate increases with the stellar mass within our sample. We find that this trend differs from lower redshift systems in that the star formation rates are higher compared to most studies, including SFRs that are a factor of $\sim 5$ higher than the $z = 9$ main sequence from HST observations \citep[][]{Bhatawdekar2021}. 

The lower mass galaxies have relatively large specific star formation rate, as well as having young ages. This is an indication that we are seeing a trend such that the highest mass galaxies are forming early and that the lower mass galaxies at $z > 8.5$ still continue to undergo galaxy formation with young ages and high sSFRs at this epoch. 

We find that using a rigorous selection criteria for determining which galaxies are at high redshift means that each field we study has different biases in redshifts. As discussed in \autoref{sec:z_dist} and shown in \autoref{fig:magz}, we find that different fields have certain redshift and magnitude biases that depend upon the filters being used within each field. This implies that using only a single field for the derivation of evolution  will contain significant biases due to missing or containing fainter imaging at certain wavelengths.  This is in addition to variations due to cosmic variance, which is very significant at the level of a single deep JWST imaging field. It is much better to use the different fields to carry out a more complete survey of the population of galaxies at different redshifts. However, while we find that there are biases in which redshifts we find galaxies, we do not find a bias in the colors of our galaxies within different fields. That is, overall each field is finding similar galaxies when viewed over broad redshifts, albeit at different locations of redshift/magnitude, depending on the depth and filter coverage of the data. 

We also see a generally good agreement between some galaxy properties, namely stellar mass, color, and star formation rates compared with simulations, including those from FLARES \citep[e.g.,][]{FLARES-V}.    These simulations do a good job of predicting both the colors of our galaxies up to the highest redshifts, as well as giving good agreement between stellar mass and color.  This is due to galaxies that are more massive in the simulation being relatively older systems, such that their color is redder than the more recent star forming systems at the lower masses. 

We also find that there is an apparent excess of $z > 12$ galaxies compared with models. In particular, we find that at lower redshifts, especially at $z < 10$, we find a good agreement with some simulations, although those with the highest predicted abundances are better fit than the lower value predictions. Even without a correction for incompleteness, which we do not attempt here, we are finding many more distant galaxies than are predicted in models. However, to make any stronger statements will require future JWST spectroscopy of our candidate galaxies to confirm that they are distant high redshift galaxies. 

Our results, and previous similar ones, clearly show that JWST is a powerful tool for exploring the universe and uncovering the earliest objects. Our study of high redshift galaxies is just the beginning as deeper and wider imaging with JWST will greatly increase our understanding of the these fundamental aspect of the universe.  We are also now acquiring Euclid data on extremely large fields, which will allow us to study the rarely brighter galaxies at the epoch of reionization \citep[][]{Weaver2024}. The great number of high redshift candidates which are robust using our methods shows that further spectroscopic and imaging data on these and similar galaxies in other fields will be very revealing.

%\clearpage

\begin{longrotatetable}
\global\pdfpageattr\expandafter{\the\pdfpageattr/Rotate 90}
\begin{deluxetable}{ccccccccrc}
\tablecomments{Upper right corner of full EPOCHS galaxy catalogue. Full machine-readable catalogue is available online. A `*' (`+') indicates that the stellar mass (SFR) has been corrected to the full size of the galaxy, based on the fitted \sextractor{} Kron ellipse.}
\tablehead{
\multicolumn{1}{c}{Name} & \multicolumn{1}{c}{R.A.} & \multicolumn{1}{c}{Dec.} & \multicolumn{1}{c}{f$_{\rm F444W}$} & \multicolumn{1}{c}{f$_{\rm F277W}$} & \multicolumn{1}{c}{$z_{\rm \eazy}$} & \multicolumn{1}{c}{$z_{\rm \bagpipes}$} & \multicolumn{1}{c}{Stellar Mass} & \multicolumn{1}{c}{SFR$_{100}$} & \multicolumn{1}{c}{$\cdots$} \\
 & \textit{deg.} & \textit{deg.} & \textit{nJy} & \textit{nJy} & & & \textit{$\log_{10}(\rm M_{\star}/\rm M_{\odot})$} & \textit{$\rm M_{\odot} \ \textrm{yr}^{-1}$} & \multicolumn{1}{c}{$\cdots$} \\}
\startdata
\multicolumn{1}{c}{3652\_JADES-Deep-GS+} & \multicolumn{1}{c}{53.18209} & \multicolumn{1}{c}{-27.81816} & \multicolumn{1}{c}{$30.80 \pm 3.08$} & \multicolumn{1}{c}{$27.31 \pm 2.73$} & \multicolumn{1}{c}{$6.50^{+0.04}_{-0.09}$} & \multicolumn{1}{c}{$6.45^{+0.07}_{-0.08}$} & \multicolumn{1}{c}{$7.6^{+0.1}_{-0.1}$} & \multicolumn{1}{c}{$0.5^{+0.1}_{-0.1}$} & \multicolumn{1}{c}{$\cdots$}  \\
\multicolumn{1}{c}{3285\_CEERSP3*+} & \multicolumn{1}{c}{214.83849} & \multicolumn{1}{c}{52.88520} & \multicolumn{1}{c}{$331.84 \pm 33.18$} & \multicolumn{1}{c}{$144.53 \pm 14.45$} & \multicolumn{1}{c}{$6.50^{+0.07}_{-0.35}$} & \multicolumn{1}{c}{$6.65^{+0.08}_{-0.34}$} & \multicolumn{1}{c}{$10.2^{+0.2}_{-0.3}$} & \multicolumn{1}{c}{$239.4^{+21.5}_{-24.8}$} & \multicolumn{1}{c}{$\cdots$}  \\
\multicolumn{1}{c}{13266\_NEP-1} & \multicolumn{1}{c}{260.76351} & \multicolumn{1}{c}{65.78839} & \multicolumn{1}{c}{$24.06 \pm 2.41$} & \multicolumn{1}{c}{$22.95 \pm 2.29$} & \multicolumn{1}{c}{$6.50^{+0.06}_{-0.21}$} & \multicolumn{1}{c}{$6.52^{+0.08}_{-0.12}$} & \multicolumn{1}{c}{$8.1^{+0.3}_{-0.3}$} & \multicolumn{1}{c}{$1.0^{+0.2}_{-0.3}$} & \multicolumn{1}{c}{$\cdots$}  \\
\multicolumn{1}{c}{4168\_CEERSP4+} & \multicolumn{1}{c}{214.79533} & \multicolumn{1}{c}{52.79046} & \multicolumn{1}{c}{$18.89 \pm 2.02$} & \multicolumn{1}{c}{$18.78 \pm 1.88$} & \multicolumn{1}{c}{$6.50^{+0.04}_{-0.20}$} & \multicolumn{1}{c}{$6.39^{+0.11}_{-0.12}$} & \multicolumn{1}{c}{$7.2^{+0.1}_{-0.0}$} & \multicolumn{1}{c}{$0.2^{+0.0}_{-0.0}$} & \multicolumn{1}{c}{$\cdots$}  \\
\multicolumn{1}{c}{23488\_JADES-Deep-GS+} & \multicolumn{1}{c}{53.16618} & \multicolumn{1}{c}{-27.76435} & \multicolumn{1}{c}{$7.91 \pm 0.79$} & \multicolumn{1}{c}{$8.24 \pm 0.82$} & \multicolumn{1}{c}{$6.50^{+0.04}_{-0.14}$} & \multicolumn{1}{c}{$6.47^{+0.09}_{-0.11}$} & \multicolumn{1}{c}{$6.9^{+0.2}_{-0.1}$} & \multicolumn{1}{c}{$0.1^{+0.1}_{-0.0}$} & \multicolumn{1}{c}{$\cdots$}  \\
\multicolumn{1}{c}{12137\_JADES-Deep-GS*} & \multicolumn{1}{c}{53.18464} & \multicolumn{1}{c}{-27.77930} & \multicolumn{1}{c}{$10.54 \pm 1.05$} & \multicolumn{1}{c}{$13.98 \pm 1.40$} & \multicolumn{1}{c}{$6.50^{+0.05}_{-0.11}$} & \multicolumn{1}{c}{$6.50^{+0.07}_{-0.09}$} & \multicolumn{1}{c}{$7.4^{+0.1}_{-0.1}$} & \multicolumn{1}{c}{$0.2^{+0.1}_{-0.0}$} & \multicolumn{1}{c}{$\cdots$}  \\
\multicolumn{1}{c}{5743\_CEERSP10*+} & \multicolumn{1}{c}{214.83846} & \multicolumn{1}{c}{52.77877} & \multicolumn{1}{c}{$41.93 \pm 4.19$} & \multicolumn{1}{c}{$37.40 \pm 3.74$} & \multicolumn{1}{c}{$6.50^{+0.05}_{-0.25}$} & \multicolumn{1}{c}{$6.48^{+0.13}_{-0.19}$} & \multicolumn{1}{c}{$8.1^{+0.2}_{-0.2}$} & \multicolumn{1}{c}{$1.7^{+0.7}_{-0.5}$} & \multicolumn{1}{c}{$\cdots$}  \\
\multicolumn{1}{c}{4530\_CEERSP8} & \multicolumn{1}{c}{215.05798} & \multicolumn{1}{c}{52.91688} & \multicolumn{1}{c}{$8.37 \pm 2.17$} & \multicolumn{1}{c}{$10.79 \pm 1.57$} & \multicolumn{1}{c}{$6.50^{+0.04}_{-0.44}$} & \multicolumn{1}{c}{$6.51^{+0.18}_{-0.27}$} & \multicolumn{1}{c}{$7.3^{+0.3}_{-0.3}$} & \multicolumn{1}{c}{$0.2^{+0.2}_{-0.1}$} & \multicolumn{1}{c}{$\cdots$}  \\
\multicolumn{1}{c}{16783\_NEP-2*} & \multicolumn{1}{c}{260.76476} & \multicolumn{1}{c}{65.86070} & \multicolumn{1}{c}{$22.30 \pm 2.43$} & \multicolumn{1}{c}{$26.46 \pm 2.65$} & \multicolumn{1}{c}{$6.51^{+0.04}_{-0.18}$} & \multicolumn{1}{c}{$6.41^{+0.10}_{-0.11}$} & \multicolumn{1}{c}{$7.4^{+0.1}_{-0.0}$} & \multicolumn{1}{c}{$0.2^{+0.0}_{-0.0}$} & \multicolumn{1}{c}{$\cdots$}  \\
\multicolumn{1}{c}{15297\_JADES-Deep-GS} & \multicolumn{1}{c}{53.16238} & \multicolumn{1}{c}{-27.80330} & \multicolumn{1}{c}{$10.81 \pm 1.08$} & \multicolumn{1}{c}{$8.96 \pm 0.90$} & \multicolumn{1}{c}{$6.51^{+0.04}_{-0.28}$} & \multicolumn{1}{c}{$6.25^{+0.26}_{-0.06}$} & \multicolumn{1}{c}{$7.4^{+0.2}_{-0.1}$} & \multicolumn{1}{c}{$0.3^{+0.2}_{-0.1}$} & \multicolumn{1}{c}{$\cdots$}  \\
\multicolumn{1}{c}{559\_CEERSP3+} & \multicolumn{1}{c}{214.80648} & \multicolumn{1}{c}{52.87883} & \multicolumn{1}{c}{$64.48 \pm 6.45$} & \multicolumn{1}{c}{$64.58 \pm 6.46$} & \multicolumn{1}{c}{$6.51^{+0.04}_{-0.30}$} & \multicolumn{1}{c}{$6.36^{+0.17}_{-0.19}$} & \multicolumn{1}{c}{$8.1^{+0.1}_{-0.1}$} & \multicolumn{1}{c}{$1.4^{+0.6}_{-0.4}$} & \multicolumn{1}{c}{$\cdots$}  \\
\multicolumn{1}{c}{8857\_NEP-1*+} & \multicolumn{1}{c}{260.74977} & \multicolumn{1}{c}{65.79742} & \multicolumn{1}{c}{$57.24 \pm 5.72$} & \multicolumn{1}{c}{$52.02 \pm 5.20$} & \multicolumn{1}{c}{$6.51^{+0.04}_{-0.19}$} & \multicolumn{1}{c}{$6.40^{+0.11}_{-0.13}$} & \multicolumn{1}{c}{$8.1^{+0.2}_{-0.2}$} & \multicolumn{1}{c}{$1.5^{+0.7}_{-0.4}$} & \multicolumn{1}{c}{$\cdots$}  \\
\multicolumn{1}{c}{16342\_NEP-3} & \multicolumn{1}{c}{260.68517} & \multicolumn{1}{c}{65.93653} & \multicolumn{1}{c}{$69.81 \pm 6.98$} & \multicolumn{1}{c}{$66.45 \pm 6.65$} & \multicolumn{1}{c}{$6.51^{+0.04}_{-0.08}$} & \multicolumn{1}{c}{$6.47^{+0.06}_{-0.05}$} & \multicolumn{1}{c}{$7.8^{+0.1}_{-0.0}$} & \multicolumn{1}{c}{$0.7^{+0.1}_{-0.1}$} & \multicolumn{1}{c}{$\cdots$}  \\
\multicolumn{1}{c}{14059\_NEP-4*+} & \multicolumn{1}{c}{260.45029} & \multicolumn{1}{c}{65.81491} & \multicolumn{1}{c}{$35.99 \pm 3.60$} & \multicolumn{1}{c}{$42.30 \pm 4.23$} & \multicolumn{1}{c}{$6.51^{+0.04}_{-0.28}$} & \multicolumn{1}{c}{$6.34^{+0.14}_{-0.14}$} & \multicolumn{1}{c}{$7.9^{+0.2}_{-0.2}$} & \multicolumn{1}{c}{$0.9^{+0.3}_{-0.2}$} & \multicolumn{1}{c}{$\cdots$}  \\
\multicolumn{1}{c}{1167\_CEERSP3*} & \multicolumn{1}{c}{214.82036} & \multicolumn{1}{c}{52.88475} & \multicolumn{1}{c}{$114.95 \pm 11.50$} & \multicolumn{1}{c}{$98.59 \pm 9.86$} & \multicolumn{1}{c}{$6.51^{+0.05}_{-0.13}$} & \multicolumn{1}{c}{$6.46^{+0.11}_{-0.11}$} & \multicolumn{1}{c}{$8.4^{+0.1}_{-0.1}$} & \multicolumn{1}{c}{$2.5^{+0.7}_{-0.6}$} & \multicolumn{1}{c}{$\cdots$}  \\
\multicolumn{1}{c}{36860\_JADES-Deep-GS*+} & \multicolumn{1}{c}{53.10976} & \multicolumn{1}{c}{-27.80747} & \multicolumn{1}{c}{$188.80 \pm 18.88$} & \multicolumn{1}{c}{$139.92 \pm 13.99$} & \multicolumn{1}{c}{$6.51^{+0.05}_{-0.37}$} & \multicolumn{1}{c}{$6.03^{+0.05}_{-0.07}$} & \multicolumn{1}{c}{$9.2^{+0.1}_{-0.1}$} & \multicolumn{1}{c}{$15.1^{+2.7}_{-3.1}$} & \multicolumn{1}{c}{$\cdots$}  \\
\multicolumn{1}{c}{17157\_NGDEEP*+} & \multicolumn{1}{c}{53.26097} & \multicolumn{1}{c}{-27.82496} & \multicolumn{1}{c}{$11.79 \pm 1.18$} & \multicolumn{1}{c}{$11.35 \pm 1.14$} & \multicolumn{1}{c}{$6.51^{+0.19}_{-0.09}$} & \multicolumn{1}{c}{$6.52^{+0.09}_{-0.14}$} & \multicolumn{1}{c}{$7.9^{+0.2}_{-0.2}$} & \multicolumn{1}{c}{$0.8^{+0.1}_{-0.1}$} & \multicolumn{1}{c}{$\cdots$}  \\
\multicolumn{1}{c}{2790\_CEERSP8} & \multicolumn{1}{c}{215.03306} & \multicolumn{1}{c}{52.89019} & \multicolumn{1}{c}{$27.03 \pm 2.70$} & \multicolumn{1}{c}{$23.51 \pm 2.35$} & \multicolumn{1}{c}{$6.51^{+0.03}_{-0.41}$} & \multicolumn{1}{c}{$6.25^{+0.21}_{-0.22}$} & \multicolumn{1}{c}{$8.2^{+0.2}_{-0.3}$} & \multicolumn{1}{c}{$1.0^{+0.2}_{-0.1}$} & \multicolumn{1}{c}{$\cdots$}  \\
\multicolumn{1}{c}{3761\_CEERSP9*+} & \multicolumn{1}{c}{214.91319} & \multicolumn{1}{c}{52.81510} & \multicolumn{1}{c}{$33.02 \pm 3.30$} & \multicolumn{1}{c}{$34.81 \pm 3.48$} & \multicolumn{1}{c}{$6.51^{+0.04}_{-0.20}$} & \multicolumn{1}{c}{$6.39^{+0.12}_{-0.13}$} & \multicolumn{1}{c}{$7.6^{+0.1}_{-0.0}$} & \multicolumn{1}{c}{$0.4^{+0.1}_{-0.0}$} & \multicolumn{1}{c}{$\cdots$}  \\
\multicolumn{1}{c}{12178\_NEP-4*+} & \multicolumn{1}{c}{260.51401} & \multicolumn{1}{c}{65.81138} & \multicolumn{1}{c}{$91.83 \pm 9.18$} & \multicolumn{1}{c}{$62.23 \pm 6.22$} & \multicolumn{1}{c}{$6.51^{+0.05}_{-0.10}$} & \multicolumn{1}{c}{$6.44^{+0.10}_{-0.09}$} & \multicolumn{1}{c}{$8.5^{+0.1}_{-0.1}$} & \multicolumn{1}{c}{$3.5^{+1.2}_{-0.7}$} & \multicolumn{1}{c}{$\cdots$}  \\
\multicolumn{1}{c}{7570\_NGDEEP} & \multicolumn{1}{c}{53.26361} & \multicolumn{1}{c}{-27.86999} & \multicolumn{1}{c}{$16.18 \pm 1.62$} & \multicolumn{1}{c}{$17.60 \pm 1.76$} & \multicolumn{1}{c}{$6.51^{+0.17}_{-0.14}$} & \multicolumn{1}{c}{$6.42^{+0.12}_{-0.12}$} & \multicolumn{1}{c}{$8.4^{+0.1}_{-0.1}$} & \multicolumn{1}{c}{$0.5^{+0.2}_{-0.1}$} & \multicolumn{1}{c}{$\cdots$}  \\
\multicolumn{1}{c}{9077\_NEP-2} & \multicolumn{1}{c}{260.90467} & \multicolumn{1}{c}{65.82584} & \multicolumn{1}{c}{$27.33 \pm 2.73$} & \multicolumn{1}{c}{$31.59 \pm 3.16$} & \multicolumn{1}{c}{$6.51^{+0.04}_{-0.32}$} & \multicolumn{1}{c}{$6.30^{+0.17}_{-0.16}$} & \multicolumn{1}{c}{$7.6^{+0.2}_{-0.1}$} & \multicolumn{1}{c}{$0.5^{+0.3}_{-0.1}$} & \multicolumn{1}{c}{$\cdots$}  \\
\multicolumn{1}{c}{12576\_NEP-1*+} & \multicolumn{1}{c}{260.74459} & \multicolumn{1}{c}{65.76737} & \multicolumn{1}{c}{$124.25 \pm 12.43$} & \multicolumn{1}{c}{$112.57 \pm 11.26$} & \multicolumn{1}{c}{$6.51^{+0.05}_{-0.09}$} & \multicolumn{1}{c}{$6.44^{+0.08}_{-0.09}$} & \multicolumn{1}{c}{$8.5^{+0.1}_{-0.1}$} & \multicolumn{1}{c}{$3.5^{+0.9}_{-0.6}$} & \multicolumn{1}{c}{$\cdots$}  \\
\multicolumn{1}{c}{$\cdots$} & \multicolumn{1}{c}{$\cdots$}  & \multicolumn{1}{c}{$\cdots$} & \multicolumn{1}{c}{$\cdots$} & \multicolumn{1}{c}{$\cdots$} & \multicolumn{1}{c}{$\cdots$}  & \multicolumn{1}{c}{$\cdots$}  & \multicolumn{1}{c}{$\cdots$}  & \multicolumn{1}{c}{$\cdots$} & \multicolumn{1}{c}{$\cdots$} \\
\enddata
\end{deluxetable}
\end{longrotatetable}
\global\pdfpageattr\expandafter{\the\pdfpageattr/Rotate 0}
%\clearpage

\section*{Acknowledgements}

We thank the JADES, GLASS, CEERS, SMACS, and NGDEEP teams for their work in designing and preparing these public and GTO observations, and the STScI staff that carried them out.   We acknowledge support from the ERC Advanced Investigator Grant EPOCHS (788113), as well three studentships from the STFC. RAW, SHC, and RAJ acknowledge support from NASA JWST Interdisciplinary
Scientist grants NAG5-12460, NNX14AN10G and 80NSSC18K0200 from GSFC.
LF acknowledges financial support from Coordenação de Aperfeiçoamento de Pessoal de Nível Superior - Brazil (CAPES) in the form of a PhD studentship. CCL acknowledges support from the Royal Society under grant RGF/EA/181016. This work is based on observations made with the NASA/ESA \textit{Hubble Space Telescope} (HST) and NASA/ESA/CSA \textit{James Webb Space Telescope} (JWST) obtained from the \texttt{Mikulski Archive for Space Telescopes} (\texttt{MAST}) at the \textit{Space Telescope Science Institute} (STScI), which is operated by the Association of Universities for Research in Astronomy, Inc., under NASA contract NAS 5-03127 for JWST, and NAS 5–26555 for HST.

This research made use of the following Python libraries: \textsc{Astropy} \citep{astropy2022}; \textsc{Morfometryka} \citep{ferrari2015}; \textsc{Pandas} \citep{pandas}; \textsc{Matplotlib} \citep{Hunter:2007}; \textsc{photutils} \citep{larry_bradley_2022}



\bibliographystyle{aasjournal}
\bibliography{main} % if your bibtex file is called example.bib

%\begin{thebibliography}{99}
%\end{thebibliography}

%%%%%%%%%%%%%%%%%%%%%%%%%%%%%%%%%%%%%%%%%%%%%%%%%%

%%%%%%%%%%%%%%%%% APPENDICES %%%%%%%%%%%%%%%%%%%%%

\appendix

\section{Table of Galaxy Properties}

In this appendix A, we give more information on the data release for our objects found as part of the EPOCHS v1 catalog. We give a description of how our objects are identified and give a description of the properties which are contained within the released catalog for the EPOCHS sample.   A list of the quantities we provide in the catalog are described in Table~\ref{tab:bigtab_description}.  Included within our released catalog (upon acceptance of the paper) are the basic properties of 1165 EPOCHS v1 galaxies.  This includes their  photometry in the JWST filters and when available HST ACS data.  We also include other derived properties based on this photometry.  This includes The photometric redshift, the stellar mass measured in different ways (see \citet{harvey2024epochs} for more information on how masses are measured), as well as the various age measurements from Bagpipes.  In terms of derived quantities we also gives values for the UV $\beta$ slope \citep[][]{Austin2024}, the dust and metallicty values, as well as derived absolute magnitudes M$_{\rm UV}$ and the rest-frame (U-V) colors.    Wiithin this table we also provide values for the uncertaintiies on these quantities. 

\vspace{0.5cm}
\begin{table}
    \centering
    \caption{Table of EPOCHS v1 catalog column names, units, descriptions and column shape, specifically for the stellar population parameters calculated using \bagpipes{}. \ext{} indicates that the column name appears multiple time with different extensions, and in this case ``\ext" can take the value of \textbf{zfix} or \textbf{zgauss}, depending on whether the redshift is fixed to the EAZY maximum likelihood result given by ``zbest", or allowed to vary within a Gaussian centered on ``zbest". Those entries with a $\star$ indicates that the column has been corrected for any flux associated with the galaxy which falls outside the extraction aperture - for masses this is done by correcting the mass by the ratio of MAG-AUTO to MAG-APER in the longest wavelength F444W band, where this exceeds unity. For star formation rates the band covering the rest-frame 1500\AA{} wavelength is used instead.}
    \begin{tabular}{ccc}
    \hline
        \large \textbf{Column Name} & \large \textbf{Unit} & \large \textbf{Description}  \\ \hline \hline

        \multicolumn{3}{c}{\textbf{Fiducial \bagpipes{} Results  (z$_{\mathrm{fix}}$ or z$_{\mathrm{gauss}}$)}} \\ \hline
        redshift\_pipes\_zgauss & & Fitted redshift (zgauss only) \\
        redshift\_pipes\_l1\_zgauss & & Lower uncertainty (50th - 16th percentile)\\
        redshift\_pipes\_u1\_zgauss & & Upper uncertainty (84th - 50th percentile)\\
        stellar\_mass\_pipes\_\ext$^{\star}$ & $\log_{10}(\mathrm{M}_{\odot})$ & Total surviving stellar mass \\
        stellar\_mass\_pipes\_l1\_\ext & $\log_{10}(\mathrm{M}_{\odot})$ & Lower uncertainty (50th - 16th percentile) \\
        stellar\_mass\_pipes\_u1\_\ext & $\log_{10}(\mathrm{M}_{\odot})$ & Upper uncertainty (84th - 50th percentile) \\
        SFR\_10Myr\_pipes\_\ext$^{\star}$ & $\mathrm{M}_{\odot}~\mathrm{yr}^{-1}$ & Average total star formation rate over a 10 Myr timescale \\
        SFR\_10Myr\_pipes\_l1\_\ext & $\mathrm{M}_{\odot}~\mathrm{yr}^{-1}$ & Lower uncertainty (50th - 16th percentile) \\
        SFR\_10Myr\_pipes\_u1\_\ext & $\mathrm{M}_{\odot}~\mathrm{yr}^{-1}$ & Upper uncertainty (84th - 50th percentile) \\
        SFR\_100Myr\_pipes\_\ext$^{\star}$ & $\mathrm{M}_{\odot}~\mathrm{yr}^{-1}$ & Average total star formation rate over a 100 Myr timescale \\
        SFR\_100Myr\_pipes\_l1\_\ext & $\mathrm{M}_{\odot}~\mathrm{yr}^{-1}$ &  Lower uncertainty (50th - 16th percentile) \\
        SFR\_100Myr\_pipes\_u1\_\ext & $\mathrm{M}_{\odot}~\mathrm{yr}^{-1}$ & Upper uncertainty (84th - 50th percentile) \\
        mass\_weighted\_age\_pipes\_\ext & Myr & Mass-weighted age of galaxy \\
        mass\_weighted\_age\_pipes\_l1\_\ext & Myr & Lower uncertainty (50th - 16th percentile) \\
        mass\_weighted\_age\_pipes\_u1\_\ext & Myr & Upper uncertainty (84th - 50th percentile) \\
        beta\_pipes\_\ext & & UV $\beta$ slope of best-fitting \bagpipes{} spectra in Calzetti filters\\
        beta\_pipes\_l1\_\ext & & Lower uncertainty (50th - 16th percentile) \\
        beta\_pipes\_u1\_\ext & & Upper uncertainty (84th - 50th percentile) \\
        Z\_star\_pipes\_\ext & Z$_\odot$ & Stellar metallicity \\
        Z\_star\_pipes\_l1\_\ext & Z$_\odot$ & Lower uncertainty (50th - 16th percentile) \\
        Z\_star\_pipes\_u1\_\ext & Z$_\odot$ & Upper uncertainty (84th - 50th percentile) \\
        A\_V\_pipes\_\ext & AB mag & Dust extinction in V band \\
        A\_V\_pipes\_l1\_\ext & AB mag & Lower uncertainty (50th - 16th percentile)\\
        A\_V\_pipes\_u1\_\ext & AB mag & Upper uncertainty (84th - 50th percentile)\\
        U-V\_pipes\_\ext & AB mag & U-V color\\
        U-V\_pipes\_l1\_\ext & AB mag & Lower uncertainty (50th - 16th percentile)\\
        U-V\_pipes\_u1\_\ext & AB mag & Upper uncertainty (84th - 50th percentile) \\
        M\_UV\_pipes\_\ext & AB mag & Absolute UV Magnitude \\
        M\_UV\_pipes\_l1\_\ext & AB mag & Lower uncertainty (50th - 16th percentile)\\
        M\_UV\_pipes\_u1\_\ext & AB mag & Upper uncertainty (84th - 50th percentile) \\
        chisq\_phot\_pipes\_\ext & & $\chi^2$ of fit \\
        
        \hline
        %\multicolumn{3}{c}{\textbf{Fiducial \bagpipes{} Results (z$_{\mathrm{gauss}}$)}} \\ 
        %\hline
        %z\_pipes & & \\
        %z\_pipes\_l1 & & \\
        %z\_pipes\_u1 & & \\
        %stellar\_mass\_pipes & $\log_{10}(\mathrm{M}_{\odot})$ & \\
        %stellar\_mass\_pipes\_l1 & $\log_{10}(\mathrm{M}_{\odot})$ & \\
        %stellar\_mass\_pipes\_u1 & $\log_{10}(\mathrm{M}_{\odot})$ & \\
        %SFR\_10Myr\_pipes & $\mathrm{M}_{\odot}~\mathrm{yr}^{-1}$ & \\
        %SFR\_10Myr\_pipes\_l1 & $\mathrm{M}_{\odot}~\mathrm{yr}^{-1}$ & \\
        %SFR\_10Myr\_pipes\_u1 & $\mathrm{M}_{\odot}~\mathrm{yr}^{-1}$ & \\
        %SFR\_100Myr\_pipes & $\mathrm{M}_{\odot}~\mathrm{yr}^{-1}$ & \\
        %SFR\_100Myr\_pipes\_l1 & $\mathrm{M}_{\odot}~\mathrm{yr}^{-1}$ & \\
        %SFR\_100Myr\_pipes\_u1 & $\mathrm{M}_{\odot}~\mathrm{yr}^{-1}$ & \\
        %mass\_weighted\_age\_pipes & Myr & \\
        %mass\_weighted\_age\_pipes\_l1 & Myr & \\
        %mass\_weighted\_age\_pipes\_u1 & Myr & \\
        %beta\_pipes & & \\
        %beta\_pipes\_l1 & & \\
        %beta\_pipes\_u1 & & \\
        %A\_V\_pipes & AB mag & \\
        %A\_V\_pipes\_l1 & AB mag & \\
        %A\_V\_pipes\_u1 & AB mag & \\
        %U\_pipes & AB mag & \\
        %U\_pipes\_l1 & AB mag & \\
        %U\_pipes\_u1 & AB mag & \\
        %V\_pipes & AB mag & \\
        %V\_pipes\_l1 & AB mag & \\
        %V\_pipes\_u1 & AB mag & \\

    \end{tabular}
    
    \label{tab:bigtab_description}
\end{table}

\section{Completeness and Contamination Simulation Results}

As described in the main text of this paper, we carry out a series of simulations of our fields to determine the contamination and completeness for detecting galaxies within each of the fields we use.  Completeness is important as we want to detect and measure as many galaxies at a given brightness and redshift as we can, thus we aim to have completeness values approach unity.  Contamination is also important to minimize, as it is is possible to have a very generous selection method that results in a sample that is very complete, but is full of contamination from objects at other redshifts.  Thus, our goal is to minimize the contamination with values approaching zero while maximizing the completeness. 

As the depths and filters and other properties of our fields differ significantly between each other, carrying out these simulations is necessary to determine the limits of where we can use data in each field while retaining a high purity.   These simulations are such that we can determine whether or not a galaxy in the \texttt{JAGUAR} simulation would be detected.  This we examine as a function of galaxy brightness, which we convert into an absolute magnitude, at a given redshift.  From this we can determine using our EPOCHS criteria how many of these galaxies that are within an actual bin of redshift and absolute magnitude compared to the total number that are intrinsic bin. 

In Figure~\ref{fig:comp1} we show plots of both the completeness and contamination for each our EPOCHS fields.  These are plotted as a function of redshift and the apparent UV absolute magnitude.   We define the completeness as the number of galaxies within the \texttt{JAGUAR} simulation of each field which we retrieve through our methods of identifying high redshift galaxies, which we call N(true-positive) divided by the total number of galaxies in the true sample from \texttt{JAGUAR} with known redshifts based on the simulation, which we call N(total). This value is given by the completeness $C(z)$ :

\begin{equation}
C(z) = \frac{N(\rm true-positives(z))}{N(\rm total(z))}
\end{equation}

\noindent whereby for the contamination $K(z)$ we calculate this value by the formula, 

\begin{equation}
K(z) = \frac{N(\rm false-positives(z))}{N(\rm obs(z))}
\end{equation}

\noindent This is such that the value $N(\rm false-positives(z))$ is the number of objects detected by our methods which are not at the correct redshift as given by the \texttt{JAGUAR} catalog.  The various completeness and contamination values for each of our fields are shown in Figure~\ref{fig:comp1}.  As can be seen in this figure, our fields have a variety of completeness levels that vary with both the absolute magnitude and redshift.  In general, those which have a larger amount of darker colours in the completeness (left hand side) have a higher completeness at a fainter magnitude.  The deepest surveys JADES and NGDEEP both are complete to nearly 100\% down to apparent UV magnitudes of 29 - 29.3. Our shallower fields, including Clio and SMACS 0723 have more structure in their completeness, such that at higher redshifts, the detection process is more complete up to magnitudes of 28-28.5, but have a higher incompleteness at the lower redshifts. This is due to the nature of the lower redshift galaxies being rejection on one or more of our criteria, often due to the limited wavelength coverage around the wavelength of the Lyman-break.

The contamination shown on the right of Figure~\ref{fig:comp1} is also quite interesting and shows trends that differ slightly between fields. For the deepest fields, the contamination is very low, close to zero up to magnitude m = 29.  However, for these fields, including JADES and NGDEEP the contamination becomes quite high at about 50\% between magnitudes 29-30 for the highest redshfits, galaxies which are detected to be at redshifts $z \sim 11-13$.  Therefore at this faint galaxy range at these higher redshifts caution should be taken when analyzing galaxies within this range.   

Contamination increases at brighter magnitudes within other fields, including NEP-TDF, CEERS, and GLASS at magnitude fainter than 28.  The other PEARLS fields have a contamination which starts to become high at even fainter limits, around magnitude 27, implying that only the brightest magnitudes should be used to construct samples from these fields.

%\begin{figure}
%\centering
%\hspace{-3cm}
%\includegraphics[width=1.2\columnwidth]{NEP_Comp.pdf}
%\caption{The selection completeness simulation results conducted using the JAGUAR semi-analytical model using the conditions for the NEP survey setup. The color bar shows the fractional completeness recovered and the contamination fraction of objects with $z < 5$ entering the sample. This is used to form the completeness factor $C$
%  in Equation ??. }
%\label{fig:nep_pure}
%\end{figure}

\begin{figure}

\fig{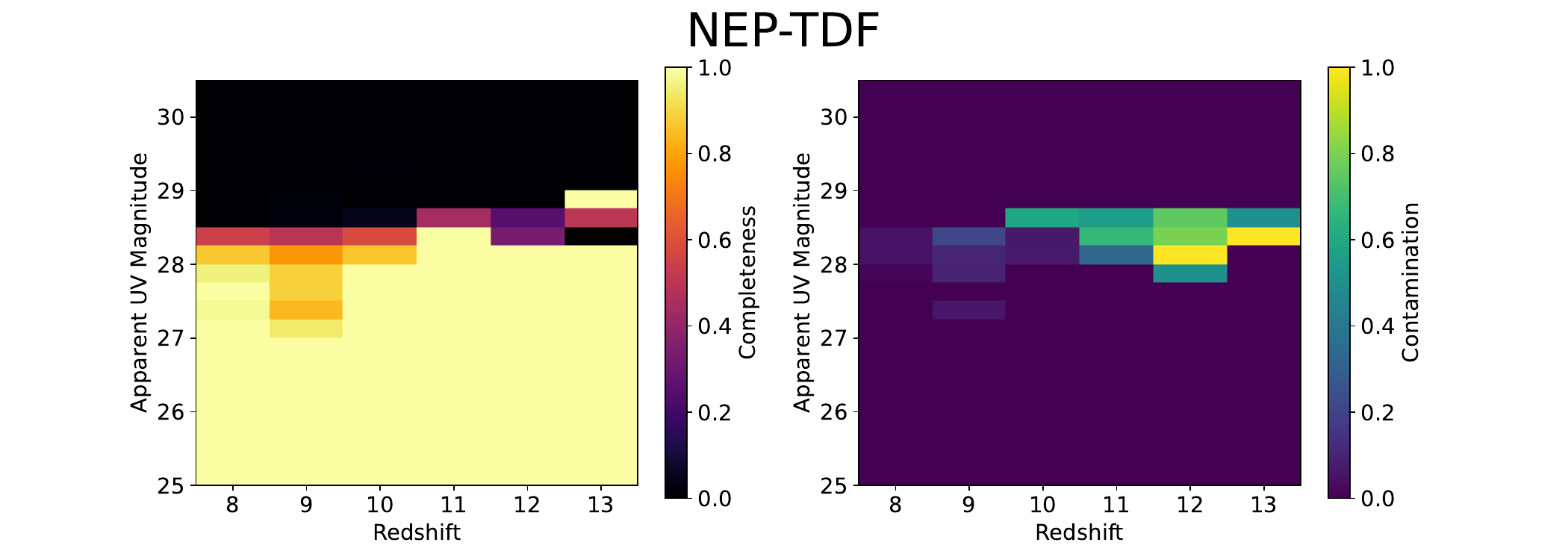}{1.05\columnwidth}{(a) Completeness and contamination for the NEP-TDF field. } 
\fig{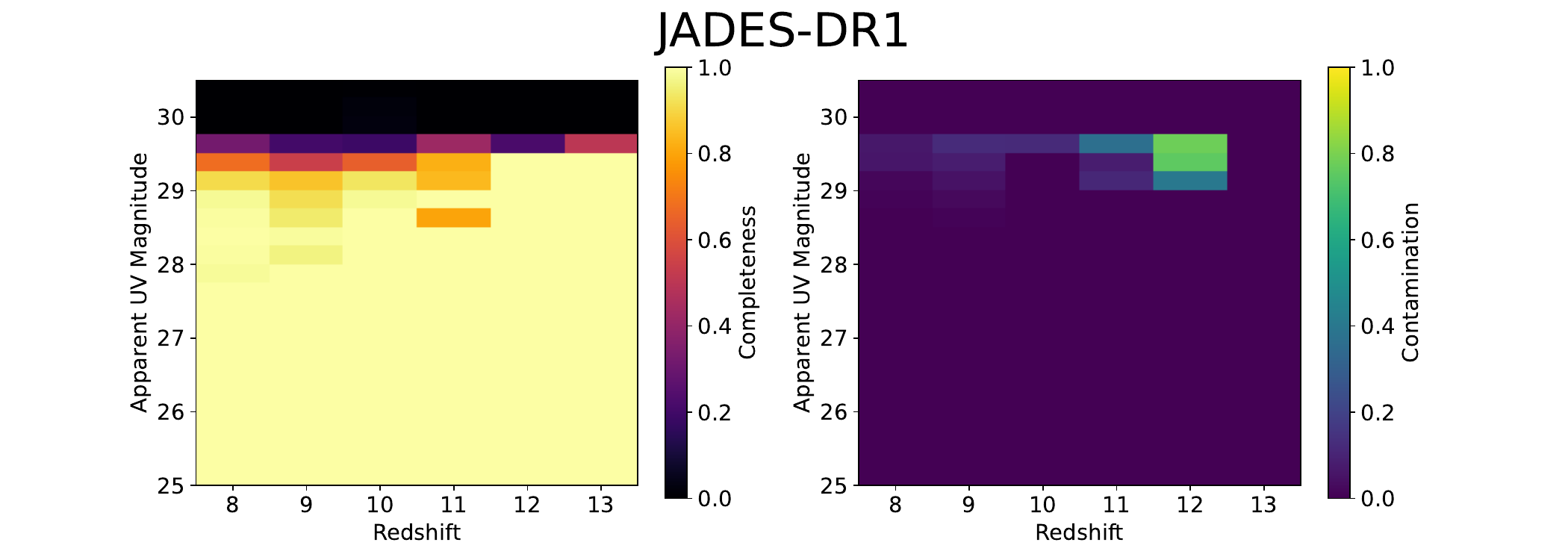}{1.05\columnwidth}{(b) Completeness and contamination for the JADES GOODS-S survey. }
\fig{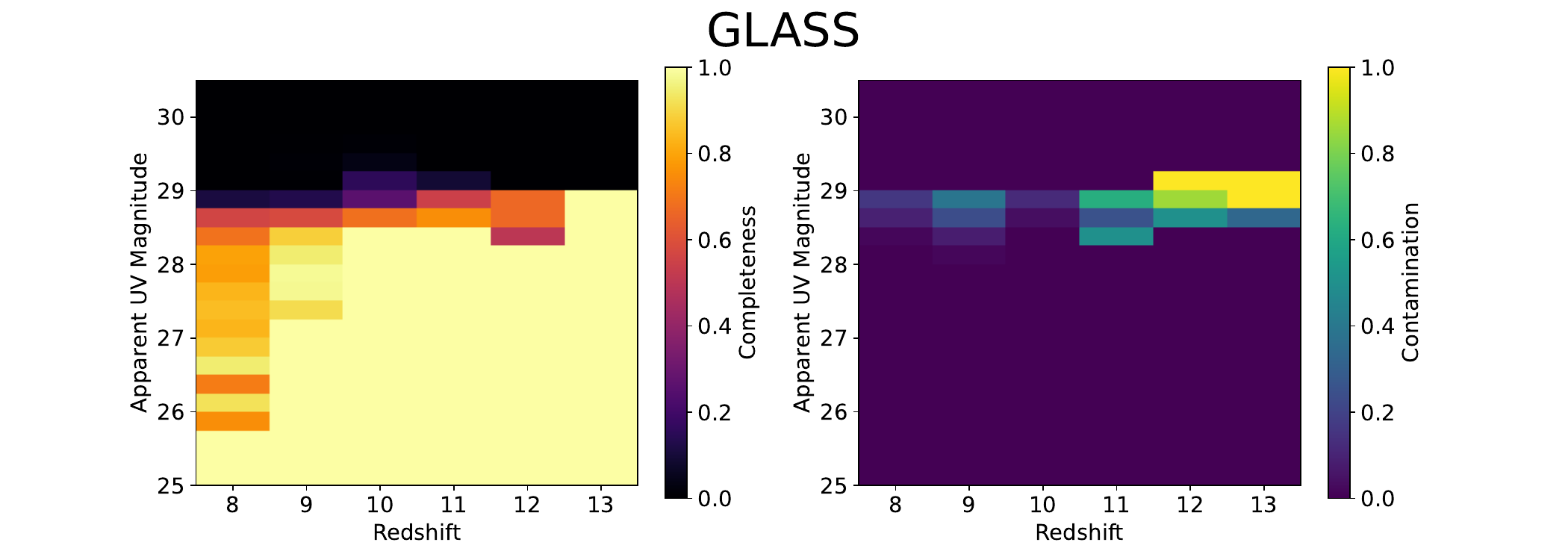}{1.05\columnwidth}{(c) Completeness and contamination for the GLASS survey. }

%\centering
%\includegraphics[width=1.2\columnwidth]{JADES1_Comp.pdf}
\caption{The selection completeness $C(z)$ and contamination $(K(z)$ simulation results conducted using the \texttt{JAGUAR} semi-analytical model using the conditions for each survey setup. The color bar shows the fractional completeness recovered and the contamination fraction of objects with $z < 5$ entering the sample. This is used to form the completeness factor $C$ in Equation B1, and $K$ in Equation B2.}
\label{fig:comp1}
\end{figure}

\begin{figure}
\figurenum{15}
\fig{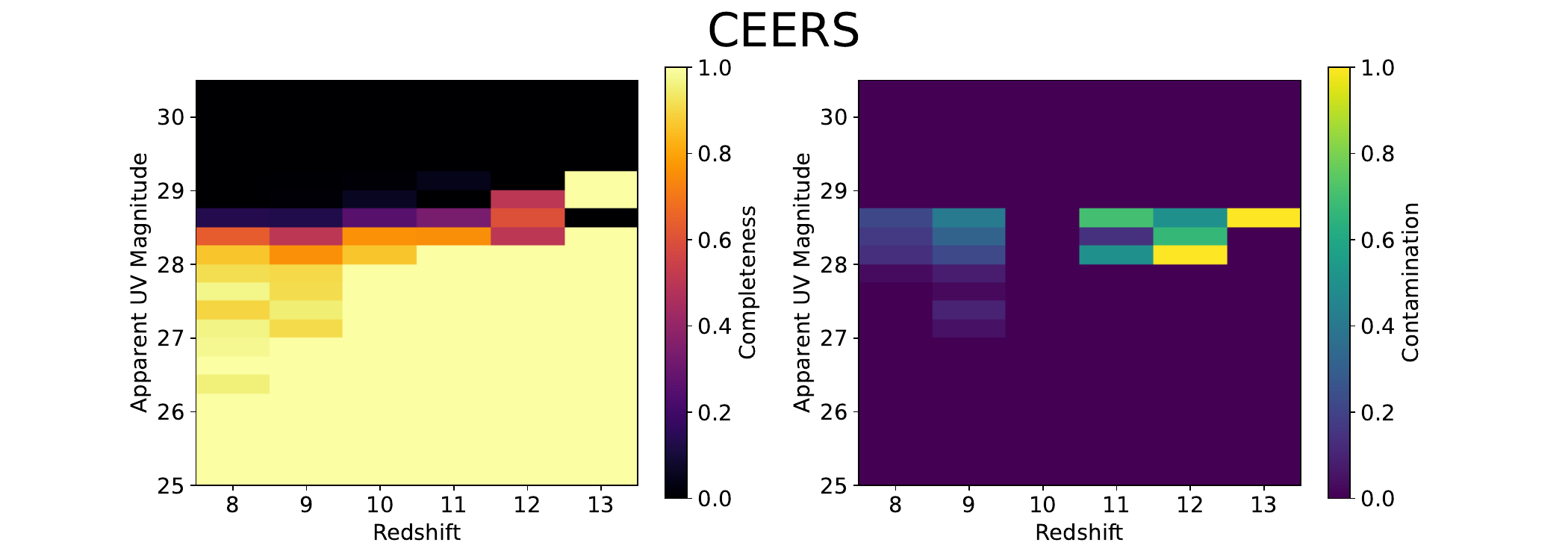}{1.05\columnwidth}{(d)  Completeness and contamination for the CEERS survey in the EGS region. }
\fig{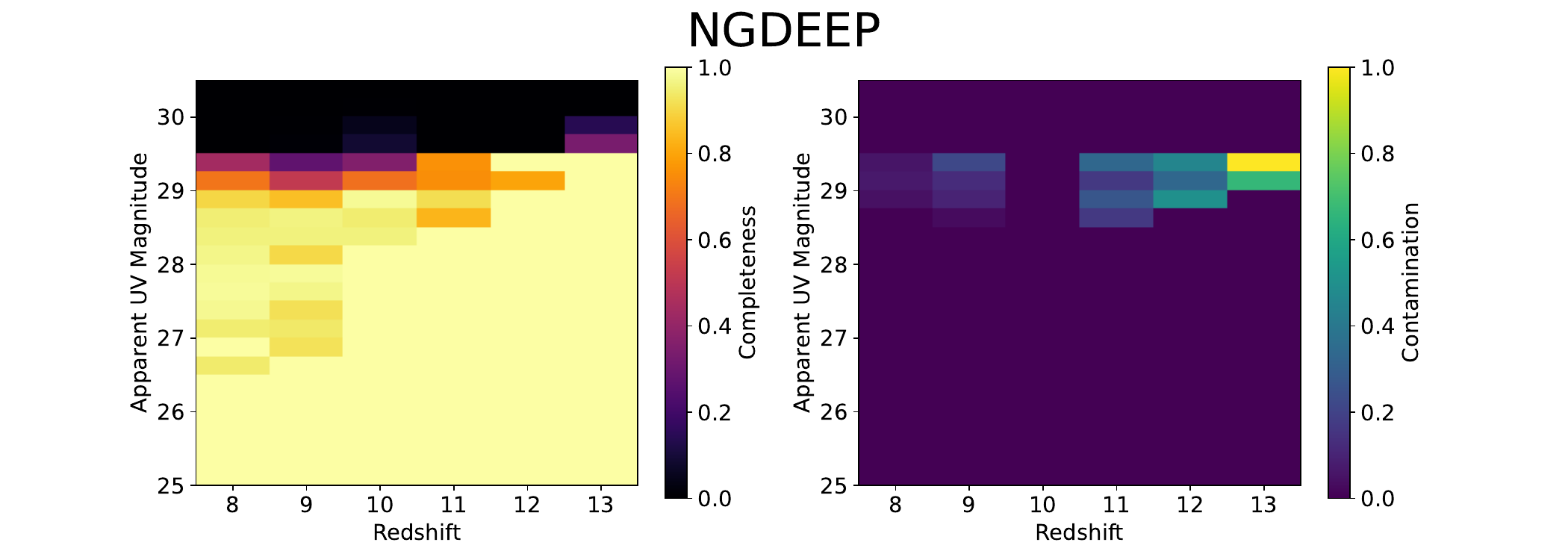}{1.05\columnwidth}{(e)  Completeness and contamination  for the NGDEEP survey. }
\fig{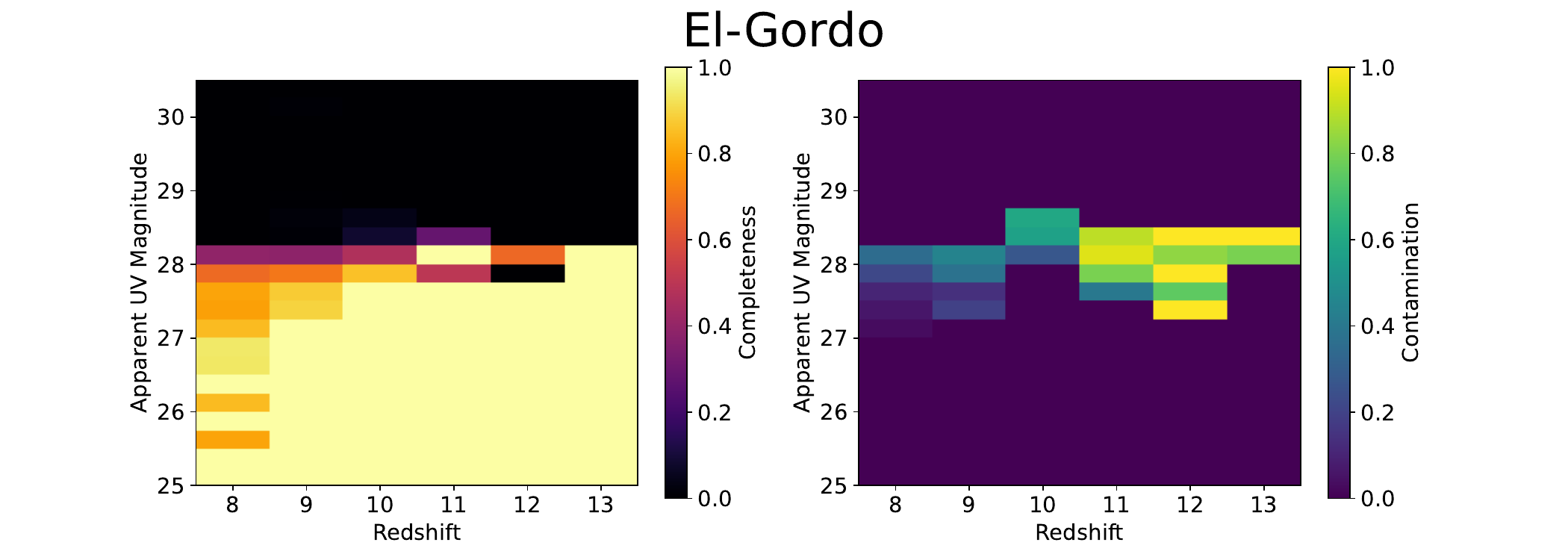}{1.05\columnwidth}{(f)  Completeness and contamination for the El Gordo field, which does not include the cluster region. }
\caption{Continued from previous page.}
\end{figure}

\begin{figure}
\figurenum{15}
\fig{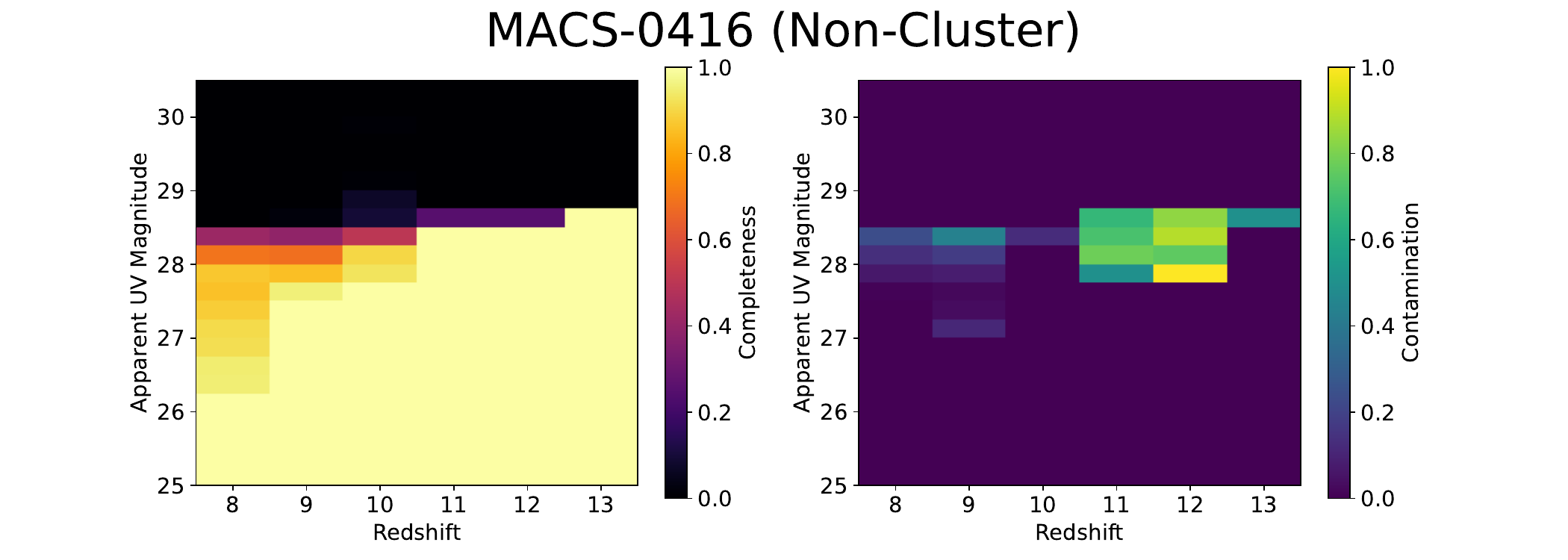}{1.05\columnwidth}{(d)  Completeness and contamination for the M0416 cluster region.  The cluster region itself is not included in this. }
\fig{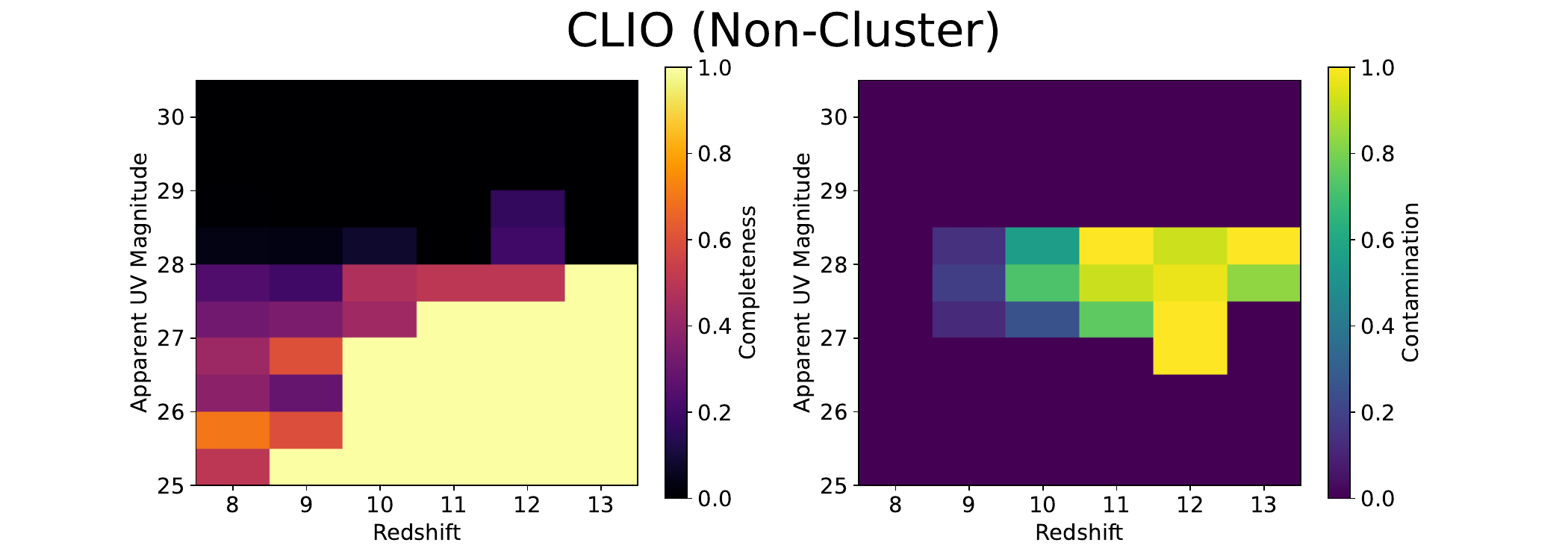}{1.05\columnwidth}{(e)  Completeness and contamination for the Clio cluster area.   The cluster region itself is not included in this.}
\fig{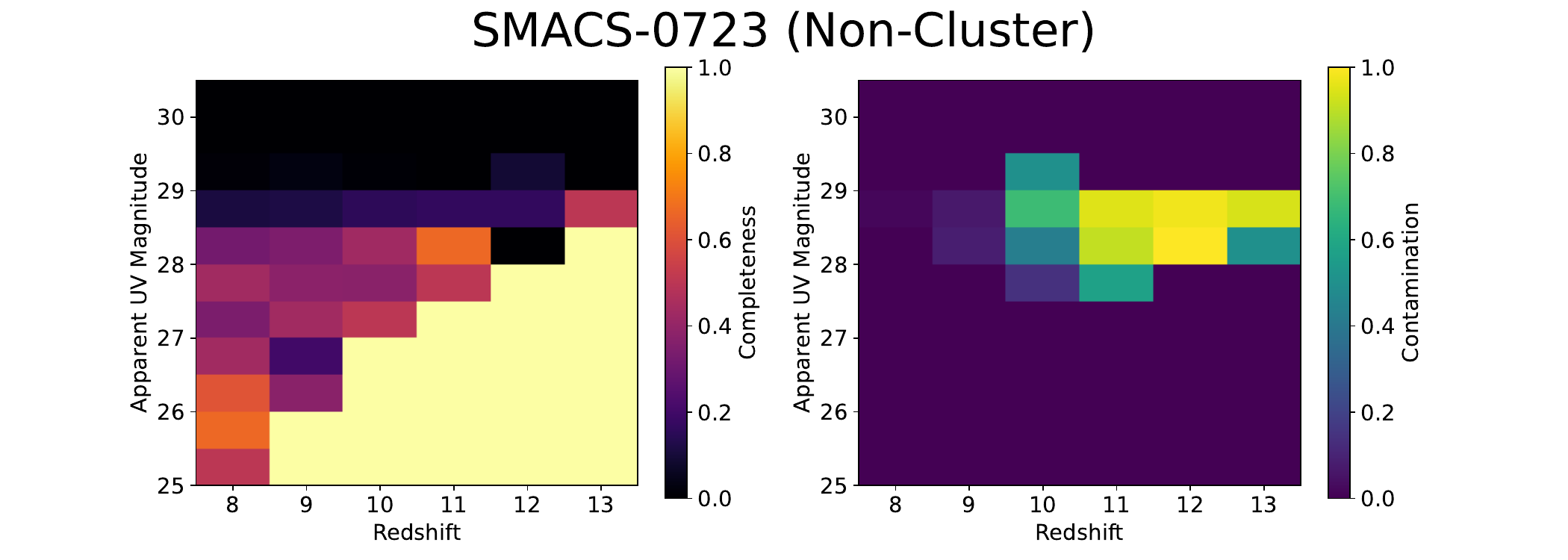}{1.05\columnwidth}{(f)  Completeness and contamination for the SMACS-0723 cluster region outside the main cluster. }
\caption{Continued from previous page.}
\end{figure}

%\begin{figure}
%\centering
%\includegraphics[width=1.0\columnwidth]{CEERS_Comp.pdf}
%\caption{ Similar to \autoref{fig:nep_pure} but for the JADES survey. }
%\label{fig:ceers_pure}
%\end{figure}

%\begin{figure}
%\centering
%\includegraphics[width=1.0\columnwidth]{GLASS_Comp.pdf}
%\caption{Similar to \autoref{fig:nep_pure} but for the GLASS field %and survey. }
%\label{fig:glass_pure}
%\end{figure}

%\begin{figure}
%\centering
%\includegraphics[width=1.0\columnwidth]{NGDEEP_Comp.pdf}
%\caption{ Similar to \autoref{fig:nep_pure} but for the NGDEEP survey. }
%\label{fig:ngdeep_pure}
%\end{figure}

%\begin{figure}
%\centering
%\includegraphics[width=1.0\columnwidth]{El-Gordo_Comp.pdf}
%\caption{Similar to \autoref{fig:nep_pure} but for the El Gordo field. }
%\label{fig:elgordo_pure}
%\end{figure}

\end{document}